\documentclass[a4paper,aps,superscriptaddress,nofootinbib]{revtex4}

\usepackage{mathtools}

\usepackage{graphicx}
\usepackage{amsmath,amssymb}
\usepackage{mathrsfs}
\usepackage{hyperref}
\usepackage[x11names]{xcolor}
\usepackage{physics}
\usepackage{ulem}
\usepackage[capitalize]{cleveref}
\usepackage{stackengine}

\usepackage{caption}
\usepackage{ragged2e}
\DeclareCaptionJustification{justified}{\justifying}
\def\hmunu{h_{\mu\nu}(t, \vec x)}
\def\Tobs{t_{\mathrm{obs}}}
\def\fmin{f_{\mathrm{min}}}
\def\fmax{f_{\mathrm{max}}}
\def\SNRthresh{\mathrm{SNR}_{\mathrm{thresh}}}
\graphicspath{
    {figures/}
}
\def\d{\mathrm{d}}
\def\be{\begin{equation}}
\def\ee{\end{equation}}
\def\bea{\begin{eqnarray}}
\def\eea{\end{eqnarray}}
\def\stat{\hat{\mathcal{S}}}

\begin{document}

CERN-TH-2024-065 \hfill ET-0343A-24

\title{Primordial gravitational wave backgrounds from phase transitions\\ with next generation ground based detectors}

\author{Chiara Caprini}
\email{chiara.caprini@cern.ch}
\affiliation{D\'epartement de Physique Th\'eorique and Center for Astroparticle Physics, Universit\'e de Gen\`eve, Quai E.~Ansermet 24, CH-1211 Gen\`eve 4, Switzerland}
\affiliation{Theoretical Physics Department, CERN, 1211 Geneva 23, Switzerland}

\author{Oriol Pujol\`as}
\email{pujolas@ifae.es}
\affiliation{Institut de F\'isica d’Altes Energies (IFAE) and The Barcelona Institute of Science and Technology (BIST)
Campus UAB 08193 Bellaterra (Barcelona), Spain.}

\author{Hippolyte Quelquejay-Leclere}
\email{quelquejay@apc.in2p3.fr}
\affiliation{Universit{\'e} Paris Cit{\'e} CNRS, Astroparticule et Cosmologie, 75013 Paris, France}

\author{Fabrizio Rompineve}
\email{frompineve@ifae.es}
\affiliation{Departament de F\'isica, Universitat Aut\`onoma de Barcelona, 08193 Bellaterra, Barcelona, Spain
\looseness=-1}
\affiliation{Institut de F\'isica d’Altes Energies (IFAE) and The Barcelona Institute of Science and Technology (BIST)
Campus UAB 08193 Bellaterra (Barcelona), Spain.}

\author{Dani\`ele A.~Steer\footnote{Corresponding author}}
\email{daniele.steer@phys.ens.fr}
\affiliation{Laboratoire de Physique de l’\'Ecole Normale Sup\'erieure, ENS, CNRS, Universit\'e PSL, Sorbonne Universit\'e, Universit\'e Paris Cit\'e, F-75005 Paris, France}
\affiliation{Universit{\'e} Paris Cit{\'e} CNRS, Astroparticule et Cosmologie, 75013 Paris, France}
\affiliation{Theoretical Physics Department, CERN, 1211 Geneva 23, Switzerland}

\vspace{1cm}
\date{\today}
\begin{abstract}

Third generation ground-based gravitational wave (GW) detectors, such as Einstein Telescope and Cosmic Explorer, will operate in the $(\text{few}-10^4)$ Hz frequency band, with a boost in sensitivity providing an unprecedented reach 
into primordial cosmology. 
Working concurrently with pulsar timing arrays in the nHz band, and LISA in the mHz band, these 3G detectors will be powerful probes of beyond the standard model particle physics {on scales $T\gtrsim 10^{7}$GeV}. Here we focus on their ability to probe phase transitions (PTs) in the early universe. 
We first overview the landscape of detectors across frequencies, discuss the relevance of astrophysical foregrounds, and provide convenient and up-to-date power-law integrated sensitivity curves for these detectors.
We then present the constraints expected from GW observations on first order PTs and  on topological defects (strings and domain walls), which may be formed when a symmetry is broken irrespective of the order of the phase transition. These constraints can then be applied to specific models leading to first order PTs and/or topological defects. 
In particular we discuss the implications for axion models, which solve the strong CP problem by introducing a spontaneously broken 
Peccei-Quinn (PQ) symmetry.  
For post-inflationary breaking, the PQ scale must lie in the $10^{8}-10^{11}$ GeV range, and so the signal from a first order PQ PT falls within reach of ground based 3G detectors. 
A scan in parameter space of signal-to-noise ratio in a representative model reveals their large potential to probe the nature of the PQ transition.  
Additionally, in heavy axion type models domain walls form, which can lead to a detectable GW background.
We discuss their spectrum and summarise the expected constraints on these models from 3G detectors, together with SKA and LISA.\footnote{Invited review for a CQG Focus Issue on the science case for next generation ground based GW detectors}

\end{abstract}

\maketitle

\newpage

\tableofcontents

\newpage

\section{Introduction on GWs from the early universe}

In contrast to photons which decouple from redshifts $z\lesssim 1100$,  gravitational waves are decoupled from the Planck time onwards, and thus a cosmological stochastic gravitational wave background (SGWB)
carries information on the universe and fundamental physics from energy scales far beyond those of particle physics accelerators and cosmic microwave background. 
Detecting and characterising the SGWB in different frequency bands from the nHz to the Hz, potentially up to the GHz, using pulsar timing arrays, space-based and ground-based experiments, as well as laboratory experiments, is therefore critically important. Their results hold promise to significantly advance primordial cosmology, gravity, and possibly particle physics beyond the standard model (BSM).  The detection of the SGWB, and its interpretation in terms of fundamental physics, constitutes one of the crucial science cases of 3G ground-based detectors.

When these 3G ground based detectors --- such as Einstein Telescope (ET) and Cosmic Explorer (CE) --- will enter operation, {Pulsar Timing Array (PTA)} experiments are expected to have detected the astrophysical stochastic GW background at nHz frequencies, Laser Interferometer Space Antenna (LISA) in the mHz band will be taking data, and the LIGO-Virgo-KAGRA (LVK) collaboration in the 1-$10^3$Hz band will have finished its O5 run (and possibly beyond).  
Until recently, only upper bounds had been established on the amplitude of a cosmological SGWB, coming from (i) BBN and CMB measurements of the total radiation energy density in the universe (see e.g.~\cite{Caprini:2018mtu}), (ii) CMB measurements of the tensor anisotropies and B-mode polarisation (see e.g.~\cite{BICEP:2021xfz}, (iii) direct detection from LVK (see e.g.~\cite{Abbott:2021xxi}).
However, the {European} PTA, NANOGrav, {Parkes PTA} and {Chinese PTA} collaborations --- analysing data collected from tens of millisecond pulsars  --- have shown strong evidence for a SGWB in the nHz band~\cite{EPTA:2023fyk, NANOGrav:2023gor,  Reardon:2023gzh, Xu:2023wog}. The precise origin of this background is currently under investigation, but amongst the best understood candidates is the  astrophysical background consisting of a population of inspiralling supermassive black hole binaries (SMBHB) in General Relativity (GR), see e.g.~\cite{EPTA:2023xxk, NANOGrav:2023hfp} and references therein. However, there is evidence that cosmological sources --- which include (see \cite{Caprini:2018mtu} for a non exhaustive list) topological defects, first order phase transitions (PTs) and vacuum fluctuations --- may even provide a better fit to the data, see~\cite{NANOGrav:2023hvm, EPTA:2023xxk}.  Further crucial experimental developments are expected in 2024 when the current data will be combined and analysed within the International PTA (see~\cite{Antoniadis:2022pcn} for the second data release). In the 2030s, the increased sensitivity of the Square Kilometer Array (SKA) \cite{Rawlings2011TheSK, Janssen:2014dka} --- including data from of order 50 pulsars or more --- will allow to precisely determine the properties of this SGWB in the nHz band. 

{GW sources operating in the early universe give rise to
SGWBs because they are homogeneously
and isotropically distributed over the entire
universe, and/or correlated on scales much
smaller than the detector resolution.}
Here we focus on early universe cosmological phenomena which can generate a SGWB in the $\text{few}-10^4$Hz frequency band, but taking into account the experimental landscape in all frequency bands. As will become clear, understanding and characterizing the cosmological SGWB is complex, 
both from a theoretical aspect (its very existence depends on unknown BSM physics) and also as regards the modelling of the source (the evolution and properties of the sources are in some cases difficult to analyse, sometimes leading to orders of magnitude uncertainties in the resulting SGWB). An additional uncertainty comes from the unknown cosmological evolution of the Universe at the high temperatures that correspond to the frequency band of interest, which would importantly affect the dilution of primordial GWs. Nonetheless, we shall see that several well-motivated BSM mechanisms will be importantly probed by 3G dectectors.

The SGWB is expressed in terms of the fraction of the critical density in GWs per logarithmic interval of (observed) frequency,
\be
    \label{eq:basic}
    \mathrm{\Omega}_{\mathrm{GW}}(t_0,f) = \frac{8 \pi G}{3H_0^2 } ~ \frac{\d \rho_{\mathrm{GW}}}{\d \ln f} (t_0, f)\,,
\ee
where $H_0$ is the Hubble constant, and ${\d \rho_{\rm{GW}}}/{\d \ln f} $ is the power spectrum, i.e.~the energy density in GWs per logarithmic unit frequency observed today. 
Its shape depends on the properties of the GW source. ``Short duration'' 
sources emit GWs during less than one Hubble time, or up to a few Hubble times, and include first order PTs \cite{Witten:1984rs,Hogan:1986qda,Kosowsky:1991ua,Kosowsky:1992vn,Kamionkowski:1993fg}, annihilation of cosmic domain walls~\cite{Preskill:1991kd, Chang:1998tb, Gleiser:1998na} (see~\cite{Ferreira:2024eru} for a recent update), and the collapse of large density perturbations~\cite{Matarrese:1993zf, Matarrese:1997ay, Noh:2004bc, Carbone:2004iv, Nakamura:2004rm, Baumann:2007zm} (see also~\cite{DeLuca:2019ufz, Inomata:2019yww, Yuan:2019fwv}). Correspondingly, the SGWB spectrum is peaked at a characteristic frequency (see e.g.~\cite{Caprini:2018mtu})
\begin{equation}
    f = \frac{a_*}{a_0} f_* = H_{*,0} \frac{f_*}{H_*}\,, 
    ~~~~~~
    \mathrm{with}~~
    H_{*,0}  = \frac{a_*}{a_0} H_* \simeq 1.65 \times 10^{3} \:\text{Hz}\: \left( \frac{T_*}{10^{10} \: \text{GeV}} \right) \left( \frac{g_*}{100} \right)^{\frac{1}{6}}\,,
    \label{eq:s}
\end{equation}
where $f_* \geq H_*$ is the frequency with which the GW was produced at time $t_*$ 
when the universe had temperature 
$T_*$. $H_*$ and $a_*$ are 
respectively the Hubble parameter and scale factor at that time, and $H_{*,0}$ is $H_*$ redshifted to today. 
The expression for $H_{*,0}$ assumes the standard $\Lambda$CDM cosmology with $g_*$ the effective number of relativistic degrees of freedom. Below this frequency 
there is a model-independent tail {$\propto f^3$}
(see~\cite{Caprini:2009fx}, and~\cite{Franciolini:2023wjm} for a recent discussion of SM effects when this tail is in the 
nHz frequency band), {while above this frequency the spectrum decays in source-dependent way, in general with a negative power law. Short duration sources may provide signals peaked in the frequency band of a specific detector, turning it into a direct probe of the early universe physics happening at energy scale $T_*$.} 
``Long duration'' sources, which include cosmic strings,  emit GWs continuously from their formation 
$t_f$ until today $t_0$,
and generate a SGWB which can cover decades in frequency. In this case, constraints in different frequency bands are particularly important. For example, as we will see below, if (stable) cosmic strings transpire to be a viable source for the PTA signal, then in ET/CE the corresponding SGWB signal should be observed with a signal-to-noise ratio (SNR) of over 1000.  A non-observation of such a signal by 3G detectors would lead to crucial information on the properties of symmetry breaking PTs --- during which cosmic strings may form.

According to Eq.~(\ref{eq:s}), the $\mathcal{O}(\mathrm{nHz})$ frequencies to which PTAs are sensitive coincide with those of GWs possibly sourced at the epoch of the chiral symmetry breaking and quark-gluon confinement (QCDPT), happening around $\sim$150 MeV. 
The LISA frequency band, on the other hand, include the energy scale of the electroweak (EW) symmetry breaking $\sim$100 GeV, and 3G ground-based interferometers correspond to scales $10^6 \: \mathrm{GeV} \lesssim T_* \lesssim 10^{10} \: \rm{GeV}$.  
The interest in high-frequency GHz GW detectors \cite{Aggarwal:2020olq} is not only that no astrophysical GHz sources are known, but furthermore that they correspond to cosmological sources operating at Grand Unification (GUT) scales.    
The concrete, well motivated BSM model on which we will focus in the second part of this paper is the Peccei Quinn (PQ) model, proposed to solve the strong CP problem. 
This model can feature multiple GW sources, as
a global U(1) symmetry is spontaneously broken in a possibly strong first order PT, further leading to the formation of topological defects including cosmic strings and (in a further PT) domain walls. {As we will see, the typical energy scales of this model render it the perfect candidate, frequency-wise, for detection with 3G Earth-based interferometers.} 
{Furthermore, it can also produce SGWB  signals spanning across  frequency bands, perfect candidates for coincident detection at several detectors.
Another well motivated extension of the SM that could give rise to a SGWB at 3G Earth-based interferometers is the $U(1)_{B-L}$ extension, see e.g.~\cite{Okada:2018xdh,Hasegawa:2019amx} for examples in which this symmetry is broken at high scale, compatible with 3G interferometers.}

Throughout this review we use the SNR as a shortcut with which to assess whether different primordial SGWBs can be detected. The calculation of the SNR, however, depends on the detector/detector network, as well as on the relevant astrophysical foregrounds in the given detector frequency band. For this reason, we begin in section \ref{section:SNR} with a brief review of the different detectors, their foregrounds, and also give explicit expressions for the computation of SNR which will be used throughout this work. We also determine the power-law integrated sensitivity curves for these detectors, including astrophysical foregrounds.

\begin{figure}[t]
   \centering
    \includegraphics[width=0.8\textwidth]{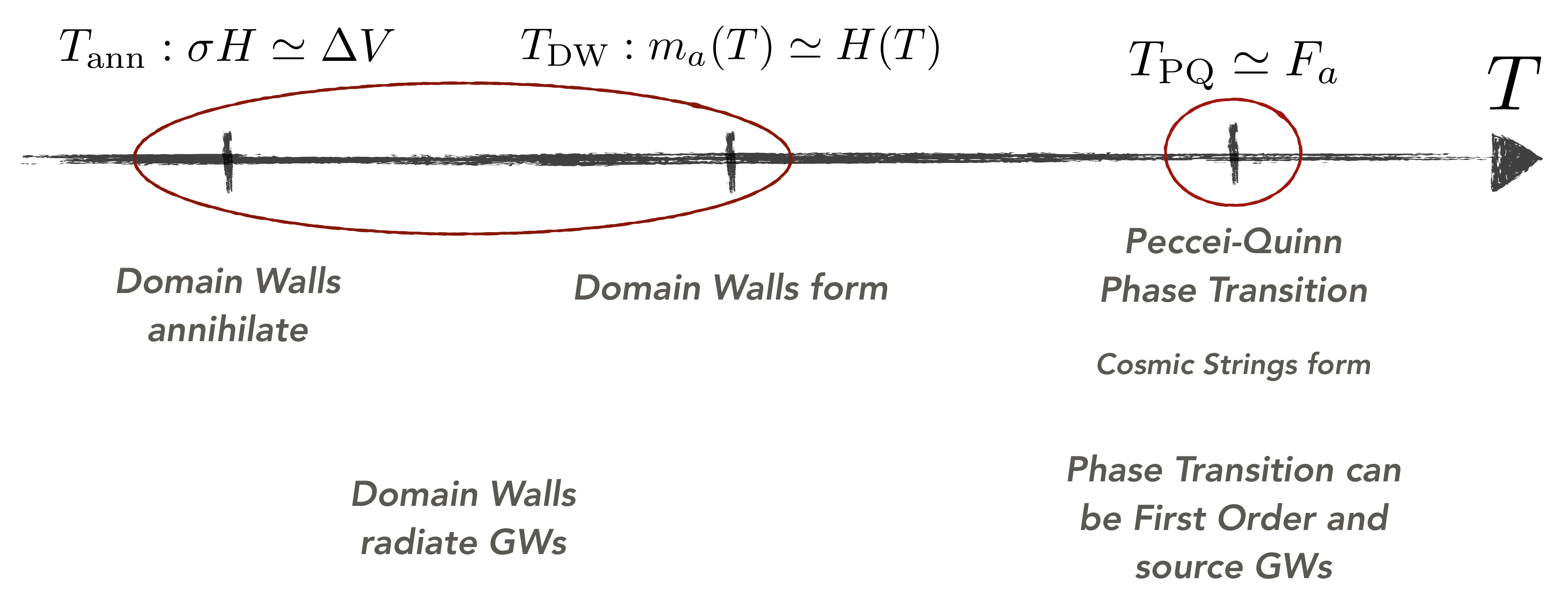}
    \caption{Sketch of the cosmological timeline of the (post-inflationary) PQ mechanism, highlighting the different possible sources of GWs, namely first order PT and topological defects. Here $F_a, m_a$ are the axion decay constant and mass, $\sigma$ and $\Delta V$ are parameters of domain wall annihilation, which will be described in Section~\ref{sec:DW}.}
    \label{fig:PQrecap}
\end{figure}

Our aim is then to give a detailed overview of two relevant cosmological SGWB sources, namely first order PTs (section \ref{sec:PT}), and topological defects (section \ref{sec:Defects}). 
{In section \ref{sec:PT}, we review the PT dynamics and the associated GW sources, namely bubble and/or relativistic fluid shells collisions, sound waves, and turbulent bulk fluid motion. We motivate the physics determining the shape of the SGWB spectrum, and provide placeholder templates for the SGWB signals from the aforementioned PT-related GW sources. 
In all cases, these templates have to be considered preliminary, as they are the subject of ongoing research.}
Detection of such a signal would constitute a significant test of new physics beyond the standard model of particle physics, {and we provide results on the potential of 3G detectors, but also of LISA and SKA, to probe the parameter space of first order PTs.}

Irrespective of the order of the phase transition, if a symmetry is broken in the PT, then topological defects may form. In section \ref{sec:Defects} we review the physics of these extended objects which may be line-like strings or domain walls.  Line-like local cosmic strings are the most well studied, and are thought to form in many BSM models, see e.g.~\cite{Jeannerot:2003qv}. 
Global strings and domain walls may form in e.g.~the PQ model. Determining the SGWB spectra from topological defects is a complex task because of the huge disparity of scales between the defect width and the Horizon size, as well as their non-linear evolution: we review some of the difficulties and uncertainties involved in calculating the SGWB spectra in section \ref{sec:Defects}.

In the last section \ref{sec:PQ}, we apply all these results to axion models, which solve the strong CP problem of QCD by introducing a spontaneously broken PQ symmetry.  Figure \ref{fig:PQrecap} sketches the cosmological timeline of the (post-inflationary) PQ mechanism, and highlights the two possible sources of GWs (details will be given in section \ref{sec:PQ}). First, the PQ phase transition can be of first order type. Then at a second PT, domain walls may form and source GWs which may be abundant enough to be observable (in heavy QCD axion scenarios). Finally, global strings are also formed in these models, but as we discuss in section \ref{subsec:global}, they mainly decay into axions rather than GWs.  In section \ref{sec:PQ} we introduce different realisations of the PQ mechanism, and then scan parameter space of SNR
in these models, thus showing the large potential of 3G GW detectors to probe the nature of the PQ transition. 
We conclude in section \ref{sec:conc}.

{Note: this review has a broad scope, and thus some topics are necessarily treated in less detail than could be done otherwise. Throughout the review, we 
provide many references, where more details may be found. Here we highlight some books and other reviews which have overlap with the different 
sections of this review. For general books covering GWs see e.g.~\cite{Maggiore1,Maggiore2,poisson}. Reference e.g.~\cite{Caprini:2018mtu} is a detailed review on the cosmological SGWB, as well as different sources of that background. SGWB detection methods for different detectors are discussed in e.g.~\cite{Romano:2016dpx, Thrane_Romano, vanRemortel:2022fkb}. Our review of the physics of first order phase transitions leading to a SGWB can be complemented with e.g.~\cite{Hindmarsh:2020hop,Caldwell:2022qsj,Athron:2023xlk,Roshan:2024qnv} for more details. 
Topological defects are the subject of different books e.g.~\cite{Vilenkin:2000jqa} and reviews e.g.~\cite{Hindmarsh:1994re,Vachaspati:2015cma}. The SGWB generated by standard local cosmic strings is discussed in detail in \cite{Auclair:2023brk}.  {For global cosmic strings and domain walls the situation is more uncertain and there has been renewed interest and progress in recent years (see e.g.~\cite{Saikawa:2017hiv} for an older review on GWs from domain walls). Our review also emphasizes the opportunity of extracting implications for QCD axion models from primordial GW backgrounds at 3G detectors.}}


\section{\label{section:SNR}Context: ET, PTA and other detectors}

We begin by discussing current and future GW detectors, and in particular their ability --- quantified in terms of SNR --- to detect cosmological SGWBs.  Astrophysical sources that are not individually resolvable by a given GW detector, either because they have low SNR, or because they overlap in time and frequency domains, constitute a foreground that must be accounted for before any predictions can be made on the cosmological SGWB detection. 
These are, e.g.~populations of black hole binaries or neutron star binaries, whose merger rates have been estimated by LVK and PTA \cite{PhysRevD.100.061101, Agazie_2023_SMBHB, EPTA:2023xxk}. 
We explain how to include these foregrounds, and hence update predictions for the sensitivity of different experiments.

In the following, we first briefly review the general methods used to infer the presence of SGWBs \cite{Romano:2016dpx, Thrane_Romano, vanRemortel:2022fkb}. Then we focus on the specifics of each type of GW detector (3G Earth-based, LISA and then PTAs) and give read-to-use formulae for the SNR in each case, taking into account the relevant astrophysical foregrounds.

\subsection{SGWB detection methods}

Our starting point is the standard decomposition (see e.g.~\cite{Romano:2016dpx}) of the transverse-traceless (TT) metric perturbation $\hmunu$ as a superposition of plane waves propagating in direction $\hat k$:
\begin{equation}
    \label{eq: h for SGWB}
    \hmunu = \int \d^2 \Omega_{\hat k} \int_{-\infty}^{+\infty} \d f \: \Tilde{h}_P (f, \hat k) e^P_{\mu\nu}(\hat k) e^{i 2\pi f (t - \hat k \cdot \vec x / c)},
\end{equation}
where 
$P \in [+, \times]$ labels the polarizations  with corresponding tensors $e^P_{\mu\nu}$, repeated $P$ symbols mean summation, and due to the stochastic nature of the signal, the Fourier components $\Tilde{h}_P (f, \hat k)$ are random variables. For the SGWB from cosmological sources, it is reasonable to assume that the signal will be stationary, isotropic, unpolarized and Gaussian-distributed \cite{Caprini:2018mtu}, in which case it is uniquely defined by the second moment 
\begin{equation}
    \label{eq: definition of S_h}
    \langle \Tilde{h}_P (f, \hat k), \Tilde{h}_{P'} (f', \hat k') \rangle = \frac{1}{16\pi} S_h(f) \delta(f - f') \delta_{PP'} \delta^{(2)} (\hat k - \hat k').
\end{equation}
Here $S_h(f)$ is the one-sided strain power spectral density (PSD) of the SGWB, related to the
fractional energy density in GWs, see Eq.~\eqref{eq:basic}, by
\begin{equation}
    \label{eq: fractional energy density def}
    h^2 \Omega_{\rm{GW}}(f) = \frac{2 \pi^2}{3 H^2} f^3 S_h(f)
\end{equation}
where $H = 100 \: \rm{km / s / Mpc}$ with $H_0 = h \times H$. In the following we use the Planck 2018 value \cite{Aghanim:2018eyx} $h=0.67$.

A GW detector does not directly measure $\hmunu$. Its response $h(t)$ is a contraction of $\hmunu$ with a tensor response function which depends on the geometry of the detector (see Eq.~\ref{eq: general detector response} and \cite{Romano:2016dpx}). Hence (see Appendix \ref{app: ORF})
\begin{eqnarray}
       \label{eq: signal PSD def}
    \langle \tilde h (f) \tilde h^* (f') \rangle &= \frac{1}{2} \delta(f - f') S_h (f) \mathcal{R}(f),
\end{eqnarray}
where the transfer function $\mathcal{R}(f)$ is the sky and polarization averaged response of the detector to plane waves of frequency $f$ (see Eq.~\eqref{eq: response function def} and \cite{Finn:2008vh}). 
For detectors with one channel, the data take the form $d(t) = n(t) + h(t)$, where we assume that the noise in the detector channel $n(t)$ 
is stationary and Gaussian, characterised by
\begin{align}
    \label{eq: noise PSD def}
    \langle \tilde n (f) \tilde n^* (f') \rangle &= \frac{1}{2} \delta(f - f') P_n (f), 
\end{align}
where $P_n$ is the one-sided detector noise PSD. We now turn to the search for a SGWB in the data.

\subsubsection{\label{subsec:SNRxcorr} SNR with cross correlations: two or more detectors with uncorrelated noise.}

The initial idea from \cite{Michelson_1987, Flanagan_PhysRevD.48.2389}, subsequently 
used in the LVK detector network, is to consider the cross correlation between two (or more) detectors \cite{Allen:1997ad}
\begin{equation}
    \label{eq: signal corr with 2 detectors}
    \stat = \int_{-\frac{\Tobs}{2}}^{\frac{\Tobs}{2}} \d t \int_{-\frac{\Tobs}{2}}^{\frac{\Tobs}{2}} \d t' d_1(t) d_2(t') Q(t - t'),
\end{equation}
where $\Tobs$ is the common observation time for detectors 1 and 2, and $Q$ is a filter function that will be chosen to maximise the SNR of the SGWB.  
{\it If} the noise in the two detectors is uncorrelated, then 
\begin{equation}
    \label{eq: correlation 12 expectation value}
    \langle \stat \rangle = \int_{-\frac{\Tobs}{2}}^{\frac{\Tobs}{2}} \d t\int_{-\frac{\Tobs}{2}}^{\frac{\Tobs}{2}} \d t' \langle h_1(t) h_2(t') \rangle Q(t - t') =  \frac{\Tobs}{2} \int_{-\infty}^{+\infty} \d f \: \Gamma_{12}(f) S_h(f) \tilde Q (f),
\end{equation}
where to get the last equality we have substituted \cref{eq:SgammaIJ}, and we have assumed one can send the bounds of the integral over $t-t'$ to $\pm \infty$, {as one does not expect a contribution to the statistics at large temporal separations compared to the light travel time between the two detectors \cite{Allen:1997ad}.}
The overlap reduction function (ORF) $\Gamma_{12}(f)$ between detectors 1 and 2 quantifies the correlation between the detector responses across frequency (see Eq.~\eqref{eq: ORF definition}). 

One can then compute the standard deviation of the statistic $\sqrt{\langle \stat^2 \rangle - \langle \stat \rangle^2}$ and determine the filter function $\tilde Q$ that maximises the SNR \cite{Allen:1996vm}. 
Working in the weak signal approximation\footnote{One might question the validity of finding very high SNR for certain SGWB sub-parameter spaces later in the review. However, even if the value of the SNR is incorrect due to the weak signal regime not being applicable in this case, it still accurately represents the detectability of the signal that will stand out over the detector noise without the need for any time integration \cite{Allen:1996vm}.} and for two detectors, gives \cite{Romano:2016dpx}
\begin{equation}
    \label{eq: two detectors SNR}
    \mathrm{SNR}_{(2)} = \sqrt{2 \,\Tobs \int_{\fmin}^{\fmax} \d f \frac{\Gamma_{12}(f)^2 S_h(f)^2}{P_{n_1}(f)P_{n_2}(f)}},
\end{equation}
where the integral is over the (positive) detector frequency band.
Proceeding analogously for a network of $N$ detectors with uncorrelated noises gives
\begin{equation}
    \label{eq: N detectors SNR}
    \mathrm{SNR}_{(N)} = \sqrt{2 \,\Tobs \int_{\fmin}^{\fmax} \d f \sum_{I=1}^N \sum_{J > I}^N \frac{\Gamma_{IJ}(f)^2 S_h(f)^2}{P_{n_I}(f)P_{n_J}(f)}},
\end{equation}
namely the sum of the squared SNR of each detector pair \cite{PhysRevD.100.104028}. Note that the assumption of uncorrelated noise in $N$ detectors applies to PTA experiments\footnote{{This assumption applies well to intrinsic pulsar noises. However, some noise sources, such as clock, planetary ephemeris errors or poorly modeled solar wind, could imprint common correlated noises in pulsar data \cite{Tiburzi2015ASO}.}}, where each pulsar can be considered as a detector.  It does not apply to the three co-located interferometers composing the Einstein Telescope, which we treat separately below. Note also that the SNR scales with the square root of the observing time $\Tobs$.

\subsubsection{\label{subsec:SNRautocorr}SNR with auto-correlations: one detector}

Some future GW detectors like LISA may not operate simultaneously with an other detector scanning over the same frequency band. In this case, the method proposed to estimate a theoretical SNR is to use the auto-correlation of the detector with itself, assuming perfect knowledge of the instrumental noise, characterised uniquely by (if Gaussian and stationary)
\begin{equation}
    \label{eq: corr matrix one detector}
    \langle n_1(t) n_1(t') \rangle = \frac{1}{2} C_n(t' - t).
\end{equation}
As explained in \cite{Smith:2019wny} one then builds a new statistic $\stat_{\rm{auto}}$, evaluating the excess of auto-correlated power in the data streams
\begin{equation}
    \label{eq: autocorr stat}
    \stat_{\rm{auto}} = \int_{-\frac{\Tobs}{2}}^{\frac{\Tobs}{2}} \d t \int_{-\frac{\Tobs}{2}}^{\frac{\Tobs}{2}} \d t' \left[ d_1(t) d_1(t') - \frac{1}{2}C_n(t' - t)\right]Q(t - t').
\end{equation}
Then following the same methodology as above, the associated SNR is given by \cite{Thrane_Romano, Smith:2019wny},
\begin{equation}
    \label{eq: SNR single detector}
    \mathrm{SNR}_{(1)} = \sqrt{\Tobs \int_{\fmin}^{\fmax} \d f \: \frac{\mathcal{R}(f)^2 S_h(f)^2}{P_n(f)^2}}.
\end{equation}
It is important to note here that this SNR is primarily theoretical, as it assumes perfect knowledge of the noise, which may not be achievable in practice in GW detectors. However, for LISA, it will be possible to combine measurements from different links to construct 
{Time-Delay Interferometry combinations that are less sensitive to the signal, as for example the Sagnac interferometer \cite{Hogan:2001jn}, allowing thereby to characterise the instrument
noise \cite{Prince:2002hp,Muratore:2021uqj}. 
Such \textit{null channels} may then be utilized to characterize the detector noise in the corresponding frequency band. 
The strong assumption of perfect noise knowledge could then be more justified \cite{vanRemortel:2022fkb}, see section \ref{subsec:LISA}. However, realistic detector configurations may compromise the use of null channels \cite{Muratore:2022nbh}.}

\subsubsection{\label{subsec:unifiedSNR}Unified SNR expressions and PLISC}

The different expressions Eqs.~\eqref{eq: two detectors SNR}, \eqref{eq: N detectors SNR}, and \eqref{eq: SNR single detector} can be combined and written as
\begin{equation}
    \label{eq: general SNR expression}
    \mathrm{SNR}_{(N)} = \sqrt{2 \,\Tobs \int_{\fmin}^{\fmax} \d f \: \frac{S_h(f)^2}{S_{\mathrm{eff}}^{(N)}(f)^2}},
\end{equation}
where $S_{\mathrm{eff}}^{(N)}$ is the {\it effective} strain noise PSD  of the GW detector network, composed of $N$ (independent) individual detectors, given by
\begin{align}
    \label{eq: effective strain noise PSD of 1 detectors}
    S_{\rm{eff}}^{(1)}(f) &= \frac{\sqrt{2} \: P_n(f)}{\mathcal{R}(f)}, \text{\: for \:} N = 1 \\
    \label{eq: effective strain noise PSD of N detectors}
    S_{\mathrm{eff}}^{(N)}(f) &= \left[ \sum_{I=1}^N \sum_{J > I}^N \frac{\Gamma_{IJ}(f)^2}{P_{n_I}(f)P_{n_J}(f)}\right]^{-1/2}, \text{\: for \:} N > 1.
\end{align}
It is clear from Eq.~\eqref{eq: general SNR expression} that the detectability of a SGWB depends on both the detector sensitivity $S_{\mathrm{eff}}^{(N)}$, and the amplitude of the GW signal $S_h$ (which itself typically depends on the number of parameters of a given SGWB source). Furthermore, it depends on the observing time.
In order to estimate the detection prospects of potential SGWBs, rather than working with $S_{\mathrm{eff}}^{(N)}$, it is therefore standard practice to introduce the Power Law Integrated Sensitivity Curve \cite{Thrane_Romano} (PLISC) of a given detector (network), which accounts for the integration over time and frequency inherent in the SNR computation formula Eq.~\eqref{eq: general SNR expression}. The principle is to compute, for a set of power law spectra $S_{\rm{PL}}(f;\gamma) = S_\gamma \left(f / f_{\rm{ref}} \right)^\gamma$, the amplitude $S_\gamma$ associated with each spectral index $\gamma$
that would result in a given SNR threshold value $\SNRthresh$. 
Then, the PLISC is simply given by the envelope of the superposition of all these power law spectra, namely
\begin{equation}
    \label{eq: PLISC def}
    S_{\rm{PLISC}}(f) = \max_i \left[ S_{\gamma_i} \left(\frac{f}{f_{\rm{ref}}} \right)^{\gamma_i} \right].
\end{equation}
Note that the PLISCs depends on the observation time $\Tobs$ chosen for the specific detector.
In this paper we scan over $\gamma \in [-50, 50]$, and fix $\SNRthresh = 5$ for all GW detectors.

\subsection{\label{sec:detectors}Future GW detectors and astrophysical foregrounds}

We now briefly describe the different next-generation GW detectors we consider their  $S_{\mathrm{eff}}^{(N)}$, as well as how the relevant astrophysical foregrounds are included. The resulting PLISCs for each detector are shown in \cref{fig:CS+DW_spectra_PTA_values}.

\begin{table}[h]
\centering
\begin{tabular}{|c||c|c|c|c|}
\hline
Detector & $\Tobs$ [years] & SGWB detection method & AFG & $\rm{SNR}_{\rm{AFG}}$ \\
\hline
 CE & 1 & Cross-correlation of $2$ L-shape interferometers & CBM & 72 \\
 ET & 1 & Cross-correlation excess of $3$ co-located V-shape interferometers & CBM & 50 \\
 LISA & $0.82 \times 4.5$ & Auto-correlation excess of $A$ and $E$ TDI variables & WDB $\|$ CBM & SSR $\|$ {448} \\ 
 SKA & 15 & Cross-correlation of pulsar pairs & SMBHB & SSR \\
\hline
\end{tabular}
\caption{
Summary of the future GW detectors considered in this review, the observation time we have chosen in our analyses, and the detection method we have adopted to detect a cosmological SGWB. 
The relevant astrophysical foregrounds (AFG), which will impact each detector's sensitivity and their corresponding SNRs in the weak signal regime (WSR) are also provided. 
ET and CE are primarily influenced by the compact binary merger (CBM) foreground, predominantly from binary neutron stars \cite{Branchesi_2023, Bellie:2023jlq}, with an amplitude several orders of magnitude lower than the detector noise, validating the WSR assumption. Conversely, LISA is expected to directly observe the galactic white dwarf binary foreground (WDB), surpassing the detector noise without any temporal integration, placing this signal in the Strong Signal Regime (SSR) \cite{Peter_Bender_1997, Adams:2013qma}. Instead, the foreground originating from extra galactic CBM --- from both the populations of white dwarf binaries and binary black holes --- is expected to be approximately one order of magnitude lower than the LISA noise (WSR) \cite{Staelens:2023xjn, Babak:2023lro}. For SKA, the foreground arising from the population of Super Massive Black Hole Binaries (SMBHB) is expected to dominate the pulsar noises at low frequencies, constraining the prospects for detecting a cosmological SGWB \cite{Babak:2024yhu}. }
\label{tab:detectors}
\end{table}

\subsubsection{Earth-based detectors}

In the few to thousands of Hertz band, the next generation GW detectors we consider on Earth are Einstein Telescope (ET) \cite{Sathyaprakash:2012jk} and Cosmic Explorer (CE) \cite{Reitze:2019iox}, proposed to be operational around 2035. Their sensitivity is increased by about one order of magnitude with respect to Advanced LVK detectors \cite{Sathyaprakash:2011bh, LIGOScientific:2016wof}: this
should allow the identification and subtraction of numerous individual mergers, thereby mitigating the astrophysical foregrounds \cite{Zhou:2022otw}.
For both detectors, when calculating SNRs, we choose an observation time span of $\Tobs = 1$ year.

Regarding ET, we work with the current baseline configuration consisting of a triangular arrangement of three Michelson interferometers with 10km arms, situated underground to minimize seismic noise. 
Since the three V-shape interferometers are co-located, one cannot actually use Eq.~\eqref{eq: N detectors SNR}, as the noises will be partially correlated among the three interferometers. To take into account these partial correlations, we follow a similar approach to \cite{Branchesi_2023} to compute the effective strain noise PSD. {Note, however, that our resulting sensitivity differs slightly from the one calculated in \cite{Branchesi_2023}, see Appendix \ref{app: ET SNR} for more details. In brief, we take the noise power spectral density $S_{\mathrm{ET-L}}$ for one ET L-shape interferometer from \cite{ET-data}, including a factor of $\sin^2(\pi / 3)$ due to the triangular shape. We then use an excess cross-correlation power statistic between the three interferometers pairs, assuming a 20\% correlation amongst interferometers as in \cite{Branchesi_2023}{, which assumes a perfect characterization of the correlated noise. This constitutes a major experimental challenge that will have to be addressed to achieve the best possible sensitivity.} In the weak signal regime (see Appendix \ref{app: ET SNR}), {keeping only the linear terms in the noise correlation coefficient $\Gamma^{(n)}_{12}=0.2$ (see Eq.~\ref{eq: interf pair PSD}),} we then find
\begin{equation}
    \label{eq: S_eff ET}
    S_{\rm{eff}}^{\rm{ET}}(f) \simeq \sqrt{\frac{1 + 2 \: \Gamma^{(n)}_{12}}{3 \:  \Gamma_{12}^2}} \times  \frac{S_{\mathrm{ET-L}}(f)}{\sin^2\left(\frac{\pi}{3}\right)}.
\end{equation}
The (constant) overlap reduction function between two co-located V-shape interferometers is $\Gamma_{12} = - 0.075$.

For CE, we consider two equal-arm $L$-shape interferometers of length 40km and 20km \cite{Evans:2021gyd}, located in the Hanford and Livingston sites. The expected sensitivity of each detector are provided in \cite{hereCE,CEwebsite}. 
In this configuration, one can cross-correlate the output of the two detectors and estimate the SGWB SNR using Eq.~\eqref{eq: two detectors SNR}, {since it is reasonable to assume that the noise is uncorrelated between the two detectors}. 
The ORF between the two sites can be computed using Eq.~\eqref{eq: ORF definition} and is shown in \autoref{fig: ORF CE}. It rapidly falls to zero from about 80 Hz, limiting the sensitivity of the CE network at high frequency (see Eq.~\eqref{eq: effective strain noise PSD of N detectors}).

We now turn to the question of the expected astrophysical foregrounds, which must be incorporated in order to properly infer the detectability of cosmological SGWBs.  Despite the order-of-magnitude sensitivity increase they offer, both ET and CE are likely to observe a foreground from unresolved compact binary mergers (CBM), in conjunction with the imperfect subtraction (due to non-perfect Bayesian parameter estimation) of the resolvable mergers \cite{Zhou:2022otw,Zhou:2022nmt}. An estimation of the amplitude of the corresponding foreground for ET and CE detectors can be found e.g.~in \cite{Bellie:2023jlq, Branchesi_2023}. It takes the form
\begin{equation}
    \label{eq: omg_foreground}
    h^2\Omega_{\rm{foreground}}=\Omega_A \left(\frac{f}{f_*}\right)^{2/3}.
\end{equation}
For CE at reference frequency $f_* = 25$ Hz, $\Omega_A / h^2 = 2.5 \times 10^{-11}$. For ET,\footnote{{Note that since our sensitivity differs slightly from \cite{Branchesi_2023}, in principle this amplitude would need to be recomputed to ensure consistency.}} at reference frequency $f_* = 10$ Hz, $\Omega_A / h^2 = 6 \times 10^{-11}$.  Using Eq.~\eqref{eq: fractional energy density def}, this can be converted to a strain noise PSD, and it is straightforward to verify that its amplitude is orders of magnitude below the detector noise. Hence it will only be observable through a proper SGWB search method. 
To address the limitations that this  foreground might impose on the prospect of detecting cosmological SGWBs, we computed the expected SNR of this signal for CE and ET, which we later compare with the one expected from cosmological backgrounds. We find $\rm{SNR}^{\rm{CE}}_{\rm{CBM}} \simeq 72$ and $\rm{SNR}^{\rm{ET}}_{\rm{CBM}} \simeq 50$.

\subsubsection{\label{subsec:LISA}Space-based detector}

The Laser Interferometer Space Antenna (LISA), planned to be operational by 2037, is a space-based detector composed of three free-falling spacecraft positioned at the vertices of an equilateral triangle, with each side measuring $2.5 \times 10^9$ meters \cite{2017arXiv170200786A}. These spacecraft will orbit the Sun at a distance of 1 AU, trailing the Earth by 20 degrees. This configuration allows the exploration of the frequency range between 0.1 mHz and 0.1 Hz.

The data employed for GW search in LISA will be obtained from Time Delay Interferometry (TDI). This mitigates laser frequency noise by delaying and recombining individual measurements along spacecraft links (for further details, see \cite{Prince:2002hp, Tinto:2020fcc}). This process yields three Michelson TDI channels, often denoted as $X$, $Y$, and $Z$, obtained by combining measurements in each pair of arms. 
{Assuming that the length of the arms is constant in time,}
it is then possible to construct uncorrelated combinations of these variables which are the $A$, $E$ and $T$ channels \cite{Estabrook:2000ef, Vallisneri:2012np}. 
Clearly one cannot cross-correlate the output from these three channels to search for an isotropic SGWB, as the signal as well as the noise are uncorrelated amongst the channels. Furthermore, the $T$ combination, which corresponds to a Sagnac interferometer, is insensitive to the GW signal at low frequency \cite{Hogan:2001jn}.
It has been proposed to use the auto-correlation of the $A$ and $E$ channels to look for excess power due to the presence of a SGWB.\footnote{Many other methods are proposed in the literature, see \cite{Boileau:2020rpg} and references therein.} As explained earlier, this method is not optimal since it assumes that we have a perfect estimation of the noise in the $A$ and $E$ channels. {As a result, LISA's ability to detect cosmological backgrounds will strongly depend on the precision of noise characterization across its sensitivity frequency band.} Thanks to the uncorrelated nature of these two combinations, one obtains for the total GWB SNR in LISA
\begin{equation}
    \label{eq: SNR in LISA}
    \mathrm{SNR}_{\rm{LISA}}^2 = \mathrm{SNR}_A^2 + \mathrm{SNR}_E^2 = 2\,\Tobs \int_{\fmin}^{\fmax} \d f \: \frac{\mathcal{R}_A(f)^2 S_h(f)^2}{P_A(f)^2},
\end{equation}
where we have assumed $\mathcal{R}_A = \mathcal{R}_E$ and $P_A = P_E$ which leads to $S_{\rm{eff}}^{\rm{LISA}} = P_A / \mathcal{R}_A$.
The LISA sensitivity is computed using \cite{Babak:2021mhe} and the LISA Data Challenge software \cite{Lisa-data-challenge}.
Concerning the observation time, we use the nominal mission duration of $4.5$ years, considering a 82\% duty cycle leading to $\Tobs = 0.82 \times 4.5$ years.

The detection of cosmological GWB with LISA will be affected by astrophysical foregrounds. The first we consider originates from the galactic population of white dwarf binaries. An estimation of the resulting foreground PSD after (perfect) substraction of individual bright binaries can be found in \cite{Babak:2021mhe,Karnesis:2021tsh}. The level of this noise is expected to exceed the detector noise around 1mHz. Consequently, we combine the PSD of this foreground with that of the detector in quadrature.
{The second foreground we consider is due to the extra galactic population of unresolvable white dwarf binaries and stellar-mass black hole (BH) binaries. We model the former using Eq.~(25) of \cite{Staelens:2023xjn} and the latter using Eq.~\eqref{eq: omg_foreground}, with $\Omega_A=7.87 \cdot 10^{-13}$ and $f_*=0.003$ Hz \cite{Babak:2023lro}.} In the same way as for the Earth-based detectors, we compute the SNR of this foreground using Eq.~\eqref{eq: SNR in LISA}.
We find $\mathrm{SNR}^{\mathrm{LISA}}_{\mathrm{BBH+WDB}} \simeq {448}$
, which is dominated by the population of extra galactic WDB since $\mathrm{SNR}^{\mathrm{LISA}}_{\mathrm{BBH}} \simeq {56}$ 
in agreement with \cite{Pieroni_2020}.

\subsubsection{Pulsar Timing Arrays}

Pulsar Timing Arrays rely on millisecond pulsar spin stability to detect GWs \cite{2005ApJ...625L.123J, Verbiest2020, 10.1093/nsr/nwx126}. These pulsars emit quasi-periodic radio pulses observed by Earth-based radio telescopes, enabling prediction of their arrival times. Using a set of pulsar timing residuals, which are the difference between the measured and expected times of arrival of the radio pulses, GWs can be detected through the characteristic correlation pattern they imprint, known as the Hellings\&Down curve \cite{1983ApJ_HD}. Current PTA collaborations are already observing a correlated signal common to their pulsars, providing strong support for a GW-origin \cite{NANOGrav:2023gor, EPTA:2023fyk, PPTA3-SGWB, CPTA-SGWB}.

Here we focus on the future SKA facility \cite{Rawlings2011TheSK, Janssen:2014dka}. The huge effective collecting area is expected to decrease the root mean square (RMS) error of each pulsar timing residuals by a factor 10. In the following, we assume that the SKA dataset will include data from 50 millisecond pulsars, distributed isotropically across the sky, with a timing residual RMS of 50 ns, assuming a cadence of observation of 4 per month for each pulsar. 

The lowest frequency accessible by PTA experiment is of order $1 / \Tobs$. Here, we choose an observation time span of $\Tobs = 15$ years.

Regarding the computation of the effective strain noise PSD of SKA, we employ Eq.~\eqref{eq: N detectors SNR} using the $\texttt{hasasia}$ Python package to produce realistic sensitivity curves \cite{Hazboun2019, PhysRevD.100.104028}. Indeed, PTA experiments are not only limited by the white noise from the radio telescopes and pulse jitter \cite{10.1093/mnras/staa3411}, but also by three main components: (i) the uncertainty in the timing model parameters used to construct the expected times of arrival of the radio pulses \cite{TM_Haasteren, 1984JApA....5..369B}, (ii) the potential intrinsic red noise that may impact the stability of pulsar spin at low frequencies \cite{Shannon2010ASSESSINGTR}, and (iii) the chromatic noise induced by electron density fluctuations in the interstellar medium along the path of the radio pulses to Earth \cite{Jones_2017, 2016ApJ...821...66L}. 
The latter can, in principle, be mitigated using multi-band observation but it is still anticipated to affect the sensitivity of the experiment \cite{Janssen:2014dka}.
For (i) we consider the pulsar spin rate, its derivative, its sky position and distance in the timing model, see \cite{PhysRevD.100.104028} for their effects on the PTA sensitivity. For (ii) and (iii), using the results from currently observed millisecond pulsars \cite{Agazie_2023, EPTA:2023akd}, we add an intrinsic achromatic red noise component to one-third of the pulsars in the array. We parameterize this noise by a power law PSD with an amplitude (at the reference frequency of 1/year) and spectral index, in terms of characteristic strain, of $\log_{10} A = -13.5$ and $\alpha = 1/2$. {This simplification highlights our current uncertainty regarding the intrinsic noise spectra and the number of SKA pulsars that will be significantly affected by non-GW noise, and, consequently, its ability to probe cosmological backgrounds.}

Concerning the astrophysical foreground within the scope of PTA experiments, it is anticipated to be produced by the population of Super Massive Black Hole Binaries (SMBHBs) situated at the cores of galaxies resulting from galaxy mergers. To account for this foreground in our study, we simplify by assuming circular, GW-driven binaries, yielding a power-law spectrum\footnote{This pure power-law spectrum likely overestimates the foreground power at low frequencies (below 10 nHz), as eccentricity and interactions with surrounding stars are expected to flatten the spectrum \cite{Amaro-Seoane:2009ucl, Kelley:2017lek}.} with a spectral index of $\alpha = - 2/3$ for the characteristic strain \cite{Phinney:2001di}. For the amplitude at a frequency of 1/year, we adopt a value of $\log_{10} A = -15.4$, consistent with a large fraction of estimates derived from cosmological simulations of this signal \cite{Agazie_2023_SMBHB}. Note that this foreground amplitude assumes no individual binary has been detected and subtracted from the data. However, it is highly probable that at least one individual SMBHB will be resolved with a 15-year SKA dataset \cite{10.1093/mnras/stv1098}, potentially decreasing the foreground amplitude at the source frequency.

\autoref{fig:CS+DW_spectra_PTA_values} shows the PLISCs of SKA when considering  timing model and white noise (WN) only or the combination of WN, timing model, intrinsic red noise and the astrophysical foreground. {Note that the sensitivity we consider for the SKA SNR contours in the following corresponds to the second configuration.}

Having reviewed the sensitivity of different detectors to a cosmological SGWB, we now turn to potential sources of this signal and how well they can be detected by the combination of future experiments.

\begin{figure}[htbp]
    \centering
    \includegraphics[width=\textwidth]{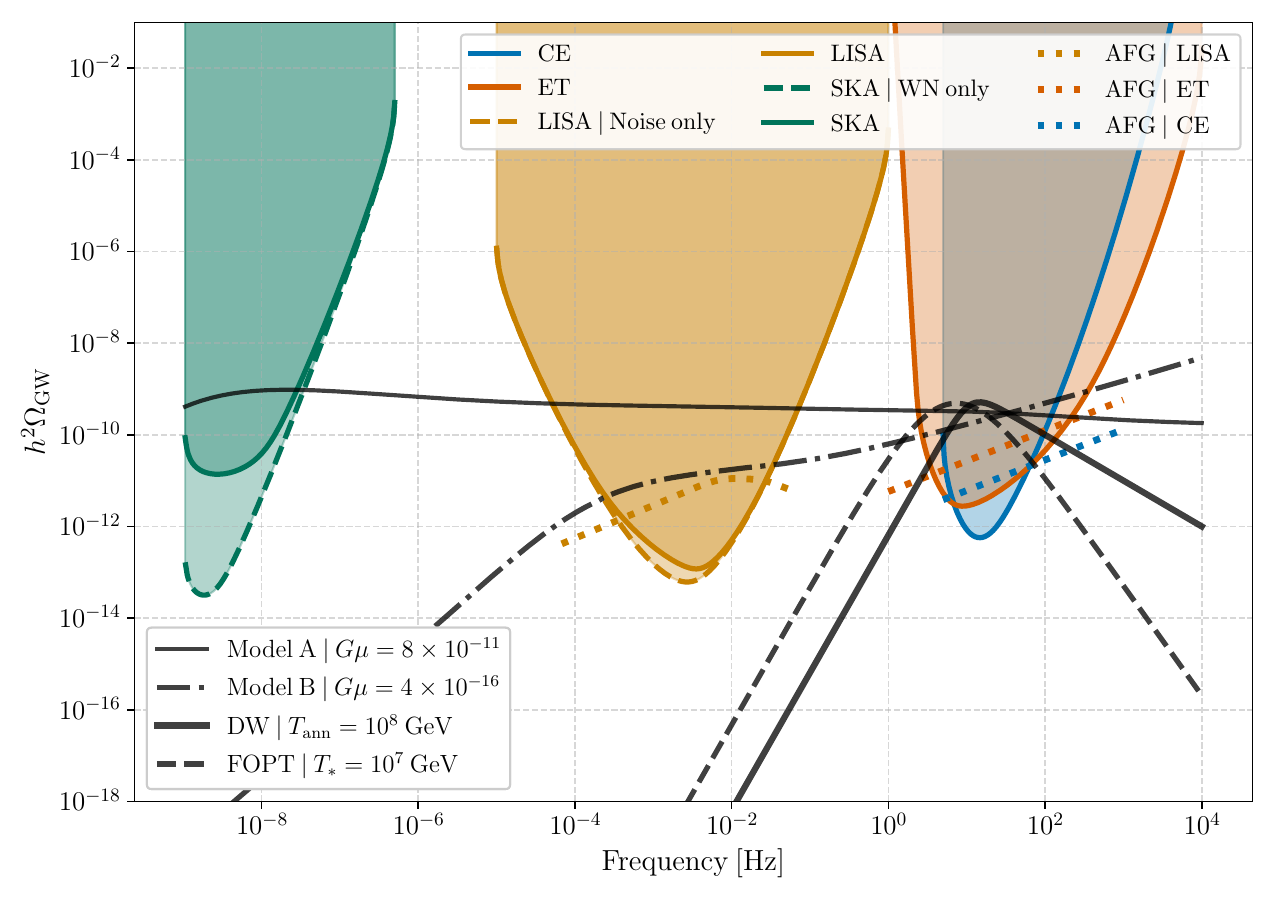}
    \caption{Expected SGWB spectra from local cosmic strings (Models A and B), domain walls (DWs) and strong first order phase transition (FOPT). Cosmic string tension for Model A has been chosen to fit the current PTA constraints \cite{NANOGrav:2023hvm, EPTA:2023xxk}, whereas the one for Model B represents detectable signal within the Earth-based detector band. In both cases, loops are assumed to have two cusps (see section \ref{sec:Defects}). For the DW spectra, we have fixed $\alpha_{\rm{ann}} = 0.04$  and $\tilde{\epsilon}=0.7$ in Eqs.~\eqref{eq:omegadw} and \eqref{eq:pfreq}, while $\beta / H_* = 40$ for the strong first order PT spectrum, Eqs.~\eqref{eq:BPL} and \eqref{eq:amp_and_freq_strongPT}. Power law integrated sensitivity curves (PLISC), corresponding to a $\mathrm{SNR} = 5$ threshold, of future GW detectors are shown, considering astrophysical foregrounds (solid lines, see \cref{tab:detectors}) or not (dashed lines). {For ET and CE, the two PLISC overlap: as discussed in section \ref{sec:detectors}, this is due to the comparatively low level of the astrophysical foregrounds with respect to the detectors noises. Therefore, to facilitate comparison with the cosmological SGWBs, we also show the expected astrophysical foregrounds as dotted lines. 
    {In LISA, two foregrounds are present (see section \ref{sec:detectors}): the one from galactic binaries is higher than the instrument noise and clearly affects the PLISC (solid line); that from extra galactic white dwarf binaries and stellar mass black hole binaries is smaller than the detector noise. It is shown by the dotted line.}}
    Details on the chosen specifications for the detectors are given in \autoref{tab:detectors}. Note that the vertical axis should be limited to $\sim 10^{-6}$ due to the bound on the amplitude of the SGWB from BBN and CMB; here we extend it for visualisation purposes.}
    \label{fig:CS+DW_spectra_PTA_values}
\end{figure}

\section{\label{sec:PT}Primordial sources: overview on first order phase transition}

According to the Standard Model (SM), at least two PTs have occurred in the early universe: the QCDPT around $\sim$150 MeV, and the EWPT around $\sim$100 GeV. 
Furthermore, the fundamental theory underlying the universe is still unknown at energy scales above those tested at the Large Hadron Collider (LHC), and PTs are a generic prediction of quantum field theories: it is therefore reasonable to expect other PTs to have taken place, linked to the breaking of higher symmetries in the context of high energy completions of the SM. 
Establishing the occurrence of primordial PTs in the early universe constitutes a significant test of new physics, 
and it is therefore important to investigate their possible observational signatures, among which, possibly, a SGWB \cite{Witten:1984rs,Hogan:1986qda,Kosowsky:1991ua,Kosowsky:1992vn,Kamionkowski:1993fg}. 

During a PT, the order parameter (generally a scalar field in the cosmological case) changes from a metastable vacuum into a more energetically favoured one.
If the system evolves maintaining thermal equilibrium, as predicted by the SM for both the QCD \cite{Aoki:2006we,Stephanov:2007fk} and the EW PTs \cite{Gurtler:1997hr,Aoki:1996cu,Kajantie:1995kf,Kajantie:1996mn,Laine:1998jb}, no significant GW emission is expected.
While GWs are indeed produced by a plasma in equilibrium, their amplitude is too small to be of observational relevance if the plasma is the one of the SM~\cite{Ghiglieri:2020mhm}, and even when hidden sectors are present \cite{Drewes:2023oxg} (note, however, that the question remains open in the context of thermal inflation \cite{Klose:2022knn,Klose:2022rxh}). 
On the other hand, if the PT proceeds via tunnelling or thermal fluctuations over a potential barrier \cite{Linde:1981zj,Hogan:1984hx,Turner:1992tz}, 
bubbles of the new phase are nucleated and expand, and ultimately percolate to complete the transition. 
A SGWB is then sourced by the anisotropic stresses generated by the scalar field configuration when the bubbles collide \cite{Kosowsky:1991ua,Kosowsky:1992vn,Caprini:2007xq,Huber:2008hg,Jinno:2016vai,Cutting:2018tjt,Cutting:2020nla}, and/or by the bulk fluid motion induced by the bubble expansion and collision \cite{Witten:1984rs,Hogan:1986qda,Kamionkowski:1993fg,Hindmarsh:2013xza,Hindmarsh:2015qta,Hindmarsh:2017gnf,Cutting:2019zws,Hindmarsh:2019phv,Kosowsky:2001xp,Dolgov:2002ra,Caprini:2006jb,Kahniashvili:2008pe,Caprini:2009yp,Brandenburg:2017neh,Niksa:2018ofa,RoperPol:2019wvy}. 
{Note that the phenomenology linked to the presence of bubble walls and to their collisions is very rich: for example, particles can be produced at the bubble walls, leading to additional GW production \cite{Jinno:2022fom} and to dark matter candidates \cite{Giudice:2024tcp}; and furthermore leptogenesis can occur at bubble wall collisions \cite{ Cataldi:2024pgt}. In the following, we will focus on GW production from anisotropic stresses due to the aforementioned scalar field gradients at bubble collisions,  and to bulk fluid motion. 
GW production from topological defects, which may form when a symmetry is spontaneously broken independently of the order of the PT, is discussed in \cref{sec:Defects}.}

The evaluation of the SGWB signal, from the PT dynamics to the modelling of the anisotropic stress source, is a complicated, though constantly progressing, endeavour, involving many assumptions and approximations (for recent reviews, see e.g.~\cite{Hindmarsh:2020hop,Caldwell:2022qsj,Athron:2023xlk,Roshan:2024qnv}).  

\subsection{Dynamics of a first order phase transition}

The starting point to determine the PT dynamics within a given model is the effective potential of the system. 
In many scenarios the barrier is not present at tree level, but is it induced by radiative and/or finite-temperature corrections. 
An important source of uncertainty is therefore linked to the fact that a perturbative treatment of the finite temperature potential can be significantly inaccurate. 
The best resummation techniques are based on matching to a three-dimensional effective field theory, but they can be heavy to implement, and must be carried out on a model-by-model basis (see \cite{Athron:2023xlk} for a review and {e.g.~\cite{Ekstedt:2022bff,Ekstedt:2024etx} for recent progress}). 
This uncertainty can severely affect SGWB predictions, via the inaccurate evaluation of the PT parameters entering the GW signal
\cite{Kainulainen:2019kyp,Gould:2019qek,Niemi:2018asa,Croon:2020cgk,Kierkla:2023von,Niemi:2021qvp,Schicho:2021gca,Gould:2023ovu,Lewicki:2024xan,Banerjee:2024qiu}. 
Correspondingly, it can also compromise the reconstruction of the model parameters, in case a SGWB signal from a first order PT is detected \cite{Bian:2019bsn,Hashino:2016xoj,Friedrich:2022cak,Caprini:2024hue}.

Once the effective potential has been determined, one must focus on the probability of tunneling 
per unit volume and time, $\Gamma(t)$.
The vacuum decay can proceed through zero-temperature quantum tunneling \cite{Coleman:1977py,Callan:1977pt}, or finite temperature quantum tunneling/thermal fluctuations \cite{Linde:1980tt,Linde:1981zj}, such that:
\begin{equation}
    \Gamma(t)\simeq \mathrm{max} \left\{ \frac{1}{R_0^4}\left(\frac{S_4(\varphi)}{2\pi}\right)^2\exp[-S_4(\varphi)],T^4\left(\frac{S_3(\varphi,T)}{2\pi \,T}\right)^{3/2}\exp\left[-\frac{S_3(\varphi,T)}{T}\right]\right\}\,,
    \label{eq:gamma_PT}
\end{equation}
where $R_0$ denotes the bubble radius at nucleation, $S_4(\varphi)$ is the Euclidean action in four dimensions and $\varphi$ the field undergoing the PT, $T$ is the temperature, $S_3(\varphi)$ is the Euclidean action in three dimensions, and the prefactors are approximated on dimensional grounds \cite{Quiros:1999jp,Laine:2016hma}. 
Both actions in \cref{eq:gamma_PT} are evaluated at the field configurations which minimises them, i.e.~the $O(4)-$ and $O(3)-$symmetric bubble, respectively. 
In general, when $T\gg 1/R_0$ the action of the $O(3)-$symmetric solution drives the tunneling, while if $T\lesssim 1/R_0$ one should compare the two actions and use the smaller \cite{Linde:1981zj}. 
The bubble radius at nucleation can be estimated, for example, in the thin wall approximation, when the difference of the potential in the minima $V_0(\varphi_f)-V_0(\varphi_t)$ (where $V_0$ denotes the potential at zero temperature, and $f$ and $t$ stand for false and true vacua) is small compared to the height of the barrier, i.e.~the transition is weak. 
In this case, $R_0=3 S_1/V_0(\varphi_f)$, with $S_1$ the surface energy of the bubble, and we have set the zero of the potential in the true vacuum phase, $V_0(\varphi_t)=0$. 
The cosmological PTs we are interested in take place during the radiation dominated era, at least at their beginning, and in most cases $S_3(\varphi,T)$ dominates in \cref{eq:gamma_PT}. 
There can be exceptions e.g.~if the barrier persists at zero temperature: in this case, a minimum of $S_3(\varphi,T)$ can be reached at temperature higher than the one at which one bubble is nucleated per horizon volume, rendering thermal tunneling impossible \cite{Megevand:2016lpr}. 
Quantum tunneling can then take over at very low temperature; examples of cosmological PTs in which this can happen in regions of their parameter space will be discussed in section~\ref{sec:PQ}, and can also be found e.g.~in \cite{Konstandin:2010cd,Konstandin:2011dr,Randall:2006py,GarciaGarcia:2016xgv,Nardini:2007me,Bunk:2017fic,Megias:2018sxv,Dillon:2017ctw,Ares:2020lbt}.

From now on we focus on a PT happening at finite temperature, in a cosmological thermal environment (for reviews, see e.g.~\cite{Quiros:1999jp,Laine:2016hma}). 
At high temperature the symmetry is restored and the equilibrium value of the field corresponds to the minimum of the finite temperature effective potential $V(\varphi,T)$, even if the latter differs from the minimum of the zero-temperature effective potential $V_0(\varphi)$. 
As the temperature decreases, the shape of $V(\varphi,T)$ changes and the PT happens. 
Assuming that the effective potential can be approximated by including the leading order thermal correction only, it becomes  \cite{Enqvist:1991xw,Ignatius:1993qn,Quiros:1999jp,Wang:2022lyd} 
\begin{equation}
V(\varphi,T)\simeq V_0(\varphi)+\left[\sum_i (-1)_{(F)} N_i^{(B,F)}J_{(B,F)}\left(\frac{m_i^2(\varphi)}{T^2}\right)\right]\frac{T^2}{2\pi^2}\simeq V_0(\varphi)-\frac{\pi^2}{90} g(T)\,T^4 +...    
 \label{eq:VphitT}
\end{equation}
The functions $J_{(B,F)}$ are the thermal bosonic and fermionic functions, given in Eqs.~(173) and (199) of \cite{Quiros:1999jp}.
These functions admit a high-temperature expansion for small argument: their expansions can again be found in Eqs.~(174) and (200) of \cite{Quiros:1999jp}.
The $\propto  g(T) T^4$ term in the second equality of \cref{eq:VphitT} corresponds indeed to the lowest term of the high-temperature expansion, accounting only for particles with $m\ll T$, which behave as a relativistic gas and do not interact with the scalar field $\varphi$. 
The number of these light degrees of freedom is $g(T)=\sum_i N_i^B+(7/8)N_i^F$. 
The following terms, not shown in the second equality of \cref{eq:VphitT} (see e.g.~\cite{Quiros:1999jp,VonHarling:2019rgb,Wang:2022lyd}), are given by the particles with $m\sim T$ that acquire a mass during the PT, such as for example the $W^{\pm}$ and $Z$ bosons at the EWPT in the SM, or the top quark.
These terms drive the evolution of the shape of the potential with the temperature, in particular the change of the potential minimum and the appearance of the barrier, if the latter is not present already in $V_0(\varphi)$ \cite{Quiros:1999jp}.  

If a large number of light particles dominates the potential \eqref{eq:VphitT}, the two phases of the thermodynamic system, composed effectively of a thermal bath and of the scalar field, can be described by the bag equation of state \cite{Espinosa:2010hh}. 
While not necessarily accurate, this approximation allows to calculate many aspects of the PT dynamics and of the bubble evolution \cite{Enqvist:1991xw,Ignatius:1993qn,Espinosa:2010hh}.
The free energy of the system is given by the effective potential: $\mathcal{F}=-p=V(\varphi,T)$, where $p$ denotes the pressure.  
Within the bag equation of state, 
the energy density, enthalpy density and entropy density of the system can be evaluated 
considering only the first two terms in the second equality of \cref{eq:VphitT}, giving
\begin{align}
 \rho=T\frac{\partial p}{\partial T}-p 
 & ~~~\xrightarrow[\text{(bag)}]{}
~~~\rho_f=\frac{\pi^2}{30} g_f(T)\,T_f^4 
+V_0(\varphi_f)\,,~~\rho_t=\frac{\pi^2}
{30} g_t(T)\,T_t^4\,, 
\label{eq:energy_bag}\\
w=T\frac{\partial p}{\partial T} 
& ~~~\xrightarrow[\text{(bag)}]{}
~~~ w_{f,t}=\frac{4\pi^2}{90} g_{f,t}(T)
\,T_{f,t}^4=\rho_{f,t}+p_{f,t}\,, 
\label{eq:w_bag}\\
s=\frac{\partial p}{\partial T} 
& ~~~\xrightarrow[\text{(bag)}]{}
~~~ s_{f,t}=\frac{4\pi^2}{90} g_{f,t}(T)
\,T_{f,t}^3\,, \label{eq:entropy_bag}
\end{align}
where we recall that the zero of the potential is set in the true vacuum phase, $V_0(\varphi_t)=0$.  

The PT proceeds through several stages, from the moment at which 
the first bubble nucleates to  when the universe is fully converted to the most energetically favourable state. 
These stages can be defined from the probability $P(t)$ that a given point in space is still in the false vacuum at time $t$, obtained by exponentiating the volume of true vacuum bubbles per unit volume, $I(t)$ \cite{Turner:1992tz}:
\begin{align}
      P(t)&=\exp[-I(t)]\,,  \label{eq:prob_vac}\\
    I(t)&=\int_{t_c}^{t} \d t'\Gamma(t')a^3(t') \frac{4\pi}{3} r^3(t,t')\,\label{eq:vol_vac} \\
    r(t,t') &= \int_{t'}^{t}\d t'' \frac{v_w(t'')}{a(t'')} \label{eq:rad_bub_t}
\end{align}
where $R_0$ has been neglected, the FLRW metric $\d s^2=-\d t^2+a^2(t)\d\mathbf{x}^2$ has been used, $t_c$ denotes an initial time at which the probability of tunneling is still negligible, $r(t,t')$ is the coordinate radius at time $t$ of a bubble nucleated at time $t'$, and $v_w$ represents the bubble wall velocity.  
The latter is defined as the speed of the phase boundary after nucleation, in the rest frame of the thermal plasma, far from the wall. 

The first notable moment of the PT is when the values of the potential at the two minima become equal, $V(\varphi_f,T_c)=V(\varphi_t,T_c)$: this occurs at the critical temperature $T_c$, which we set to define
also $t_c$ in \cref{eq:vol_vac}. 
Note that the time and the temperature can be related using the conservation of the entropy $\d(a^3s)/\d t=0$, giving $\d T/\d t=-3H (\partial V/\partial T)/(\partial^2 V/\partial T^2)$, from definition \eqref{eq:entropy_bag} \cite{Athron:2023rfq}. 
Within the bag equation of state, the relation between the time and the temperature in the universe takes the familiar form $\d T/\d t=-3T\,H(T)$.

Since at $T_c$ the two vacua still have the same free-energy, no bubble has nucleated yet. 
The nucleation temperature $T_n<T_c$, or equivalently the nucleation time $t_n>t_c$, can be defined as the moment when one bubble is nucleated on average per Hubble volume \cite{Enqvist:1991xw,Athron:2023xlk}:
\begin{equation}
    N(t_n)=\int_{t_c}^{t_n}\d t'\Gamma(t')\frac{4\pi}{3 H^3(t')}\,P(t')=1.
    \label{eq:nucl_one_bubble}
\end{equation}
The above condition can be simplified to $\Gamma(T_n)/H(T_n)^4\simeq 1$, assuming that $\Gamma(t)$ dominates around $t_n$, that $P(t<t_n)\simeq 1$, and interpreting the time integral as a multiplication by one Hubble time.
In the case of weak PTs, when one can neglect the contribution of the vacuum energy and set $H^2(T)\simeq (\pi^2/30) g(T) \,T^4/(9M_\mathrm{Pl}^2)$ ($M_\mathrm{Pl}$ denoting the reduced Planck mass), the simplified nucleation condition gives approximately $S_3(T_n)/T_n\simeq 4\log\left(M_\mathrm{Pl}/T_n\right)$. 

Since the source of the GW signal is the anisotropic stresses linked to the bubble collision, the relevant stage as far as GW production is concerned is towards the end of the PT, when bubbles collide and the entire universe settles in the true vacuum. 
We identify this with the stage of percolation \cite{Athron:2022mmm}, reached when a connected group of bubbles spans the entire universe. 
The condition for percolation reads \cite{Athron:2022mmm}
\begin{equation}
    P(t_*)\simeq 0.71\,.
    \label{eq:percolation_T}
\end{equation} 
However, percolation might not occur if the volume still in the false vacuum keeps expanding, as can be the case if the PT occurs in a vacuum dominated background (see also section~\ref{sec:PQ}), as in this case bubbles can never meet. 
To ensure percolation, and thereby GW production, one also needs to ensure that the fractional physical volume which is still in the false vacuum, $V_\mathrm{phys}(t)=a^3(t)P(t)$ is decreasing at percolation time \cite{Turner:1992tz,Athron:2022mmm}:
\begin{equation}
    \left.\frac{1}{V_\mathrm{phys}(t_*)}\frac{\d V_\mathrm{phys}(t)}{\d t}\right|_{t_*}=\left.3H(t_*)-\frac{\d I(t)}{\d t}\right|_{t_*}<0\,.
    \label{eq:volume_decrease}
\end{equation}

When the PT has completed, the universe is back to a thermal state. 
If the PT is strong and significant supercooling has taken place, a reheating process must occur after percolation, reestablishing thermal equilibrium. 
The reheating temperature can be calculated e.g.~by imposing entropy conservation between a time at which the entropy is known (for example, the thermal state before the onset of the PT) and the time of reheating: $T_\mathrm{reh}=(g_s/g_\mathrm{reh})^{1/3} (a_s/a_\mathrm{reh}) T_s$, where the first generic time is denoted with a subscript $s$.  
However, this implies solving the Friedmann equations to determine the ratio $a_s/a_\mathrm{reh}$. 
The evaluation can be simplified assuming that the reheating process is instantaneous in time and occurs just after percolation, when the vacuum energy contribution is already subdominant and does not influence any longer the universe's expansion. 
In this case, it is possible to set $H(T_\mathrm{reh})\simeq H(T_*)$. 
Within the bag equation of state \eqref{eq:energy_bag}, this leads to
\begin{equation}
    \label{eq:Treh}
    T_\mathrm{reh}\simeq \left(\frac{g_*}{g_\mathrm{reh}}\right)^{\frac{1}{4}}\left[1+\frac{V_0(\varphi_f)}{(\pi^2/30)g_*\,T^4_*}\right]^{1/4}\,T_*    ~.
\end{equation}

Another relevant quantity is the distribution of bubble sizes, i.e.~the number density of bubbles with radius $\rho(t)=a(t)r(t,t_r)$, where $t_r$ denotes the nucleation time of a bubble with coordinate radius $r(t,t_r)$ at time $t$.  The distribution of bubble sizes is given by the product of the nucleation rate per unit time and volume at $t_r$, with the fraction of space in the false vacuum at $t_r$ \cite{Turner:1992tz,Ellis:2018mja}:
\begin{equation}
    \frac{\d n(t,t_r)}{\d\rho}=\Gamma(t_r)\left(\frac{a(t_r)}{a(t)}\right)^3 P(t_r) \frac{\d t_r}{\d\rho}=\Gamma(t_r)\left(\frac{a(t_r)}{a(t)}\right)^4 P(t_r) \frac{1}{v_w(t_r)} \label{eq:bubble_sizes_distrib}
\end{equation}
where the second equality follows from \cref{eq:rad_bub_t}. 
The 
number density of bubbles at time $t$ is then obtained upon integration over time 
\begin{equation}
    n(t)=\int_{t_c}^{t} \d t_r\Gamma(t_r)\left(\frac{a(t_r)}{a(t)}\right)^3 P(t_r)\,. \label{eq:bublle_numb_dens}
\end{equation}
The mean bubble separation at time $t$ is given by $R(t)=n(t)^{-1/3}$ \cite{Hindmarsh:2020hop}, while the mean bubble radius at time $t$ becomes
\begin{equation}
    \bar R(t)=\frac{1}{n(t)}\int_{t_c}^{t} \d t_r\frac{\d n(t,t_r)}{\d t_r} \rho(t,t_r) \,. \label{eq:mean_bub_rad}
\end{equation}

In the thermal case, very often the action decreases slowly due to the gradual change in temperature as the universe expands.  
It is then assumed that the action can be approximated with a Taylor series in time up to first order, $S(t)\simeq S(t_*)+(\d S(t)/\d t)_{t_*}(t-t_*)$, where we have taken the expansion around percolation time (defined by \cref{eq:percolation_T}). 
This is an arbitrary choice (for example, to simplify the equations, one could rather choose the time at which $P(t_e)=1/e$ \cite{Hindmarsh:2020hop}).  

The linear expansion for the action
allows to simplify considerably the treatment of the PT dynamics. 
Rewriting the nucleation rate in \cref{eq:gamma_PT} as a function of time as $\Gamma(t) =A(t)e^{-S(t)}$, one can see that it becomes exponential \cite{Turner:1992tz,Megevand:2016lpr,Hindmarsh:2020hop}:
\begin{equation}
    \Gamma(t) = \Gamma(t_*)\exp[(\beta(t-t_*)],~~\mathrm{with}~~~\beta\equiv -\left.\frac{\d S(t)}{\d t}\right|_{t_*}. \label{eq:gamma_time_exp}
\end{equation}
Therefore, the probability of tunneling increases very rapidly near $t_*$. 
In this case one assumes that essentially all of the bubbles are nucleated within less than a Hubble time, and therefore that the expansion of the universe can be neglected. 
The transition rate parameter $\beta$ expresses how rapidly the nucleation rate changes, and it therefore provides the timescale for the duration of the PT, as we now show. 
The fractional volume of true vacuum bubbles \cref{eq:vol_vac} becomes, in the exponential nucleation case,  $I(t)= 8\pi (v_w^3/\beta^4) \Gamma(t_*) \exp [\beta(t-t_*)] $, 
assuming that both the scale factor $a(t)$ and the bubble wall velocity $v_w$ are constant in time, and that therefore the lower bound of the integration in \cref{eq:vol_vac} can be set to $t_c\rightarrow -\infty$ without too much error. 
The probability that a point is in the false vacuum at time $t$, \cref{eq:gamma_PT}, becomes then \cite{Hindmarsh:2020hop}
\begin{equation}
    P(t)= \exp \left[-8\pi \left(\frac{v_w^3}{\beta^4}\right) \Gamma(t)\right]= \exp \{\ln(P(t_*))\exp[\beta(t-t_*)]\}\,,
\end{equation}
where $P(t_*)=0.71$ at percolation. 
The above equation shows that $\beta$ provides the time-scale over which the fractional volume in the metastable phase has dropped to $\sim 0.39$, and can therefore be interpreted as the timescale for the duration of the PT. 
This is an important parameter, entering the GW signal in the dimensionless form \cite{Grojean:2006bp}
\begin{equation}
    \frac{\beta}{H(t_*)}=T_*\left.\frac{\d}{\d T}\frac{S_3(T)}{T}\right|_{T_*}\,, \label{eq:betaoverH}
\end{equation}
where we have expressed it in terms of temperature assuming $\d T/\d t=-3T\,H(T)$.
The time interval between $t_*$ and nucleation $t_n$ is in general larger than $1/\beta$: one finds  \cite{Turner:1992tz,Hindmarsh:2020hop}
\begin{equation}
    t_*-t_n=-\frac{1}{\beta}\ln\left[\frac{8\pi}{-\ln(P(t_*))}\, v_w^3\left(\frac{H(t_n)}{\beta}\right)^4\right]\,,
    \label{eq:tstarmentn}
\end{equation}
where we have used $\Gamma(t_n)\sim H^4(t_n)$, so that $P(t_n)=\exp[-8\pi v_w^3(H(t_n)/\beta)^4]$, and we remind that under assumption \eqref{eq:gamma_time_exp} in general $(H(t_n)/\beta)^4<1$. 
At nucleation, the universe is still in the metastable phase, and the phase transition completes very rapidly around $t_*$ with duration set by $1/\beta$.

Another parameter entering the GW signal is the length scale associated with the bubble size towards the end of the phase transition, when the bubbles collide and the GW sourcing starts. 
This can be taken as the mean bubble separation at $t_*$  \cite{Hindmarsh:2020hop}, which one evaluates from the number density of bubbles \cref{eq:bublle_numb_dens}:
\begin{equation}
  R(t_*)=(n(t_*))^{-1/3}\simeq 
    \left[\frac{8\pi}{(1-P(t_*))}\right]^{1/3} \frac{v_w}{\beta}\,.\label{eq:bub_separ}
\end{equation}
In the simple case of exponential nucleation, the mean bubble separation and the mean bubble radius are about the same: from \cref{eq:mean_bub_rad}, $\bar R(t_*)\simeq 1.1 v_w/{\beta}$. 
For the level of precision of the GW signal, the two expressions can be used interchangeably. 
Another possibility is to extract the characteristic length scale from the peak of the distribution of the bubble sizes at $t_*$, \cref{eq:bubble_sizes_distrib}, or from the peak of the same quantity multiplied by $\rho^3$, emphasising the role of large bubbles \cite{Ellis:2018mja,Athron:2023xlk}.

If a potential barrier between the symmetric and broken phases is still present at zero temperature, the action $S_3(\varphi, T)/T$ has a minimum at some temperature $T_m$.
If this occurs at higher temperature than the one corresponding to $t_*$, the linear expansion leading to \cref{{eq:gamma_time_exp}} is no longer appropriate \cite{Megevand:2016lpr,Megevand:2021llq,Eichhorn:2020upj}. 
The action has then to be expanded to second order around the time at which it reaches its minimum, such that the nucleation rate takes a Gaussian form $\Gamma(t)=\Gamma(t_m)\exp[-\beta_2^2(t-t_m)^2/2]$, 
with $\beta_2=\sqrt{\d^2 S(t)/\d t^2|_{t_m}}$. 
Nucleation is then concentrated around $T_m$. 
We don't expand on this situation further, and refer to \cite{Eichhorn:2020upj,Ellis:2018mja} for the expressions of the bubble number density and length scale. 

From now on, we fix a ``phase transition temperature'' $T_*$ and corresponding Hubble rate $H_*$. 
If the PT is characterised by negligible supercooling, we assume that this is the temperature at percolation.  
If the PT is characterised by important supercooling, we further assume instantaneous reheating, {such that the Hubble rate during the vacuum dominated phase (thus in particular at percolation) is roughly the same as the Hubble rate after reheating, i.e. $H(T_\text{reh})\simeq H_*$. However, notice that the percolation temperature $T_*$ is generically much smaller than $T_\text{reh}$ in this scenario.}

\subsection{Gravitational wave generation from a first order phase transition}

Tensor metric perturbations are sourced by the presence of tensor anisotropic stresses in the early universe components, such as the relativistic fluid and/or the scalar field.  
In this context, a first order PT offers a rich phenomenology, with several processes potentially leading to tensor anisotropic stresses:
\begin{itemize}
	\item Bubbles and spherical fluid configurations collision. At percolation, the bubbles walls collision breaks the bubble spherical symmetry, leading to non-zero tensor anisotropic stresses from the scalar field gradients \cite{Kosowsky:1991ua,Kosowsky:1992vn,Huber:2008hg,Cutting:2018tjt,Cutting:2020nla,Caprini:2007xq,Jinno:2016vai}. 
    Furthermore, if the field undergoing the transition is coupled to the surrounding plasma, which is generically the case when the PT occurs in a thermal environment, spherical shells where the fluid has non-zero velocity are produced around the bubbles by the field-fluid coupling \cite{Kamionkowski:1993fg,Espinosa:2010hh}. 
    In strongly supercooled PTs, GWs are also sourced by the collision of the thin shells of relativistic fluid velocity that form around the bubbles walls \cite{Kamionkowski:1993fg,Konstandin:2017sat,Ellis:2019oqb,Lewicki:2022pdb}. 
    The characteristic duration of both these GW sourcing processes is of the order of $\beta/H_*$, see \cref{eq:betaoverH} and \cref{eq:tstarmentn}. 

	\item Sound waves. If the PT is weakly first order, the compression and rarefaction waves formed around the bubbles and overlapping towards the end of the PT lead to the development of bulk fluid motion in the form of sound waves. These lead to tensor anisotropic stresses in the fluid and thereby source GWs \cite{Hogan:1986qda,Hindmarsh:2015qta,Hindmarsh:2017gnf,Cutting:2019zws,Hindmarsh:2019phv}. 
    As the kinetic viscosity of the primordial universe is typically very small \cite{Caprini:2009yp}, this ``acoustic'' phase can generally subsist for several Hubble times after bubble percolation, before it gets dissipated by the fluid viscosity. However, its lifetime can also be cut off by the production of shocks and the development of non-linearities in the bulk fluid motion \cite{Pen:2015qta,Caprini:2015zlo}.  

	\item (Magneto)Hydrodynamic (M)HD turbulence: the development of non-linearities in the fluid velocity might lead to vorticity in the flow, evolving into a fully developed turbulent stage. 
    This is expected since the Reynold's number of the primordial universe (measuring the ratio of the advection and viscous terms in the fluid equation of motion) is typically very high \cite{Caprini:2009yp}. Magnetic fields might accompany the turbulence and reach equipartition with the kinetic energy \cite{Durrer:2013pga}. 
    Tensor anisotropic stresses sourcing GWs are then present because of the chaotic distribution of the velocity field and because of the large scale magnetic field \cite{Dolgov:2002ra,Caprini:2006jb,Gogoberidze:2007an,Kahniashvili:2008pe,Caprini:2009yp,Brandenburg:2017neh,Niksa:2018ofa,RoperPol:2019wvy}. This phase is expected to last several eddy turnover times, again because of the smallness of the kinetic viscosity in the early universe \cite{RoperPol:2022iel,Auclair:2022jod}.  
\end{itemize}
The amplitude of the SGWB produced by these sourcing processes depends on the amount of energy in the form of tensor anisotropic stresses available for each of them. 
This is related to the dynamics of the PT and of the fluid, and can in principle be evaluated in the context of specific models. 
In practice, this evaluation is a very complicated problem involving many steps, and uncertainties still subsist in the determination of the anisotropic stresses and thereby on the SGWB signal, as we will see.  

{For the three sourcing processes described above, the tensor anisotropic stress $\Pi_{ij}=T_{ij}^{\rm{TT}}$ i.e.~the transverse traceless component of the spatial part of the source energy momentum tensor $T_{ij}$, is a random variable, since the bubble nucleation process is stochastic. 
$\Pi_{ij}$} also has a finite lifetime $\delta\tau_{\mathrm{fin}}=\tau_\mathrm{fin}-\tau_*$, where $\tau$ denotes conformal time, we assume radiation domination and set $\tau_*=\mathcal{H}_*^{-1}=(a_*H_*)^{-1}$ as the initial time of action of the source. 
$\delta\tau_{\mathrm{fin}}$ can be short compared to the Hubble time (e.g.~for bubble collision), but can also be of several Hubble times (e.g.~for turbulence). 
After the decay of the source, the GWs become freely propagating plane waves, redshifting with the expansion of the universe. 
The SGWB power spectrum today satisfies then:
\begin{equation}
h^2\mathrm{\Omega}_{\rm{GW}}(\tau_0,f) = \frac{h^2}{\rho_c}\left(\frac{a_\mathrm{fin}}{a_0}\right)^4\rho_\mathrm{fin} \,\mathrm{\Omega}_{\rm{GW}}(\tau_\mathrm{fin},f)=  h^2F_{\mathrm{GW},0}\, \mathrm{\Omega}_{\rm{GW}}(\tau_\mathrm{fin},f)\,,~~\mathrm{with}~~h^2F_{\mathrm{GW},0}\simeq 1.64 \cdot 10^{-5} \left(\frac{100}{g_\mathrm{fin}}\right)^{\frac{1}{3}}.
    \label{eq:OM_generic}
\end{equation}
{In order to give a compact expression for $\mathrm{\Omega}_{\rm{GW}}(\tau_\mathrm{fin},f)$, we assume that the GW source  operates during the radiation dominated era, and is such that the total energy density in the Universe is $\rho_\mathrm{tot}(\tau)\propto a^{-4}(\tau)$.
This assumption will always be valid in the case of weak PTs, for which the vacuum energy is subdominant throughout the PT. 
It is also adapted in general to sound waves and MHD turbulence, which continue to source GWs long after the bubble collisions are over and the Universe returns to a radiation dominated expansion.  
It is clearly not suited to the case of strong PTs, for which vacuum energy may dominate the Universe expansion for a while. However, in this case, the SGWB is mainly sourced by bubble collisions, which is a fast process occurring around percolation time. The usual assumption for this kind of source is therefore to neglect the expansion of the Universe altogether, and to assume that GW generation occurs in flat space-time. }

{We define for later convenience the energy fraction $K(\tau)=\rho_s(\tau)/\rho_\mathrm{tot}(\tau)$, where $\rho_s(\tau)$ denotes the energy density of the GW sourcing process (i.e., kinetic energy of the scalar field and/or of the fluid motion), and $\rho_\mathrm{tot}(\tau)$ denotes the total energy density in the universe at the PT time, which we assume to be radiation. 
It is reasonable to expect that the source  $\rho_s(\tau)$ also redshifts with the Universe expansion, like radiation. 
However, superimposed on the radiation-like scaling, one also expects a peculiar time evolution, characteristic of the source itself: for example, a growing phase followed by an overall decay.
This is captured by the residual time dependence of $K(\tau)$. 
One can then rewrite a normalised anisotropic stress 
$\tilde\Pi_{ij}=\Pi_{ij}/(K\rho_\mathrm{tot})$.}
Further neglecting possible changes in the relativistic degrees of freedom, and averaging over the irrelevant, rapid oscillations in time of the plane waves, the GW energy density fraction when the source has stopped operating 
becomes then (where $k$ denotes the comoving wave-number) \cite{Caprini:2007xq,Caprini:2009yp,Caprini:2018mtu}
\begin{equation}
    \mathrm{\Omega}_{\rm{GW}}(\tau_\mathrm{fin},k)= \frac{3}{4\pi^2} k^3 \int^{\tau_\mathrm{fin}}_{\tau_*} {\frac{\d \tau'}{\tau'}}\int^{\tau_\mathrm{fin}}_{\tau_*}{\frac{\d \tau''}{\tau''} \cos[k(\tau'-\tau'')]\,K^2(\tau',\tau'')\,P_{\tilde\Pi}(k,\tau',\tau'')}\,,\label{eq:OmGW_int}
\end{equation}
where $P_{\tilde\Pi}(k,\tau',\tau'')$ is the unequal time correlator (UETC) of the normalised anisotropic stress $\tilde \Pi_{ij}(\tau)$ in Fourier space,
\begin{equation}
    \label{eq:PIspec}
    \langle \tilde\Pi_{ij}(\mathbf{k}, \tau') \, \tilde\Pi_{ij}^*(\mathbf{q}, \tau'') \rangle = (2\pi)^3 \,
\delta^{(3)}(\mathbf{k} - \mathbf{q}) \,  P_{\tilde\Pi}(k, \tau', \tau'')\,.
\end{equation}
{In \cref{eq:OmGW_int}, the radiation-like dependence of the GW source has been absorbed in the factors $1/\tau'$ and $1/\tau''$, while $K^2(\tau',\tau'')$ contains a possible overall time dependence of the source (such as an overall growth and/or decay), and $P_{\tilde\Pi}(k, \tau', \tau'')$ contains the residual time dependence, typically a time decorrelation.  
Note that, for a source effectively operating in flat space-time such as e.g.~bubble collision, one can just substitute the factors $1/\tau'$ and $1/\tau''$ in \cref{eq:OmGW_int} with a factor $H_*^2$ \cite{Caprini:2009yp}.}

{In order to find the SGWB signal from a first order PT, one therefore needs to determine the time dependence of $K^2(\tau',\tau'')$ together with the anisotropic stress UETC $P_{\tilde\Pi}(k, \tau', \tau'')$}, for each of the the sourcing processes described above: bubbles and spherical fluid configurations collision, sound waves, MHD turbulence.
This is a complicated task, which can be tackled both analytically and numerically.  
However, it turns out that, by making use of two simple models for the UETC, it is possible to understand the scaling with the sources' parameters provided by more refined analytical and numerical evaluations.
We present these models below, as an update of what discussed in sections 3.1 and 3.2 of \cite{Caprini:2019egz}.
In the following we will then clarify the connection between the general scalings provided by these simple models, and what found in specific analytical and numerical evaluations on a source-by-source basis. 

\begin{itemize}
    \item The first model consists in assuming that the UETC is simply constant in time, {$K^2(\tau',\tau'')P_{\tilde\Pi}(k, \tau', \tau'')=K^2\,P_{\tilde\Pi}(k)$} \cite{RoperPol:2022iel}. 
    As we will see, the constant-in-time approximation can be applied to bubble collision and to (M)HD turbulence. 
    
    The constant-in-time assumption is justified as follows. The characteristic time for the GW production is $\tau_\mathrm{GW}\sim 1/k$. 
Therefore, for all wave-numbers satisfying $k>1/\tau_s$, where $\tau_s$ is a characteristic time for  the source (e.g.~its duration $\delta\tau_\mathrm{fin}$), one has $\tau_\mathrm{GW}<\tau_s$, and the source can therefore be considered constant in time as far as the GW production is concerned. 
Most importantly, the wave-numbers for which this holds include the relevant region where the SGWB spectrum peaks.
Indeed, one can also associate a characteristic length scale to the source, $\mathcal{R}_{*}\sim v\tau_s$. 
In general, the anisotropic stress power spectrum $k^3P_{\tilde\Pi}(k)$, and consequently $\mathrm{\Omega}_{\rm{GW}}(\tau_\mathrm{fin},k)$, grow as $k^3$ and then peak around this characteristic scale, i.e.~at $k_*\sim 1/\mathcal{R}_{*}$. 
Therefore, the constant approximation can be applied around the peak of the GW spectrum, as $k_*\sim 1/\mathcal{R}_{*}\geq 1/\tau_s$, since $v\leq 1$.

Assuming that the UETC is constant in time, it is particularly simple to find the SGWB spectrum by integrating \cref{eq:OmGW_int}:
\begin{eqnarray}    
    \label{eq:const_in_time}
    \mathrm{\Omega}_{\rm{GW}}(\tau_\mathrm{fin},k)\simeq \frac{3}{4\pi^2}K^2 [k^3P_{\tilde\Pi}(k)] \left\{
    \begin{array}{ll}\log^2(1+\mathcal{H}_*\delta\tau_{\mathrm{fin}}) & \mathrm{if}~k\,\delta\tau_{\mathrm{fin}}<1\,, \\\log^2(1+\mathcal{H}_*/k) & \mathrm{if}~k\,\delta\tau_{\mathrm{fin}}\geq 1\,.
    \end{array}
    \right.     
\end{eqnarray}
The SGWB spectrum as a function of $k$ grows as the anisotropic stress power spectrum $k^3P_{\tilde\Pi}(k)$, i.e.~as $k^3$, until the scale corresponding to the source duration $k\sim 1/\delta\tau_{\mathrm{fin}}$. 
The behaviour at larger wave-numbers $k> 1/\delta\tau_{\mathrm{fin}}$ then depends on whether the source lasts for shorter or longer than one Hubble time $\mathcal{H}_*^{-1}$. 
If $\mathcal{H}_*\delta\tau_{\mathrm{fin}}<1$, the second branch of Eq.~\eqref{eq:const_in_time} takes up a linear behaviour $\propto k^1$, since $\log^2(1+\mathcal{H}_*/k)\simeq (\mathcal{H}_*/k)^2$ for $\mathcal{H}_*/k<1$. 
This continues until the power spectrum $[k^3P_{\tilde\Pi}(k)]$ changes slope from $k^3$ to a source-dependent decaying behaviour, at $k>k_*\sim 1/\mathcal{R}_*$. 
If the source lasts longer than one Hubble time $\mathcal{H}_*\delta\tau_{\mathrm{fin}}>1$, the second branch of Eq.~\eqref{eq:const_in_time} shows a logarithmic growth for $1/\delta\tau_{\mathrm{fin}}<k<\mathcal{H}_*$. 
Note that the scaling at the peak of the spectrum for this constant-in-time approximation always satisfies \cite{RoperPol:2022iel}
\begin{equation}
    \mathrm{\Omega}_{\rm{GW}}^\mathrm{peak}(\tau_\mathrm{fin},k_*)\simeq \frac{3}{4\pi^2}K^2 [k_*^3P_{\tilde\Pi}(k_*)]\left(\frac{\mathcal{H}_*}{k_*}\right)^2\,.
    \label{eq:const_peak}
\end{equation}

\item The second model consists in assuming that the UETC is stationary in time, {$K^2(\tau',\tau'') P_{\tilde\Pi}(k, \tau', \tau'')=K^2 f(k,|\tau'-\tau''|)$} \cite{Hindmarsh:2013xza,Hindmarsh:2015qta}. 
This assumption has been used mainly for SGWB generation by sound waves {(see \cite{Caprini:2024gyk} for an alternative proposal)}.

Since the source is necessarily dissipated at some point, it is ``stationary'' but only within its lifetime $\delta\tau_\mathrm{fin}$. 
Upon changing variable {$\tau_-=\tau''-\tau'$}, \cref{eq:OmGW_int} takes the form:
\begin{equation}
\mathrm{\Omega}_{\rm{GW}}(\tau_\mathrm{fin},k)\simeq
\frac{3}{4\pi^2}K^2 k^3
{\left[\int_{\tau_*}^{\tau_\mathrm{fin}} \frac{\d\tau'}{\tau'}
    \int_{\tau_*-\tau'}^{\tau_{\mathrm{fin}}-\tau'} \d\tau_-\frac{\cos(k\tau_-)f(k,\tau_-)}{\tau'+\tau_-}
     \right]\,.}
     \label{eq:stationary_itermediate}
\end{equation}
The evaluation can be simplified if the integrals can be decoupled. This occurs, for example, if one can send the limits of the integral over $\tau_-$ to $\pm \infty$.
The common assumption is that this is doable when the autocorrelation time of the fluid perturbations, i.e.~the time for which $f(k,\tau_-)$ is non-zero, is small compared with the source duration $\delta \tau_\mathrm{fin}$ \cite{Hindmarsh:2015qta,Hindmarsh:2019phv,Auclair:2022jod}. 
Naturally, it is strictly speaking impossible that the function $f(k,\tau_-) $ be negligibly small for all {$\tau_-<\tau_*-\tau'$ and $\tau_->\tau_\mathrm{fin}-\tau'$, and for every $\tau'\in [\tau_*,\tau_\mathrm{fin}]$}. 
However, there are situations in which integrating the second integral in \cref{eq:stationary_itermediate} over $\pm \infty$ is justified: see e.g.~\cite{RoperPol:2023dzg} for a discussion of this assumption in the context of GW production by sound waves. 
Assuming this holds, and further assuming that $\tau_-$ can be neglected in the denominator of \cref{eq:stationary_itermediate}, gives (we remind that $\tau_*=\mathcal{H}_*^{-1}$):
\begin{equation}
    \mathrm{\Omega}_{\rm{GW}}(\tau_\mathrm{fin},k)\simeq \frac{3}{4\pi^2}K^2 k^3 \int_{\tau_*}^{\tau_\mathrm{fin}}
    {\frac{\d\tau'}{(\tau')^2}}
    \int_{-\infty}^{+\infty} \d\tau_-\cos(k\tau_-)f(k,\tau_-)\simeq \frac{3}{4\pi^2}K^2 \frac{\mathcal{H}_*\delta\tau_\mathrm{fin}}{1+\mathcal{H}_*\delta\tau_\mathrm{fin}}[\mathcal{H}_*k^3F(k)]\,,
\label{eq:omega_stationary}
\end{equation}
where we have generically indicated with $\mathcal{H}_*k^3F(k)$ the dimensionless function representing the SGWB spectral shape, to be determined within specific source contexts. 
Note that under the assumptions we have made, the function $\mathcal{H}_*k^3F(k)$ contains no residual time dependence.  
\end{itemize}

After these general considerations on the scalings of the SGWB spectrum, which will become useful later on, we now focus on the sourcing processes specific to PTs. 
As far as the GW production is concerned, it is possible to think of the system undergoing the PT as composed by the scalar field and a perfect relativistic fluid.  
The energy momentum tensor of the system is 
\begin{equation}
    T_{\mu\nu}=\partial_\mu\varphi\partial_\nu\varphi-g_{\mu\nu}\left( \frac{1}{2}\partial_\alpha\varphi\partial^\alpha\varphi+V_0(\varphi)\right)+w\, u_\mu u_\nu+g_{\mu\nu} p\,,
\label{eq:Tmunu}
\end{equation}
where $g_{\mu\nu}$ is the background FLRW metric $\d s^2=-\d t^2+a^2(t)\d\mathbf{x}^2$,
{$w$, $p$ are the enthalpy density and pressure of the fluid, $u_\mu=\gamma (1,a\mathbf{v})$} is the fluid four-velocity.\footnote{Note that we account for the presence of bulk  motion in the fluid, which is therefore out of equilibrium. 
In principle, we should therefore include scalar and vector perturbations in the FLRW metric. 
However, we neglect their presence, and only focus on the tensor (GW) perturbations generated by the fluid velocity and by the scalar field. This is because the scalar and vector metric perturbations generated by the PT dynamics would be at too small scales to bear any observable significance today.} 
Upon definitions \eqref{eq:energy_bag}-\eqref{eq:entropy_bag}, one obtains for the energy density $\rho_{\mathrm{tot}}=T_{00}$ \cite{Athron:2023xlk}: 
\begin{equation}
    \rho_{\mathrm{tot}}=\frac{1}{2}\left[\dot \varphi^2+\frac{(\partial_i\varphi)^2}{a^2}\right]+V(\varphi,T)-T\frac{\partial V(\varphi,T)}{\partial T}\gamma^2= \rho_\varphi+\rho + w\,\gamma^2 v^2\,. \label{eq:rho_tot_1}
\end{equation}
This shows that the energy density of the system is composed by the kinetic energy of the scalar field $\rho_\varphi=[\dot \varphi^2+(\partial_i\varphi)^2/a^2]/2$, the fluid energy density including the potential contribution $\rho= V(\varphi,T)-T\frac{\partial V(\varphi,T)}{\partial T} $ (c.f.~Eqs.~\eqref{eq:VphitT} and \eqref{eq:energy_bag}), and the fluid kinetic energy $\rho_v=w\,\gamma^2 v^2$.

The trace of the energy momentum tensor $\theta=-T^{\mu}_{\mu}/4=(\rho-3p)/4$ measures how the system's equation of state differs form the one of a relativistic fluid. 
For example, in the case of the bag equation of state given in Eqs.~\eqref{eq:energy_bag} to \eqref{eq:entropy_bag}, $\theta$ corresponds to the false vacuum potential energy, $\theta=V_0(\varphi_f)$. 
Introducing it in \cref{eq:rho_tot_1}, the contributions from the potential and the relativistic fluid energy densities become explicitly separated:
\begin{equation}
    \rho_{\mathrm{tot}}= \rho_\varphi+\theta + \frac{3}{4}w + w\,\gamma^2 v^2. \label{eq:rho_tot_2}
\end{equation}
Using the trace $\theta$, it is customary to define the parameter $\alpha$, representing the relative importance of the potential to the relativistic fluid energy \cite{Kamionkowski:1993fg,Hindmarsh:2015qta,Hindmarsh:2020hop}: 
\begin{equation}
     \alpha=\frac{4(\theta_f-\theta_t)}{3\,w_f}\,,~~~\xrightarrow[\text{(bag)}]{}
~~~\alpha_{\mathrm{bag}} =\frac{V_0(\varphi_f)}{\pi^2 g_f\,T^4_f/30}\,.
    \label{eq:alpha}
\end{equation}
In the bag equation of state, \cref{eq:alpha} takes a more self-evident form, i.e.~the ratio between the false vacuum energy (remember we have set $V_0(\varphi_t)=0$) and the radiation energy density in the false vacuum dominated phase. 
$\alpha$ can also be written in terms of the potential energy difference 
$\alpha_{\Delta V}=4[\Delta V(\varphi,T)-T/4(\partial \Delta V(\varphi,T)/\partial T)]/(3\,w_f)$, where $\Delta V=V(\varphi_f,T_f)-V(\varphi_t,T_t)$.  
Note that several definitions are given in the literature for the parameter $\alpha$, e.g.~in terms of the pressure difference $\alpha_{p} =-4(p_f-p_t)/(3w_f)$, or the energy density difference $\alpha_{\rho} =4(\rho_f-\rho_t)/(3w_f)$. 
Refs.~\cite{Giese:2020rtr,Giese:2020znk}, on the other hand, propose a definition of $\alpha$ which can be generalised to cases in which the speed of sound in the broken phase is different than $c_s^2=1/3$: it is given in terms of the pseudotrace $\bar\theta=\rho-p/c_s^2$, as $\alpha_{\bar\theta}=4(\bar\theta_f-\bar\theta_t)/(3w_f)$. 

In general, the parameter $\alpha$ can be interpreted as a measure of the PT strength. 
The PT is a dynamical process in which the energy components in \cref{eq:rho_tot_2} transform one into the others: the false vacuum potential energy gets converted into kinetic energy of the bubble walls, kinetic energy of the surrounding fluid bulk motion, and thermal energy of the environment.
The anisotropic stresses associated to the first two components source the GW signal. 
It is customary to define the efficiency factors 
\begin{equation}
    \kappa_\varphi=\frac{\rho_\varphi}{\theta_f-\theta_t}~~ \mathrm{and}~~ \kappa_v=\frac{\rho_v}{\theta_f-\theta_t}\,,
\end{equation}
quantifying how much of the available energy (i.e.~false vacuum energy, represented here by $\theta$) is converted into kinetic energy of the bubble walls and of the fluid bulk motion, potentially sourcing GWs. 
Then, the energy fraction $K=\rho_s/\rho_\mathrm{tot}$, appearing e.g.~in \cref{eq:OmGW_int}, gets rewritten as 
\begin{equation}
    K=(\kappa_\varphi+\kappa_v)\frac{\alpha}{1+\alpha+\frac{4\theta_t}{3w_f}}~~\stackunder{=}{(bag)}~~(\kappa_\varphi+\kappa_v)\frac{\alpha_\mathrm{bag}}{1+\alpha_\mathrm{bag}}\,,
    \label{eq:K_kappa}
\end{equation}
where the denominator is the energy density in the symmetric phase before bubble nucleation, and the second equality holds in the bag equation of state. 
The relative importance of the two GW sources represented by $\kappa_\varphi$ and $\kappa_v$ is related to the PT strength.

\begin{itemize}

\item PTs with $\alpha \gtrsim \mathcal{O}(1)$ are classified as strong: the vacuum energy is equal or larger than the relativistic fluid energy, and the PT is characterised by supercooling. 
The vacuum energy leads to a phase of accelerated expansion and, after tunneling, it gets converted mostly into kinetic
energy of the bubbles, the thermal component being subdominant.
What happens at this point depends on the balance between the vacuum energy and the friction on the bubble wall, due to the interaction with the fluid particles \cite{Jinno:2017fby,Ellis:2020nnr,Lewicki:2022pdb}. 
In the EWPT, for example, the masses of the fluid particles depend on the value of the Higgs field undergoing the PT. 
The energy momentum tensors of the scalar field and the surrounding fluid are  therefore coupled via an effective friction term, representing the fact that the gradient of the scalar field drives the fluid out of equilibrium by changing the particle masses \cite{Ignatius:1993qn,KurkiSuonio:1995vy,KurkiSuonio:1996rk,Espinosa:2010hh}. 
The analysis of the out-of-equilibrium dynamics of this coupled system allows in principle to determine 
the bubble wall speed $v_w$ on a model-by-model basis (as it depends on the specific particle content and interactions of the theory)
\cite{Moore:1995si,Moore:1995ua,Bodeker:2009qy,Konstandin:2014zta,Kozaczuk:2015owa,Hoeche:2020rsg,BarrosoMancha:2020fay}. 
This is a rather difficult endeavour.
Schematically, if the friction is constant (e.g., $\Delta P_{\mathrm{LO}}\sim \Delta m^2 T^2/24$ at linear order for the EW theory, where $\Delta m^2$ is the difference in masses of the particles that become heavy during the phase transition), then, whenever the false vacuum energy is larger than the ``negative'' pressure exerted on the bubble wall by the friction (i.e.~$V_0(\varphi_f)>\Delta P_{\mathrm{LO}}$), the bubble wall continuously accelerates until $v_w\rightarrow 1$  \cite{Moore:1995si,Moore:1995ua,Bodeker:2009qy}. 
In this so-called runaway regime, very little kinetic energy is transferred to the fluid, and the SGWB is mainly sourced by bubble collision, from the $\Pi_{ij}^\varphi=(\partial_i\varphi\partial_j\varphi)^{TT}$ component of \cref{eq:Tmunu}  \cite{Ellis:2019oqb,Ellis:2020nnr}. 
However, next-to-leading order processes such as particle splitting at the bubble wall can induce friction terms increasing with {$\gamma_w=1/\sqrt{1-v_w^2}$} 
(both behaviours $\Delta P_{\mathrm{NLO}} \propto\gamma_w$ \cite{Bodeker:2017cim,Azatov:2020ufh,Gouttenoire:2021kjv,Azatov:2023xem} and $\Delta P_{\mathrm{NLO}}\propto \gamma_w^2$ \cite{Hoche:2020ysm} have been proposed). 
In this case, there exists a $v_w$ for which the total pressure on the bubble wall balances to zero, $V_0(\varphi_f)-\Delta P_{\mathrm{LO}}-\Delta P_{\mathrm{NLO}}(\gamma_w)=0$. As soon as the wall reaches this terminal velocity, it stops accelerating. 
If this occurs before collision, 
the majority of the released energy is transferred to fluid motion \cite{Jinno:2017fby,Jinno:2019jhi,Ellis:2019oqb,Ellis:2020nnr,Lewicki:2022pdb}. 
In a strong PT, the fluid motion takes the form of thin shells of relativistic velocity around the bubbles walls \cite{Kamionkowski:1993fg}. 
The SGWB is therefore mainly sourced by the fluid kinetic energy, from the $\Pi_{ij}^v=(w\gamma v_iv_j)^{TT}$ component of \cref{eq:Tmunu}, which develops due to the collision of the fluid shells. 
In general, if the radiation component can be neglected, one has $\kappa_v=1-\kappa_\varphi$. 
Furthermore, since $\alpha \gtrsim \mathcal{O}(1)$, \cref{eq:K_kappa} implies $K\simeq 1$.

The anisotropic stresses sourcing the SGWB, from the collisions of both the bubbles and the fluid shells, depend on the collision process, and are difficult to predict. 
In particular, the evolution of the discontinuity (be it the bubble wall or the fluid shell) after collision is highly non-trivial \cite{Konstandin:2017sat,Jinno:2019jhi}, and strongly influences the SGWB power spectrum shape at small scales. 
Both analytical approaches \cite{Caprini:2007xq,Jinno:2016vai,Jinno:2017fby} and numerical simulations \cite{Huber:2008hg,Child:2012qg,Cutting:2018tjt,Konstandin:2017sat,Cutting:2020nla,Lewicki:2022pdb,Jinno:2019bxw} have been used to estimate the anisotropic stresses, and in turns the SGWB signal.  
The generic features of the SGWB power spectrum from these works are, approximately: (i) a growth like $k^3$ at large scales, possibly followed by (ii) a linear increase $k^1$ at intermediate scale \cite{Jinno:2017fby}, (iii) a peak at $k\sim \beta_c$ (we denote $\beta_c$ the comoving quantity, to be compared with comoving wave-number $k$) and (iv) a power-law decay after the peak. 
It turns out that these features 
can be well understood based on the constant-in-time model of Eq.~\eqref{eq:const_in_time}.
Indeed, bubbles and/or fluid shells collision is a fast process, whose characteristic time-scale is the duration of the PT $H_*/\beta$ 
(see Eq.~\eqref{eq:betaoverH}), and whose characteristic length-scale is the bubble size at collision $R_*\simeq 4 v_w/\beta$ (see \cref{eq:bub_separ}). 
Eq.~\eqref{eq:const_in_time} then reduces to 
\begin{align}    
    \mathrm{\Omega}_{\rm{GW}}^{\mathrm{strong}}(k)\simeq \frac{3}{4\pi^2} [k^3P_{\tilde\Pi}(k)] \left\{
    \begin{array}{ll}\left(\mathcal{H}_*/\beta_c\right)^2 & \mathrm{if}~k<\beta_c\,, \\
    \left(\mathcal{H}_*/k\right)^2 & \mathrm{if}~k\geq \beta_c\,.
    \end{array}
    \label{eq:SGWB_strong_generic}
    \right.
\end{align}
For strong PTs with $v_w\sim 1$, $\beta\sim 1/R_*$, so the linear increase characteristic of the region between the inverse time-scale and the inverse length-scale is absent, and the peak of the spectrum corresponds to $k\simeq \beta_c \simeq 1/\mathcal{R}_*$, as determined by $k^3 P_{\tilde\Pi}(k)$. 
The peak of the spectrum scales as $K^2 (H_*/\beta)^2$, as predicted by \cref{{eq:const_peak}} (we remind that we have set $K\simeq 1$ for strong PTs).
At higher $k$, the peak is followed by a power-law decay given by $k P_{\tilde\Pi}(k)$. 

Several spectral shapes with distinct features have been proposed in the literature, derived from different approaches to the problem of modelling the anisotropic stresses: e.g.,~using lattice simulations of the scalar field only \cite{Cutting:2018tjt,Cutting:2020nla}, of the thin fluid shells surrounding the bubbles \cite{Konstandin:2017sat,Lewicki:2022pdb}, or using analytical models \cite{Jinno:2016vai,Jinno:2017fby}. 
In the rest of this work, we will follow \cite{Caprini:2024hue} and take as placeholder for the SGWB spectrum today, for strong PTs, the result given in \cite{Lewicki:2022pdb}, obtained with numerical simulations modeling the contribution from the thin fluid shells. 
Similarly to \cref{eq:SGWB_strong_generic}, this can be written as a broken power law using the generic template \cite{Caprini:2024hue}: 
\begin{equation}
	\label{eq:BPL}
	h^2\mathrm{\Omega}_{\rm{GW}}^{\mathrm{str}}(\tau_0,f) = \Omega_b \left(\frac{f}{f_b}\right)^{\! n_1} \left[ \frac{1}{2}+\frac{1}{2} \left(\frac{f}{f_b}\right)^{\! a_1} \right]^{\frac{n_2-n_1}{a_1}} ,
\end{equation} 
where the peak of the spectrum in general is $f_p = f_b\, (-n_1/n_2)^{{1/ a_1}}$, and the amplitude at the peak is
$\Omega_p = \Omega_b \left\{\left[ \left(- n_2/n_1\right)^{{n_1}/{(n_1 - n_2)}} +  \left(-n_1/n_2\right)^{-{n_2}/{(n_1 - n_2)}} \right]/2\right\}^{-(n_1 - n_2)/{a_1}}$ \cite{Caprini:2024hue}. 
The analysis of \cite{Lewicki:2022pdb} finds that the amplitude at the peak and the peak frequency are (see \cref{eq:s})
\begin{equation}
    \label{eq:amp_and_freq_strongPT}
    h^2 \Omega_p  = h^2 F_{\mathrm{GW},0} \,A_\text{str}\,\left(\frac{H_*}{\beta}\right)^{\!2} \,, 
    \quad\quad
    f_p \simeq 0.11\,H_{*,0}\, \frac{\beta}{H_*}\,,
\end{equation}
with $A_{\mathrm{str}}\simeq 0.05$.
Even though setting $K\simeq 1$ should be a good approximation for very strong PTs, note that the situation can be complicated by the fact that the efficiency factors $\kappa_\varphi$ and $\kappa_v=1-\kappa_\varphi$ depend on the bubble size \cite{Lewicki:2022pdb}. 
The slopes and the form factor which are found in \cite{Lewicki:2022pdb} are $n_1 = - n_2 \simeq 2.4$ and $a_1 \simeq 1.2$, such that $\Omega_b=\Omega_p$ and $f_b=f_p$.
The power law increase at scales larger than the peak is shallower than the $f^3$ expected by causality \cite{Caprini:2009fx}. 
This shallow increase, possibly due to the fact that simulations do not account for the expansion of the universe, is likely to artificially boost the parameter space reach of interferometers (see e.g.~\cref{fig:generalFOPT}, \cref{fig:PQ_results}), especially when the spectrum peaks at high frequency compared to the frequency range of sensitivity of a given detector, as is the case for most of the parameter space explored in section \ref{sec:PQ}. 
In order to provide conservative estimates, in our analysis we have therefore decided to set $n_1 = 3$ and $ n_2 =- 2.4$.

\item PTs with $\alpha\lesssim \mathcal{O}(10^{-2})$ are classified as weak: 
the potential energy being largely subdominant, what generates the SGWB in this case is mostly the bulk fluid motion. 
Numerical simulations solving the scalar field Klein-Gordon equation coupled to the fluid continuity equations \cite{Hindmarsh:2013xza,Hindmarsh:2015qta,Hindmarsh:2017gnf,Cutting:2019zws}, have allowed to gauge the efficiency with which the kinetic energy of the scalar field and of the fluid are converted into anisotropic stresses, and ultimately into GWs. 
By studying the evolution of the root-mean-squared bulk fluid velocity $\bar{U}_f^2$ and of the equivalent for the scalar field $\bar{U}_\varphi^2$ (a bar denoting averaged quantities, and $V$ the averaging volume)
\begin{equation}\label{eq:Uf}
	\bar{U}_f^2=\frac{1}{V\bar{w}}\int_V \d^3x \,w \,\gamma^2 v^2\,,\quad \quad
	\bar{U}_\varphi^2=\frac{1}{V\bar{w}}\int_V \d^3x (\partial_i \varphi)^2\,,
\end{equation}
simulations have demonstrated that $\bar{U}_\varphi^2$ quickly becomes sub-dominant, since it is proportional to the total area of the phase boundary, while $\bar{U}_f^2$ is proportional to the volume. 
Furthermore, simulations complemented with analytical interpretation of their outcome \cite{Hindmarsh:2019phv,Hindmarsh:2016lnk}, have also demonstrated that, for weak PTs, the bulk fluid motion takes the form of sound waves, which develop when the enthalpy and velocity perturbations are small. 
Non-linearities in the fluid motion giving rise to turbulence are not expected in the case of weak PTs. 

The scalar field contribution being negligible, \cref{eq:K_kappa} takes the form
\begin{equation}
    K=\frac{\rho_v}{\rho_{\mathrm{tot}}}=\kappa_v\frac{\alpha}{1+\alpha+\frac{4\theta_t}{3w_f}}= \frac{\bar w}{\rho_{\mathrm{tot}}}\bar{U}_f^2\,.
    \label{eq:K_fluid}
\end{equation}
It is plausible to assume that 
an estimate of the kinetic energy fraction of the entire flow can be given by the kinetic energy fraction of a single expanding bubble $K_b$ \cite{Espinosa:2010hh,Hindmarsh:2015qta,Jinno:2022mie}, identified with the spherical fluid configuration set into motion by the phase boundary \cite{Kamionkowski:1993fg,Espinosa:2010hh}: \begin{equation}
	K_b= \frac{3}{\rho_\mathrm{tot} v_w^3} \int \d\xi \, \xi^2\, w\, \gamma^2 v^2 \sim \frac{\bar w}{\rho_\mathrm{tot}} \bar{U}_f^2\,,\label{eq:kin_en_frac_one_bubble}
\end{equation}
where $\xi=r/t$ such that $\d^3x=4\pi\, t^3\xi^2\d\xi$. 
The fluid flow around the bubbles can be studied by modelling the phase boundary as a combustion front in a hydrodynamical setting \cite{KurkiSuonio:1995vy,KurkiSuonio:1995pp,Espinosa:2010hh,Hindmarsh:2019phv,Giese:2020rtr}. 
This system has been studied in Ref.~\cite{Espinosa:2010hh} for a perfect fluid with a relativistic equation of state, and with speed of sound~$c_s^2 = 1/3$. 
By enforcing energy and momentum conservation across the boundary, and integrating the continuity equations imposing spherical symmetry and self-similarity,
one obtains the fluid $v(\xi)$ and enthalpy $w(\xi)$ profiles surrounding the bubble. 
Three steady steate solutions exist, satisfying appropriate boundary conditions at the bubble centre (fluid at rest), at the wall as described above, and at a distance $r>t$ from the bubble centre (fluid at rest by causality). 
In these solutions, the bubble wall settles to a steady state with constant velocity $v_w$. They are:
1. subsonic deflagrations, characterised by a compressional flow outside the bubble in the symmetric phase, terminating with a shock front, while the fluid is at rest inside the bubble, and the wall moves at subsonic speeds; 
2. detonations, characterised by a rarefaction wave behind the bubble wall in the broken phase, while the fluid outside the bubble is at rest, and the wall moves at supersonic speeds; 
3. hybrids, for which the wall speed is supersonic and the fluid is moving both behind and ahead of the wall. 
It is important to notice that, in the case of deflagrations, the characteristic scale of the bulk fluid motion has to be corrected, as the shell of moving fluid resides outside the broken phase bubble, and the wall expands at subsonic speed. 
\cref{eq:bub_separ} becomes then 
\begin{equation}
R_*= \left[\frac{8\pi}{(1-P(t_*))}\right]^{1/3}\frac{\mathrm{max}(v_w,c_s)}{\beta}\,. 
\end{equation}

The kinetic energy fraction of a single bubble \cref{eq:kin_en_frac_one_bubble} is known from the fluid $v(\xi)$ and enthalpy $w(\xi)$ profiles.
Similarly to \cref{eq:K_kappa}, one defines an equivalent efficiency factor 
\begin{equation}
 \kappa(\alpha_{\mathrm{bag}},v_w)=K_b \,\frac{1+\alpha_{\mathrm{bag}}}{\alpha_\mathrm{bag}}
    \label{eq:equiv_kappa_b}\,.
\end{equation}
In the bag equation of state, $\kappa$ is only a function of $\alpha_{\mathrm{bag}}$ and $v_w$, and Ref.~\cite{Espinosa:2010hh} provides several fits of this relation, for the different bubble propagation modes (deflagrations, detonations, hybrids). 
The simplest one, valid for high wall speed, is widely used in the literature to evaluate the GW signal from first order PTs, and reads:
\begin{equation}
\label{eq:kappav}
\kappa(\alpha_{\mathrm{bag}},v_w) \simeq 
\alpha_{\mathrm{bag}} \left(0.73+0.083\sqrt{\alpha_{\mathrm{bag}}}+\alpha_{\mathrm{bag}}\right)^{-1} ~~\mathrm{for} ~~ v_w \sim 1\,.
\end{equation}
In our analysis we will also use the fits of $\kappa(\alpha_{\mathrm{bag}},v_w)$ provided in Ref.~\cite{Espinosa:2010hh}.
However, it is far from granted that concrete particle-physics model can be easily related to the bag model.
For this reason, Ref.~\cite{Giese:2020rtr} has generalised the study of the PT energy budget beyond  the bag model, and for generic speed of sound in the broken phase. 
A new efficiency factor $\kappa_{\bar\theta}$ is provided, as a function of $\alpha_{\bar\theta}$, $v_w$ and $c_s^2$, where ${\bar\theta}$ is the pseudo trace (see also the discussion below \cref{eq:alpha}). 

Recently, from the analysis of the results of numerical simulations with the ``Higgsless'' approach \cite{Jinno:2020eqg,Jinno:2022mie}, it has been proposed that a factor should be inserted in the second equality of \cref{eq:kin_en_frac_one_bubble}, since the efficiency in producing kinetic energy in the bulk fluid motion with respect to the single bubble case is not one \cite{Hindmarsh:2015qta,Jinno:2022mie}: one would have then \cite{Caprini:2024hue} 
\begin{equation}
    K\simeq 0.6 \, K_b\,.
    \label{eq:new_K_fluid}
\end{equation}
We will account for this factor 0.6 in our analysis, since it constitutes a conservative estimate with respect to what was assumed in previous studies; however, note that it necessitates further analyses to be confirmed.

The SGWB production by bulk fluid motion for weak PTs has been studied through lattice simulations of the field-fluid system \cite{Hindmarsh:2013xza,Hindmarsh:2015qta,Hindmarsh:2017gnf,Cutting:2019zws}, through ``Higgsless'' simulations evolving the fluid only, initialised with an inhomogeneous background given by spherical regions of non-zero vacuum energy \cite{Jinno:2020eqg,Jinno:2022mie}, and through analytical studies based on overlapping sound waves \cite{Hindmarsh:2016lnk, Hindmarsh:2019phv,Cai:2023guc,RoperPol:2023dzg}. 
The latter in particular assume that the UETC of the sound waves anisotropic stress is stationary. 
The SGWB spectrum then takes the form given in \cref{eq:omega_stationary}
\begin{equation}
    \mathrm{\Omega}_{\rm{GW}}^{\mathrm{sound}}(\tau_\mathrm{fin},k)\simeq \frac{3}{4\pi^2}K^2 \frac{\mathcal{H}_*\delta\tau_\mathrm{fin}}{1+\mathcal{H}_*\delta\tau_\mathrm{fin}}[\mathcal{H}_*k^3F(k)]\,.
    \label{eq:SGWB_sound_generic}
\end{equation}
In the case under analysis of weak PTs, 
the source can last several Hubble times after bubble percolation, before the sound waves gets dissipated by the very small fluid viscosity. 
Indeed, the absence of a rapid decay of the source energy is confirmed by simulations (for which only numerical viscosity is present) \cite{Hindmarsh:2013xza,Hindmarsh:2015qta,Hindmarsh:2017gnf,Jinno:2022mie}.
Therefore, for weak PTs with $\mathcal{H}_*\delta\tau_\mathrm{fin}\gg 1$, the factor coming from the time integration simply reduces to
$\mathcal{H}_*\delta\tau_\mathrm{fin}/(1+\mathcal{H}_*\delta\tau_\mathrm{fin})\rightarrow 1$.
The situation is different in the case of intermediate PTs, treated below: the development of non-linearities in the fluid bulk motion and the consequent production of shocks are expected to cut-off the source duration \cite{Pen:2015qta,Caprini:2019egz}.
This occurs if the time of development of non linearities in the bulk flow, $\tau_{\mathrm{nl}}\simeq \mathcal{R}_*/\bar U_f$, is shorter than one Hubble time \cite{Caprini:2019egz}. 
One can therefore set the duration of the source to 
$\delta\tau_{\mathrm{fin}}\simeq \mathcal{R}_*/\bar U_f \simeq (3/4)\mathcal{R}_*/\sqrt{K}$. 
The last equality holds when the PT has completed, leaving behind the sound waves in the fluid: in this case, from \cref{eq:rho_tot_2} one has $\rho_\mathrm{tot}=(3/4) w $, and from \cref{eq:K_fluid} one gets $\bar U_f^2=(3/4) K$.  

The shape of the SGWB power spectrum is represented here by the dimensionless factor $[\mathcal{H}_*k^3F(k)]$. 
The analytical studies \cite{Hindmarsh:2016lnk,Hindmarsh:2019phv,RoperPol:2023dzg}, in rough agreement with the simulations results \cite{Hindmarsh:2013xza,Hindmarsh:2015qta,Hindmarsh:2017gnf,Cutting:2019zws,Jinno:2022mie}, find that it is proportional to $H_*R_*/c_s$, and that it has the structure of a double 
broken power law, where both the bubble size $R_*$ and the sound shell thickness $\Delta_w = |v_w-c_s|/v_w$ are imprinted. 
We provide below a formula for the SGWB power spectrum from sound waves, taken from \cite{Caprini:2024hue}, which can be applied to both weak and intermediate PTs.
Note, however, that this formula is  preliminary, since it is largely simplified with respect to what found in \cite{Gowling:2021gcy,Boileau:2022ter,Gowling:2022pzb}, and it is still the subject of ongoing evaluations \cite{RoperPol:2023dzg,Sharma:2023mao}.

\item PTs with $\alpha\sim \mathcal{O}(0.1)-\mathcal{O}(1)$ are classified as intermediate: in this case, the bulk flow remains the dominant source of GW production, as opposed to the scalar field, but the velocity and enthalpy perturbations are expected to become of order one. 
This could result in the development of shocks, leading to a turbulent phase in the fluid flow, following the sound waves phase \cite{Caprini:2015zlo,Pen:2015qta}. 
Turbulence can be both of the compressional (dilatational) and/or of the vortical (solenoidal)  type, and will in general be accompanied by magnetic fields  \cite{Niksa:2018ofa}. 
Simulating PTs with $\alpha\sim \mathcal{O}(0.1)-\mathcal{O}(1)$ is numerically challenging, but attempts exploring the mildly non-linear regime seem to indicate the onset of vortical fluid motion \cite{Cutting:2019zws}. 
Analytical evaluations of the GW signal from MHD turbulence exist \cite{Kosowsky:2001xp,Dolgov:2002ra,Caprini:2006jb,Gogoberidze:2007an,Caprini:2009yp,Niksa:2018ofa,RoperPol:2022iel,Auclair:2022jod}, as well as numerical simulations of the GW production in the turbulent regime \cite{Brandenburg:2017neh,RoperPol:2019wvy,RoperPol:2018sap,RoperPol:2021xnd,Brandenburg:2021bvg}. 
Both, however, do not link the onset of turbulence to the scalar field and to PT dynamics, but insert turbulence directly in the initial conditions.
Therefore, more insight is needed to (i) properly characterise the development of the turbulent phase, linking it to the actual PT evolution; (ii) 
quantify the contribution of the turbulent phase to the GW signal, with respect to the one of the acoustic phase; (iii)  establish the possible presence of magnetic fields in relation with the initial conditions and the nature of the PT. 

In the meantime, the common approach is to assume that the SGWB spectrum is given by the sum of the two contributions, sound waves and MHD turbulence (inserted in the initial conditions). 
The kinetic energy in the form of MHD turbulence is assumed to be a fraction of the total bulk kinetic energy. 
An unknown
parameter $\epsilon$ is inserted to represent this fraction \cite{Caprini:2015zlo,RoperPol:2023bqa,Caprini:2024hue}.
One further assumes that MHD turbulence is fully developed, so that equipartition has been reached between the kinetic and magnetic turbulent energies. 
The total energy fraction in the form of MHD turbulence becomes then
\begin{equation}
    \frac{\rho_{\mathrm{kin}}+\rho_{\mathrm{mag}}}{\rho_{\mathrm{tot}}}=\Omega_{\mathrm{kin}} + \Omega_{\mathrm{mag}} \simeq 2\, \Omega_{\mathrm{kin}}=\epsilon\, K \equiv \Omega_{\mathrm{MHD}}  \,,
    \label{Oms_turb}
\end{equation}
where $\Omega_{\mathrm{kin}}$ and $\Omega_{\mathrm{mag}}$ are the  kinetic and magnetic turbulent energy density fractions, and $K$ is given in \cref{eq:K_fluid}. 
The state of the art for SGWB spectra from both sound waves and turbulence is that they can be modelled as double broken power laws. 
Ref.~\cite{Caprini:2024hue} provides a generic template, which can be adapted to both sources, as follows:
\begin{align}
\Omega^{\mathrm{sw, MHD}}_{\mathrm{GW}}(f,\tau_0
) &=
\Omega_\text{int} \times S(f) = \Omega_2 \times S_2(f), \label{eq:DBPL}
\\
S(f) &= N \left( \frac{f}{f_1} \right)^{n_1}
\left[
1 + \left( \frac{f}{f_1} \right)^{a_1}
\right]^{\frac{- n_1 + n_2}{a_1}}
\left[
1 + \left( \frac{f}{f_2} \right)^{a_2}
\right]^{\frac{- n_2 + n_3}{a_2}}. \nonumber
\end{align}
The normalization factor $N$ is determined from $\int_{- \infty}^\infty \!\dd \ln f \, S (f) = 1$.
Alternatively, one can also define the amplitude $\Omega_2$ at the second frequency break, normalising the spectrum such that $S_2(f_2) = 1$, i.e.\ $S_2(f) = S(f)/S(f_2)$.
We stress that, as in the case of bubble/relativistic fluid shells collisions, this template and the following descriptions are placeholders, and are the subject of ongoing research.

The SGWB produced by sound waves can be related to the generic form \eqref{eq:SGWB_sound_generic} representative of a stationary source. 
Combining the input from numerical simulations and analytical studies, the amplitude can in general be written as \cite{Hindmarsh:2020hop}
\begin{align}
h^2 \Omega_\mathrm{int}^{\mathrm{sw}}
&=
h^2 F_{\mathrm{GW},0} \, A_\text{sw}\, K^2 \frac{({H}_*R_*)^2/\sqrt{K}}{4/3+{H}_*R_*/\sqrt{K}},
\label{eq:sw_amplitude}
\end{align}
where $K$ is given by Eqs.~\eqref{eq:kin_en_frac_one_bubble}, \eqref{eq:equiv_kappa_b} and \eqref{eq:kappav}, and the source duration has been set to the time of formation of non-linarities $\delta\tau_{\mathrm{fin}}\simeq \tau_{\mathrm{nl}}\simeq (3/4)\mathcal{R}_*/\sqrt{K}$ (see discussion in the weak PT case), providing the general factor $({H}_*R_*)^2/\sqrt{K}/(4/3+{H}_*R_*/\sqrt{K})$, where we have further incorporated the ${H}_*R_*$ factor coming from the spectral shape part $[\mathcal{H}_*k^3F(k)]$. 
In our analysis we will follow \cite{Caprini:2024hue}, which adopts the spectral characteristics found in \cite{Jinno:2022mie} (see, however, \cite{Gowling:2021gcy,Boileau:2022ter,Gowling:2022pzb,RoperPol:2023dzg,Sharma:2023mao}): we therefore incorporate the factor $0.6$ given in \cref{eq:new_K_fluid}, and set $A_\text{sw} \simeq 0.11$ \cite{Jinno:2022mie}.
Furthermore, the set the frequency breaks $f_1$ and $f_2$ in \cref{eq:DBPL} to the two characteristic length scales of the sound wave source, namely the the bubble size $R_*$ and the sound shell thickness $\Delta_w = |v_w-c_s|/v_w$ \cite{Hindmarsh:2016lnk, Hindmarsh:2019phv,Jinno:2022mie}
\begin{align}
f_1 & \simeq 0.2 \, H_{*,0} \, (H_* R_*)^{-1}\,, &
f_2 & \simeq 0.5 \, H_{*,0} \, \Delta_w^{-1} \,
(H_* R_*)^{-1} \,.
\label{eq:sw_shape}
\end{align}
The slopes are set to $n_1= 3$, $n_2 = 1$, $n_3 = -3$, and the form factors $a_1 = 2$, $a_2 = 4$ \cite{Jinno:2022mie}. 
The amplitude at $f_2$ is then related to $\Omega_{\mathrm{int}}^{\mathrm{sw}}$ via \cite{Caprini:2024hue}
\begin{equation}
    \Omega_2^{\mathrm{sw}} = \frac{1}{\pi}\left( \sqrt{2} + \frac{2\,f_2/f_1}{1+f_2^2/f_1^2}\right) \Omega_{\mathrm{int}}^{\mathrm{sw}}.
    \label{eq:Om2_sound}
\end{equation}

The SGWB produced by fully developed, instantaneous MHD turbulence (i.e.~inserted in the initial conditions \cite{RoperPol:2022iel,Auclair:2022jod}), instead, is well modeled by the constant-in-time case, Eq.~\eqref{eq:const_in_time}. 
This spectral shape has been validated by (M)HD simulations \cite{RoperPol:2022iel,Auclair:2022jod}. 
We refer to~\cite{Caprini:2024hue} for the derivation allowing to put the SGWB spectrum from MHD turbulence derived in \cite{RoperPol:2022iel} in the form of \cref{eq:DBPL}, and here we only provide the results.
The SGWB spectral amplitude, in terms of the amplitude at the second break $\Omega_2$, is
\begin{equation}
    h^2 \Omega_2 = h^2 F_{{\rm{GW}}, 0} \, A_{\mathrm{MHD}}
    \, \Omega_{\mathrm{MHD}}^2 \, (H_* R_*)^2 \,,
    \label{eq:Om_2_MHD}
\end{equation}
where $A_{\mathrm{MHD}}\simeq 4.37\times 10^{-3}$. 
One notes the quadratic behaviour in both the source energy $\Omega_{\mathrm{MHD}}^2$ and characteristic scale $(H_* R_*)^2$ typical of the constant-in-time scenario, see~\cref{eq:const_peak}.
Furthermore, the two frequency breaks are also determined within the constant-in-time model.
The first break corresponds to the duration of the MHD source, which is of the order of a few eddy turnover times, 
${\mathcal{N}} \tau_\mathrm{e}\simeq {\mathcal{N}} \mathcal{R}_*/(2 \pi \sqrt{\varepsilon} \bar U_f)$, with $\mathcal{N}\simeq 2$ \cite{RoperPol:2022iel,RoperPol:2023bqa,Caprini:2024hue}; the second break, instead, corresponds to the characteristic length scale of the MHD turbulence:
\begin{align}
    f_1 = \frac{\sqrt{3 \, \Omega_\mathrm{MHD}}}{2\, {\cal N}} \, H_{*,0} \, (H_* R_*)^{-1}, 
    \quad 
    f_2 \simeq 2.2 \, H_{*,0} \, (H_* R_*)^{-1}.
\label{geom_pars_turb}
\end{align}
The slopes are $n_1 = 3$, $n_2 = 1$, $n_3 = -8/3$, and the form factors 
$a_1 = 4$, $a_2 \simeq 2.15$ \cite{RoperPol:2022iel,RoperPol:2023bqa,Caprini:2024hue}.

\end{itemize}


\subsection{Constraints from 3G detectors}

Figure \ref{fig:CS+DW_spectra_PTA_values}  shows an example of the SGWB signal from a strong first order PT, given in Eqs.~\eqref{eq:BPL} and \eqref{eq:amp_and_freq_strongPT}. 
We have chosen the parameter values $\beta/H_*=40$ and $T_*=10^7$ GeV, for which the signal peaks in the frequency range of the future Earth-based interferometers ET and CE (see Eqs.~\eqref{eq:s} and \eqref{eq:amp_and_freq_strongPT}). 

Figure \ref{fig:generalFOPT} shows the reach of ET, CE, LISA and SKA in probing the parameter space of first order PTs.
To produce this figure, we have used the placeholder SGWB spectra presented above, namely: Eqs.~\eqref{eq:BPL} and \eqref{eq:amp_and_freq_strongPT} for the GW signal from bubble and/or relativistic fluid shells collisions, relevant for strong PTs; and Eq.~\eqref{eq:DBPL}, supplemented with Eqs.~\eqref{eq:sw_amplitude} and \eqref{eq:sw_shape} for the SGWB from sound waves, and with Eqs.~\eqref{eq:Om_2_MHD} and \eqref{geom_pars_turb} for the SGWB from  MHD turbulence, valid for weak and intermediate PTs. 
We have scanned over 
$\beta/H_*$ and $T_*$ for three representative values of the PT strength $\alpha$, namely $\alpha=0.05, 1$ and $\alpha\gg 1$, 
each corresponding to different GW emission mechanisms, as discussed above.  
Note that for $\alpha\gg 1$, the only parameters entering the GW spectrum are indeed $\beta/H_*$ and $T_*$; while 
for $\alpha\leq 1$, when sound waves and MHD turbulence are sourcing the signal, we need to connect $R_*$ to $\beta$ through \cref{eq:bub_separ}, $K$ to $\alpha$ through \cref{eq:new_K_fluid}, \cref{eq:equiv_kappa_b} \cref{eq:kappav}, and $\Omega_\mathrm{MHD}$ to $K$ through \cref{Oms_turb}. 
For $\alpha=1$, we have set $v_w=0.9$ and $\epsilon=0.5$. 
For $\alpha=0.05$, we have set $v_w=0.55$, and used the efficiency parameter $\kappa(\alpha_{\mathrm{bag}},v_w)$ of \cref{eq:equiv_kappa_b} given in Appendix A of \cite{Espinosa:2010hh}, for subsonic deflagrations. 
The contribution from turbulence is always absent when $\alpha=0.05$.

The colour scheme in \cref{fig:generalFOPT} shows the SNR of the signal for the different detectors, and we have highlighted the SNR=5 contours (dashed line) for each detector (see \cref{tab:detectors}). 
The dotted lines show the SNR contours for the astrophysical foregrounds, relevant for each detector.
In the region enclosed by the dotted contours, the cosmological signal is certainly louder than the expected foregrounds, so it should be clearly detectable. 
In the region enclosed by the dashed contours, detection can still be possible depending on the particular frequency shape of the SGWB signal. 
This figure shows the tremendous potential of 3G detectors across the frequency band to detect the SGWB from first order PTs, together with their complementarity with LISA and PTA, as far as the temperature of the PT $T_*$ is concerned. 
3G detectors will allow to explore new energy scales for first order PTs with respect to the classical SM cases: see Section \ref{PQ-PT} for a physically well motivated example.

\begin{figure}[ht!]
    \centering
    \begin{minipage}[h]{0.49\textwidth}
        \centering
        \includegraphics[width=\textwidth]{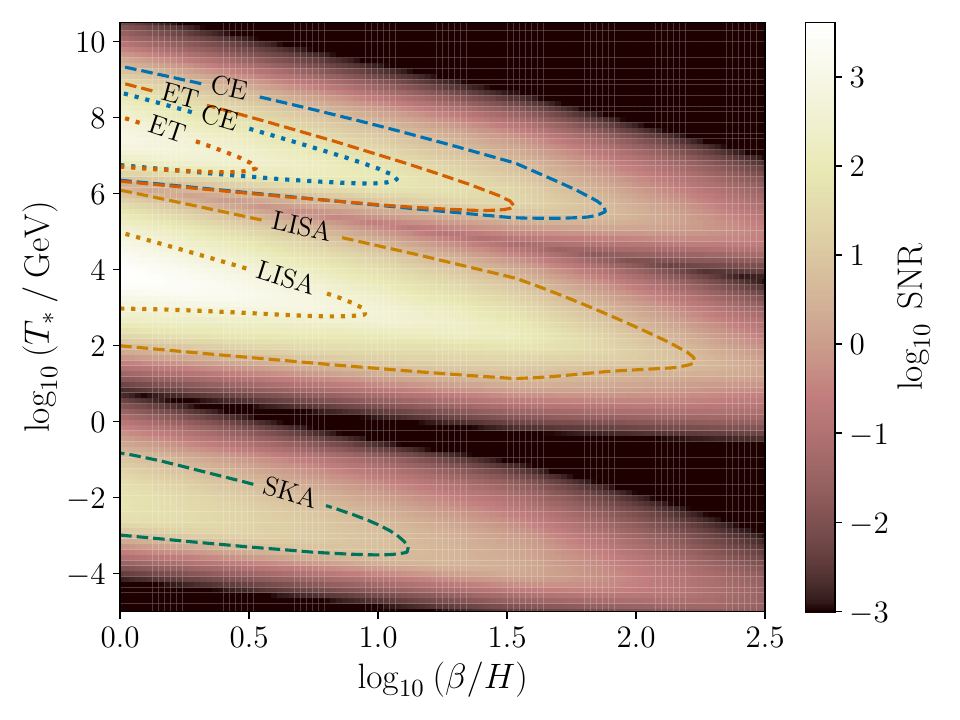}
        \captionof*{figure}{(a) $\alpha = 0.05 \:|$ Sound waves}
        \label{fig:alpha0.01}
    \end{minipage}
    \hfill
    \begin{minipage}[h]{0.49\textwidth}
        \centering
        \includegraphics[width=\textwidth]{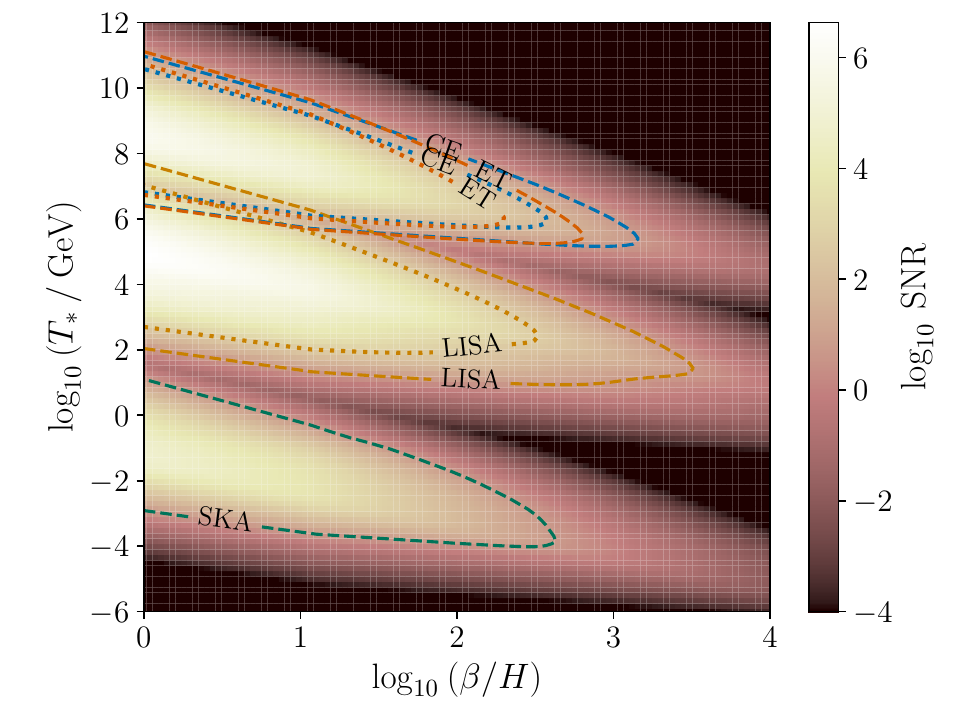}
        \captionof*{figure}{(b) $\alpha = 1 \:|$ Sound waves + Turbulence}
        \label{fig:alpha1}
    \end{minipage}
    
    \vspace{0.5cm} 
    
    \begin{minipage}[h]{0.49\textwidth}
        \centering
        \includegraphics[width=\textwidth]{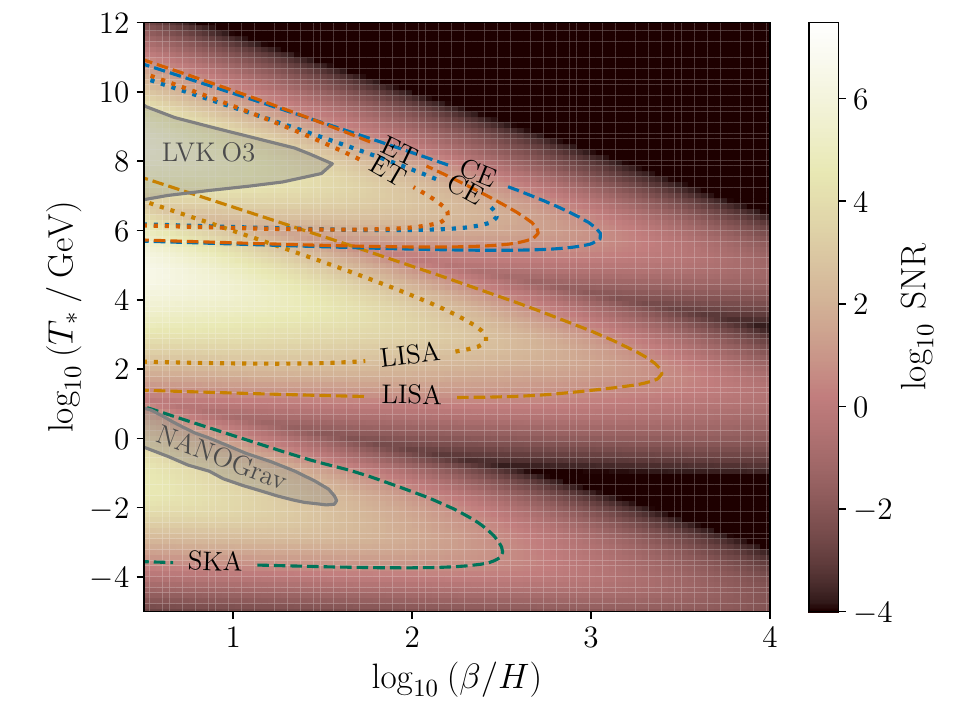}
        \captionof*{figure}{(c) $\alpha \gg 1 \:|$ Bubble collisions + relativistic fluid shells}
        \label{fig:alpha100}
    \end{minipage}
    \caption{The parameter space of a generic first order PT that can be probed by future GW detectors.
    The figure shows a $\log_{10}$ SNR grid, and the SNR value can be read from the color bar. In practice, for each point in the PT parameter space we calculate the SNR for each detector, and the figure shows the maximum value among detectors.
    We consider three values of the phase transition strength, each corresponding to a different set of GW emission mechanisms, as presented in the main text. We explicitly show the SNR=5 contours for the different detectors (dashed lines), as well as the SNR contours for the foregrounds relevant for each detector (dotted lines, see \cref{tab:detectors}). 
    In the region enclosed by the dotted contours, the cosmological signal is certainly louder than the expected foregrounds, so it should be clearly detectable. 
    For the parameter space between the dotted and the dashed contours, detection can still be possible depending on the particular frequency shape of the SGWB signal. 
    The gray shaded regions represent the 95\% confidence exclusion region taken from~\cite{Badger:2022nwo} obtained with LVK O3 data, and the 95\% confidence region of the NANOGrav 15 years posterior, explaining the common correlated signal seen in their data \cite{NANOGrav:2023hvm}. Note that these regions are included for guidance, although they are obtained with SGWB spectra that differ in the details from ours.
    The corresponding exclusion regions for $\alpha=0.05$ and $\alpha=1$ derived in \cite{Romero:2021kby} are not visible in this figure.}
    \label{fig:generalFOPT}
\end{figure}


\section{\label{sec:Defects}Primordial sources: overview of topological defects}

Topological defects of different kinds --- including local and global strings, domain walls --- may form whenever a symmetry is spontaneously broken, independently of the order of the phase transition which may or may not be first order \cite{Kibble:1976sj}. See \cite{Vilenkin:2000jqa,Hindmarsh:1994re,Vachaspati:2015cma} for reviews on topological defects in cosmology. The important point in the context of this review is that topological defects are ubiquitous in many well motivated BSM models. For instance, line-like cosmic strings are unavoidable in supersymmetric GUTs with symmetry breaking patterns which avoid the monopole problem~\cite{Jeannerot:2003qv}. In axion models that solve the strong CP problem, both strings and domain walls can form, see section \ref{sec:PQ} and Fig.~\ref{fig:PQrecap}. Since topological defects are stable and long-lived,
a network of defects formed in a phase transition has many potential observational signatures including in CMB anisotropies e.g.~\cite{Ade:2013xla,Ringeval:2010ca}, particle and photon radiation e.g.~\cite{Brandenberger:1986vj,Sabancilar:2009sq,Vachaspati:2009kq,Long:2014mxa,MacGibbon:1989kk,Steer:2010jk}, gravitational lensing e.g.~\cite{Vilenkin:1984ea,Bloomfield:2013jka,Sazhin:2006kf}. The network can also source a cosmological SGWB which, for cosmic strings, can cover many decades in frequency, see e.g.~\cite{Auclair:2019wcv,Simakachorn:2022yjy}. GW experiments operating in different frequency bands can therefore probe the properties of the network, and hence those of the underlying symmetry breaking phase transition.
In this section we outline the expected constraints on different topological defects from future ground-based 3G detectors, in combination with SKA and LISA.

Consider a symmetry breaking phase transition occurring at some energy scale $\eta$, and suppose the system under consideriation is symmetric under the action of a group $G$, whereas its ground state is only invariant under a subgroup $H$. Then the manifold of degenerate ground states is ${\cal{M}}=G/H$, and it is the topology of ${\cal{M}}$ that determines the kind of defects which may form \cite{Vilenkin:2000jqa,Hindmarsh:1994re}.  If ${\cal{M}}$ is disconnected, then domain walls may form, and they are classified by the zeroth homotopy group $\Pi_0({\cal M}) \neq 1$.
For line-like cosmic strings, ${\cal{M}}$ must contain non-contractible loops which cannot be shrunk to a point, namely $\Pi_1({\cal{M}})\neq 1$. The simplest example is the breaking of a $U(1)$ symmetry for which ${\cal M}=S_1$ and strings of different integer winding number $n$ are formed, but there are many other examples with non-abelian homotopy groups leading to more complex line-like defects. Strings are referred to as local/global if $G$ and $H$ are also local/global symmetries. In the axionic models we discuss in \ref{sec:PQ}, both global strings and domain walls are present.
If at a lower energy scale there is a second symmetry breaking phase transition $G\rightarrow H\rightarrow K$, then hybrid defect networks can form which may or may not be stable depending on the structure of $G/K$ \cite{Kibble:2015twa}.  Examples are strings bounding domain walls \cite{PhysRevD.26.435,KIBBLE1982237} (see also section \ref{sec:PQ}), and  strings bounded by monopoles \cite{Kibble:2015twa,Leblond:2009fq} which appear in certain BSM models, see e.g.~\cite{Buchmuller:2020lbh,Buchmuller:2019gfy}

The precise properties of defects --- for instance their thickness, energy per unit length/surface, dynamics and interactions --- depends on the details of the underlying action. As a simple example, consider the local $U(1)$ Abelian Higgs model \cite{Vilenkin:2000jqa,Hindmarsh:1994re} , consisting in a complex field $\Phi$ and a gauge field $A_\mu$.
The corresponding equations of motion have string solutions of the form (see {\it e.g.} \cite{Vilenkin:2000jqa,Hindmarsh:1994re} for details)
\begin{equation}
   \Phi=f(r)e^{i n\theta} \qquad A_\theta=n \, a(r)/e\,r
\end{equation}
where $r$ is the radial distance from the core, $f(r)$ and $a(r)$ smooth profile functions, $\theta$ the polar coordinate, and $e$ the gauge coupling. 
For $n=1$ and any value of $\beta =m^2_{\rm{scalar}}/m^2_{\rm{gauge}}$
the strings are topologically stable. However, if $\beta<1$, then a winding $n$ string has {\it lower} energy per unit length than $n$ winding 1 strings, and as a result the network of strings contains bound states of strings, and junctions (or zippers) between different winding strings \cite{Bettencourt:1994kc,Copeland:2006eh,Salmi:2007ah}. Such junctions are also characteristic of cosmic superstrings, namely fundamental strings of string theory stretched to cosmological scales, and which can form at the end of brane inflation \cite{Copeland:2003bj}. The vast majority of the literature on local cosmic strings, however, is devoted to the simpler $\beta\geq 1$ limit in which there are no bound states or junctions, and only $n=1$ strings with energy per unit length $\mu \simeq \eta^2$. 
Since these strings cannot have ends (they would cost infinite kinetic energy), a network of such strings formed at the symmetry breaking phase transition will contain both infinitely long strings and closed loops. As we explain below, it is these closed loops which eventually predominantly decay into gravitational radiation and source a cosmological SGWB.  
In certain cases, the string-forming field $\Phi$ may couple to other fields (fermionic or bosonic) leading to additional degrees of freedom condensing in the string core and thus to currents propagating along the strings, see \cite{Witten:1985fp,Lazarides:1986di,Carter:2000wv}. These currents may modify the string dynamics and in particular stabilise small loops: the corresponding `vortons' (namely the stable loops which appear
as point particles having different quantized charges and angular momenta) have been advocated as a dark matter candidate \cite{Auclair:2020wse}.

As opposed to their local counterparts, {\it global} $U(1)$ strings have a logarithmically divergent, time-dependent, energy per unit length, 
$\mu(t) \simeq \eta^2 \ln (m_{\rm{scalar}}/H(t))$ where $H(t)$ is the Hubble parameter. Furthermore, they predominantly decay into scalar radiation as opposed to gravitational radiation. As a consequence, the GW bounds on these strings are weaker for the same symmetry breaking scale $\eta$ (but there are other important cosmological constraints, see below).

To determine these different bounds, it is crucial to understand how a network of defects evolves.  As we  outline below, `standard' ($\beta>1$, non-current carrying) local $U(1)$ string networks reach an attractor, statistically self-similar {\it scaling} solution in which all macroscopic quantities with the dimensions of length and time are proportional to the Hubble length and time.  However, the properties of the scaling solutions are still not entirely understood on all scales, and in particular on those smallest scales in which gravitational wave backreaction (GBR) effects can be important. There are also debates concerning the loop distribution. The situation for global string networks, including their approach (or not) to a scaling solution also remains under intense debate, see subsection \ref{subsec:global}.
The case of domain walls is discussed in subsection \ref{sec:DW}.

\subsection{\label{sec:LocalCS}Local strings}

The finite energy per unit length of `standard' strings is related to the energy scale of symmetry breaking through
\begin{equation}
	G\mu \sim 10^{-6} \qty(\frac{\eta}{10^{16} ~\mathrm{GeV}})^2,
\end{equation}
where $G$ is Newton's constant. All the gravitational properties of strings are determined by $G\mu$, and in particular CMB anisotropies sourced by strings \cite{Ringeval:2010ca} lead to the constraint $G\mu \lesssim 1.7 \times 10^{-7}$ \cite{Ade:2013xla}. 
The width of a string $w\sim 1/\eta$ is tiny compared to its characteristic macroscopic size $\ell$, and for that reason usually the dynamics of strings is described by the Nambu-Goto action (see e.g.~\cite{Vilenkin:2000jqa,Hindmarsh:1994re}). During the evolution of a network, however, strings will also collide. When this happens, there are two possible outcomes: either they simply pass through each other; or different string segments join up in a process known as intercommutation.  Simulations of local $U(1)$ strings have shown that intercommutation occurs with probability\footnote{There are some exceptions at high velocities, see \cite{Achucarro:2006es}} $p=1$ \cite{Matzner}, with the subsequent formation of a discontinuity in the tangent vector of each string known as a {\it kink}.  (Cosmic superstrings have $p\ll 1$.) 

Kinks are important features on strings for GWs. Indeed, it follows from the Nambu-Goto equations of motion in flat space that kinks propagate along the string at the speed of light. Furthermore, solving Einstein's equations in weak field limit, using the stress energy tensor of a string derived from the Nambu-Goto action, it can be shown that  
sharp angle kinks source high-frequency, linearly polarised, short duration GW bursts \cite{Damour:2000wa,Damour:2001bk,Binetruy:2009vt}. There are other features on loops which also source GW bursts, namely {\it cusps} which are points where the string folds back on itself and instantaneously travels at the speed of light, and also kink-kink collisions.  For a loop of length $\ell$ at redshift $z$ and comoving distance $r(z) \gg \ell$ from the observer, the waveform of these GW burst events was calculated in
~\cite{Damour:2000wa,Damour:2001bk,Binetruy:2009vt} and is given by
\begin{equation}
    h_{b}(\ell,z,f) = A_{b}(\ell,z)f^{-q_b}.
\end{equation}
Here $f$ is the GW frequency at the observer, the index $b$ (for burst) describes either cusps (c), kinks (k) or kink-kink collisions (kk), 
\begin{equation}
    q_b=(4/3,5/3,2) \; {\text{for}} \; b={\text{(c,k,kk) }},
    \label{eq:qbdef}
\end{equation}
and the amplitude $A_b\sim G\mu \ell^{2-q_b}(1+z)^{q_b-1}/r(z)$.
For cusp and kink bursts, the emission of GWs is concentrated in a beam of half-angle
\begin{equation}
    \theta_m \sim [f(1+z)\ell]^{-1/3}.
\end{equation}
Since $\theta_m$ scales as an inverse power of $f$, these individual GW bursts should therefore be easier to detect with lower-frequency GW detectors such as LISA \cite{Auclair:2023brk}.  Currently the non-observation of cusp and kink bursts by the LVK collaboration \cite{LIGOScientific:2021nrg} puts an upper limit on $G\mu$ given a loop distribution model, see below. The increased sensitivity of 3G detectors will lead to tighter constraints on $G\mu$ from bursts, however, as we will discuss below, these constraints are not competitive with the SGWB constraints on $G\mu$.

Since loops are the main sources of GWs, it is crucial to understand their distribution, namely $n(\ell,t)d\ell$, the number density of loops with length between $\ell$ and $\ell+d\ell$ at time $t$.  As a result of the self-intercommutation of long strings, loops (with corresponding kinks) are formed by the infinite string network. 
Furthermore, once formed, these loops generally self-intersect many times and fragment into many smaller loops until eventually all loops reach stable non-self-intersecting trajectories. According to the Nambu-Goto equations,  these stable loops evolve periodically with period $T=2/\ell$, and thus  oscillate and emit GWs (possibly other radiation, see below) thus decaying. It is through this process that string network looses energy, and reaches the attractor {\it scaling solution}. 

Determining all the GW signatures of {local} cosmic strings therefore boils down to understanding: \begin{enumerate} 
\item  how a loop radiates energy, and in particular $P_{j}^{{\rm{GW}}}$, the power emitted into GWs of frequency $2j/\ell$ (the $j$th harmonic). 
\item the distribution $n(\ell,t)$ of non-self-intersecting loops of length $\ell$ at time $t$. 
\end{enumerate}

\subsubsection{Energy loss}

At large $j$ the emission from cusp and kink bursts (and also kink-kink collisions) dominates, and using the waveform given above, it follows that \cite{Binetruy:2009vt} 
\begin{equation}
    P^{{\rm{GW}},b}_j \sim j^{-q_b}.
    \label{eq:power}
\end{equation}
Often this power-law behaviour is extrapolated down to $j=1$, though gravitational backreaction effects may modify it see e.g.~\cite{Blanco-Pillado:2017oxo,Blanco-Pillado:2019nto}. The power emitted in each burst type is then $\Gamma^{b} = \sum_{j=1}^{\infty} P^{\mathrm{GW},b}_j$, where the $\Gamma^b$ are given in \cite{Siemens:2006vk,Damour:2001bk,Ringeval:2017eww}.
On a loop with $N_c$ cusps and $N_k$ kinks per oscillation period, there are $N_k^2/4$ kink-kink collisions, and the total GW power emitted is then
\begin{equation}
    \label{eq: total Gamma}
    \Gamma^{\rm{GW}} = N_c\Gamma^{c} + N_k\Gamma^{k} + \frac{N_k^2}{4} \Gamma^{kk}.
\end{equation}
Assuming that loops {\it only} decay into GWs, then it follows that the length of a loop changes with time according to
\begin{equation}
\label{eq:dotl}
   \frac{\dd \ell}{\dd t} = -\Gamma^{\rm{GW}} G \mu \equiv \gamma_d. 
\end{equation}
A non-self-intersecting loop formed with initial length $\ell_i$ at time $t_i$ therefore has lifetime $\ell_i/\gamma_d$. 
Note that, in regions of high curvature, particles and gauge boson radiation may also be emitted \cite{Blanco-Pillado:1998tyu,Olum:1998ag,Matsunami:2019fss,Auclair:2019jip,Hindmarsh:2021mnl}, leading to modifications of Eq.~(\ref{eq:dotl}), for example with terms that dominate at low $\ell \sim w$ with $w$ the string width \cite{Blanco-Pillado:1998tyu,Olum:1998ag,Auclair:2019jip}. A further consequence is that cosmic strings become multi-messenger sources, emitting both GWs and possibly a diffuse gamma ray background, see \cite{SantanaMota:2014xpw,Auclair:2019jip,Auclair:2021jud,Hindmarsh:2022awe}.

\subsubsection{Non-self-intersecting loop distribution}

The general picture is that due to intercommutation of infinite strings (and self-intercommutation of loops), loops are permanently produced over time together with numerous kinks. The build up of kinks means that the strings become more `wiggly' (small-scale structure), and the resulting distribution of kink angles as well as the inter-kink distance was determined in \cite{Copeland:2009dk} where it was shown that the backreaction of emitted GWs on the shape of a kink is expected to be a crucial effect in order for this interkink distance to scale.
The loops produced eventually reach non-self-intersecting trajectories, and it is the distribution $n(\ell,t)$ of these loops at time $t$ which we are after.

This distribution is generally assumed to satisfy a Boltzmann equation which, neglecting collisions between loops (and also self-intersections of loops), is given by \cite{Copeland:1998na, Peter:2013jj}
\begin{equation}
	\label{eq: continuity equation}
	\eval{\pdv{t}}_{\ell} \qty[a^3 n(\ell,t)] + \eval{\pdv{\ell}}_{t}\qty[\dv{\ell}{t} a^3 n(\ell,t)] = a^3 {\cal{P}}(t,\ell)\,,
\end{equation}
where $a(t)$ is the scale factor and $\dot{\ell}$ is given in Eq.~\eqref{eq:dotl} assuming only GR from loops. (See \cite{Auclair:2019jip,Auclair:2020wse} for solutions of Eq.~\eqref{eq: continuity equation} including other forms of radiation, namely different $\dot{\ell}$.) The source term on the right hand side is the loop production function, with ${\cal{P}}\dd{\ell}\dd{t}$ the number of non self-intersecting loops with length between $\ell$ and $\ell+\dd{\ell}$ which are produced at times $t$ to $t+\dd{t}$ by the infinite string network.   Since the infinite string network has been shown to scale in many simulations
\cite{Bennett:1985qt,Bennett:1986zn,Bennett:1987vf,Bennett:1989yp,Allen:1990tv,Allen:1991jh,Ringeval:2005kr,Blanco-Pillado:2013qja} then by dimensional analysis 
\begin{equation}
    t^5 {{\cal{P}}(t,\ell)} = f(\ell/t) \equiv f(\gamma).
\end{equation}
To determine the form of $f(\gamma)$ it is generally necessary to resort to numerical simulations and analytical modelling. Indeed, since Nambu-Goto simulations do not include effects from either gravitational radiation (GR) or gravitational backreaction (GBR) on the string dynamics (important for smoothing out kinks for example), they give a handle on $f(\gamma)$ only on large scales $\gamma \gg \gamma_d$, where $\gamma_d$ was defined in Eq.~\eqref{eq:dotl}. Another approach is to use numerical simulations to determine the distribution of non-self-intersecting loops at time $t$ and on scales $\gamma \gg \gamma_d$, combined with analytical forms of $f(\gamma)$ including GBR derived in \cite{Rocha:2007ni,Polchinski:2006ee,Dubath:2007mf}. The parameters in $f$ are then fixed so that the solution of \ref{eq: continuity equation} agrees with numerical results on large scales. 

In the following we consider two different loop distributions which we label as Models A and B using the convention of \cite{LIGOScientific:2021nrg}.  In model A, based on \cite{Blanco-Pillado:2013qja}, all loops are produced with initial size $\gamma=0.1 \gg \gamma_d$ at time $t$ so that $f 
=C\delta (\gamma - 0.1)$, with $C$ fit to Nambu-Goto simulations. In model B, based on \cite{Lorenz:2010sm}, loops of all sizes are assumed to be produced with a power-law form $f 
=C' \gamma^{2\chi-3}\Theta(0.1-\gamma)\Theta(\gamma - \gamma_c)$, where the exponent is negative meaning that more smaller loops are produced. The scale $\gamma_c \ll \gamma_d$ physically attempts to incorporate the effects of gravitational back-reaction: below this scale, kinks are rounded off and so the strings are straighter and less likely to intercommute and form loops. Then, once $n(\ell,t)$ is determined from (\ref{eq: continuity equation}), $C'$ and $\chi$ are fit to (different) numerical simulations \cite{Ringeval:2005kr} on scales $\gamma \gg \gamma_d$.  While models A and B agree with each other (and by construction with Nambu-Goto numerical simulations) for $\gamma \gg \gamma_d$, they differ enormously on smaller scales, where model B has many smaller loops \cite{Auclair:2019zoz,Auclair:2020oww}.\footnote{Note that ${\cal{P}}(\ell,t)$ (or rather its integral over $\ell$) in turn is related to the energy lost from the infinite string network through the formation of loops. {\it If} the formation of non-self-intersecting loops is  assumed instantaneous (i.e.~ignoring loop fragmentation for instance) then model B has been criticised for non-conservation of energy of the infinite string network in \cite{Blanco-Pillado:2019vcs}. See also \cite{Blanco-Pillado:2019tbi}. Counter-arguments include the fact that fragmentation is an important effect observed in many simulations, which changes the energy conservation argument \cite{Auclair:2021jud}. Here we are agnostic regarding these two models.}

Note that models A and B are both calibrated on Nambu-Goto simulations of cosmic string networks. However, large scale field-theory simulations of networks of Abelian-Higgs strings have also been developed \cite{Vincent:1997cx,hep-ph/0107171,astro-ph/0605018,Bevis:2010gj,Daverio:2015nva,Correia:2019bdl,Correia:2020gkj,Correia:2020yqg}. Again these  do not include gravitational radiation, but allow for classical radiation into the scalar and gauge fields. They observe that loops decay within a Hubble time to classical radiation of scalar fields. In other words, $n(\ell,t)$ appears to be essentially zero so that loops do not contribute to GWs. These results as yet need to be understood further, and tied in with the results of Nambu-Goto simulations. Indeed, note that some high resolution field theory simulations of {\it {individual}} loops do show them evolving according to approximate Nambu-Goto dynamics \cite{Matsunami:2019fss,Hindmarsh:2021mnl,Blanco-Pillado:2023sap}.
The discrepancy between these two sets of simulation results is under debate (it may be down to a question of scales which must be resolved, from the tiny string width $w$ to the Horizon size which differ by orders of magnitude).  
We work with Nambu-Goto based results in the following.

\subsubsection{Stochastic GW background from local strings}

The energy density in GWs observed at a given frequency $f$ today, sourced by a local string network, can now be obtained by summing over the GW emission produced from all the oscillating loops produced during the evolution of the string network from its formation until today, and taking into account the redshifting of frequencies from the moment of emission until today. In terms of the number density of loops ${n}(\ell,t)$ 
~\cite{Blanco-Pillado:2017oxo}

\begin{equation}\label{eqn:omega-method-1}
   \mathrm{\Omega}_{\rm{GW}}(\ln f) = \frac{16\pi (G\mu)^2 }{3 H_0^2 f}\sum_{b=(c,k,kk)}\frac{N_{b}\Gamma^{b}}{\zeta(q_b)}\; \sum_{j=1}^\infty   j^{1-q_b} \int_0^{z_f} \frac{\d z}{H(z) (1+z)^6}~{n}\left(\frac{2j} {(1+z)f},t(z)\right)\,\,.
\end{equation}
Here $\zeta(q_b)$ is the Riemann zeta function, $q_b$ is given in Eq.~\eqref{eq:qbdef}, and $z_f$ is the redshift below which friction effects on the string dynamics becomes negligible \cite{Vilenkin:2000jqa}.
An alternative but entirely equivalent approach is to calculate the incoherent superposition of burst signals from cusps, kink and kink-kink collisions emitted from all the loops in the network, removing strong infrequent bursts that can be resolved individually, see e.g.~\cite{LIGOScientific:2021nrg}.

Clearly the properties of the loop distribution ${n}(\ell,t)$ are crucial to determine the SGWB spectrum. Since model B has 2 characteristic length scales $(\gamma_c,\gamma_d)$, the resulting SGWB spectrum has two characteristic frequency scales $f_{d}\sim H_0/\gamma_d$  and $f_{c}\sim H_0/\gamma_c \gg f_d$. For model A, there is only $f_d$. As has been discussed elsewhere, the spectrum is scale invariant at high frequency ($f\gg f_d$ for model A, $f\gg f_c$ for model B) due to the properties of the loop distribution in the radiation era. Its amplitude however is model dependent, and is much larger for model B since this contains more small loops.

In the following we assume standard flat $\Lambda$CDM cosmology with Planck-2018 fiducial parameters, which determines $H(z)$ and $t(z)$. We also fix $N_c=2$ and $N_k=0$ for which $\Gamma^{\rm{GW}} \simeq 57$, 
a value of the order of that determined numerically for many different loop shapes \cite{PhysRevD.31.3052,Burden:1985md,Garfinkle:1987yw,Blanco-Pillado:2017oxo}.
Note that both the LVK and EPTA collaborations considered $N_k$ as a free parameter together with $G\mu$ \cite{LIGOScientific:2021nrg,EPTA:2023xxk}.  The explicit expressions for $n(\ell,t)$ for the two models may be found for e.g.~in the appendix of \cite{LIGOScientific:2021nrg}.

\subsubsection{Constraints from 3G detectors}

Figure \ref{fig:CS+DW_spectra_PTA_values} shows $\Omega_{\rm{GW}}(f)$ for both models, calculated using \eqref{eqn:omega-method-1}. 
{For Model A (solid, thin line), $G\mu$ has been fixed to $G\mu = 7.9\times 10^{-11}$, the value providing the best fit of the EPTA DR2 data set, assuming strings are the {\it unique source} of the SGWB detected by PTA, see \cite{EPTA:2023xxk}. 
In this case, strings formed with $G\mu= 7.9\times 10^{-11}$ would produce a signal in ET/CE with an SNR of more than {$500$}, providing a wonderful opportunity for coincident detection and thereby a confirmation of the origin of the SGWB in the PTA band.
For model B, on the other hand, the value providing the best fit of the EPTA DR2 data set would correspond to $G\mu = 2.5\times 10^{-11}$, giving a very loud signal in the Earth-based interferometers band: indeed, such large tensions are already excluded by the LVK data, for which the current constraint on Model B is $G\mu \lesssim {\rm{few}}10^{-15}$ \cite{LIGOScientific:2021nrg}.  
Therefore, for Model B, we have chosen one example of signal detectable only within the Earth-based interferometers band, with $G\mu = 4\times 10^{-16}$ (dot-dashed line).}\footnote{Note that there the spectrum is not flat since we are in the regime $f_d \ll f \ll f_c$.}

Figure \ref{fig:CE_ET_SNR_CS} shows the SNR of local cosmic strings in CE and ET as a function of $G\mu$, again for both models. As summarized in \autoref{tab:string_tension_tab}, for Model A, CE will be able to detect strings with an SNR greater than the foreground from compact binary mergers for $G\mu \gtrsim 3 \times 10^{-14}$, whereas for ET, the corresponding value is $G\mu \gtrsim 4\times 10^{-13}$. However, with a precise estimation of the astrophysical foreground, CE and ET detectors might be able to impose even more stringent constraints on $G\mu$ down to $10^{-16}$, respectively {few} $10^{-15}$, for Model A, considering a SNR threshold of 5 (see \autoref{tab:string_tension_tab}).
{This depends in particular on how the frequency shape of the SGWB spectrum differs from the one of the foregrounds.}
For comparison, the current LVK constraints on Model A are $G\mu \lesssim 9.6 \times 10^{-9}$, whereas PTA constrains $G\mu \lesssim 10^{-11}$. 
Forecasts for LISA have shown that it will be able to probe both models A and B, provided the string tension $G\mu \gtrsim {\cal{O}}(10^{-17})$ \cite{Auclair:2019wcv,Blanco-Pillado:2024aca}. Thus, next generation ground-based interferometers will be competitive with LISA.

\begin{table}[htbp]
  \centering
  \begin{tabular}{|c||c|c|c|c|}
    \hline
     & \multicolumn{2}{c|}{Model A} & \multicolumn{2}{c|}{Model B} \\
    \cline{2-5}
     & \multicolumn{1}{c|}{SNR $> 5$} & \multicolumn{1}{c|}{SNR $> \rm{SNR}_{\rm{AFG}}$} & \multicolumn{1}{c|}{SNR $> 5$} & \multicolumn{1}{c|}{SNR $> \rm{SNR}_{\rm{AFG}}$} \\
    \hline
    CE & $\log_{10} G\mu > -16.0$ & $\log_{10} G\mu > -13.5$ & $\log_{10} G\mu > -18.7$ & $\log_{10} G\mu > -16.9$ \\
    ET & $\log_{10} G\mu > -14.5$ & $\log_{10} G\mu > -12.4$ & $\log_{10} G\mu > -17.4$ & $\log_{10} G\mu > -16.2$ \\
    \hline
  \end{tabular}
  \caption{For both loop density models A and B, we provide the value of the string tension for which the SNR in the associated detector will be greater than 5, or greater than the SNR of the astrophysical foreground (AFG) expected in the detector (see \autoref{tab:detectors} for details).}
  \label{tab:string_tension_tab}
\end{table}

\subsubsection{Bursts searches}

GW detectors will also search for individual loud linearly polarised gravitational wave bursts from cusp/kink on nearby string loops. Their observation, or non-observation, can put further constraints on $G\mu$ which may --- or may not --- be competitive with the constraints coming from the SGWB above.  Furthermore, as loops are periodic in time, these bursts should repeat periodically \cite{Auclair:2023mhe}, thus increasing their detectability.  
The LVK O3 constraints on Models A and B from the non-observation of bursts is not as competitive with that derived from the stochastic search \cite{LIGOScientific:2021nrg}, and the same is true for ET where the non-detection of bursts should constrain $G\mu \lesssim 10^{-9}$ for model A and $G\mu \lesssim 10^{-11}$ for model B \cite{PAprivatecomm}. 
 The same is true for LISA: non-detection of these bursts would constrain $G\mu \lesssim 10^{-11}$ for both models \cite{Auclair:2023brk}.

\subsubsection{Beyond `standard' strings}

So far we have focused on `standard' strings and $\Lambda$CDM cosmology. The effects of non-standard cosmology, e.g.~an early matter domination or kination or a late secondary inflationary stage inside the radiation era, on such strings are very significant and may
break the scale-invariant properties of the SGWB spectrum at high frequencies, see~\cite{Gouttenoire:2019kij,Gouttenoire:2021jhk} for further details.  Indeed the loop distribution is sensitive to the evolution of $a(t)$, see Eq.~\eqref{eq: continuity equation}.

Returning to standard cosmology, if the strings also radiate particles at cusps and kinks, then this changes $\dot{\ell}$ and thus the loop distribution on small scales \cite{Auclair:2019jip}. As expected, the effect on the SGWB is to cut off the spectrum at high frequencies, $f_{c}\sim (H_0 \gamma_d/w)$ for cusps (the kink cutoff frequency is higher). Generally $f_c$ is  much larger than the Hz scale of ground-based 3G detectors for all $G\mu$ of interest \cite{Auclair:2019jip}.
As mentioned above, particle radiation opens up the possibility of other observational signatures beyond GWs from cosmic strings, see \cite{SantanaMota:2014xpw,Auclair:2019jip,Auclair:2021jud,Hindmarsh:2022awe}.

If the strings become unstable at late times, for instance due to a second phase transition creating monopole-antimonopole pairs, then this leads to a low-frequency cutoff on the SGWB meaning that strings could evade the PTA constraints \cite{Buchmuller:2019gfy,Servant:2023tua}.Finally, if strings carry currents, their evolution is altered (they no longer satisfy the Nambu-Goto action), see \cite{Rybak:2023jjn} and references therein. In \cite{Auclair:2022ylu} it was shown that if currents are present only for a certain amount of time between the instant of network formation and today, then they can give a sizable SGWB in the 3G frequency band.

\begin{figure}[htbp]
    \centering
\includegraphics[width=0.6\textwidth]{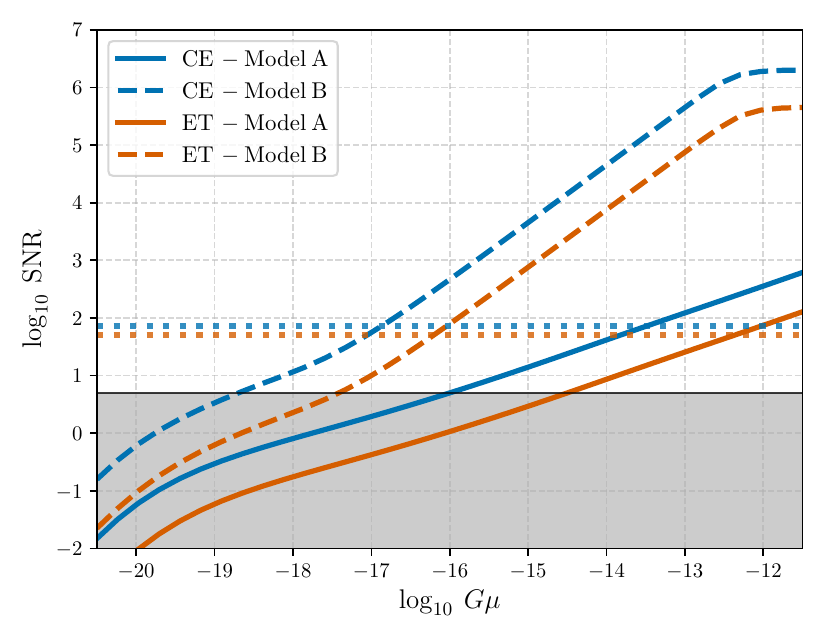}
    \caption{$\log_{10}$ SNR of the SGWB from a local cosmic string network, as a function of the string tension $G\mu$. We assume $T_{\rm{obs}}=$ 1yr for both CE and ET. The gray shaded region corresponds to SNR of the signal $< 5$, while the dotted lines correspond to the respective SNR value of the astrophysical foreground from compact binary mergers in CE (blue) and ET (orange) detectors (see \cref{tab:detectors}).}
    \label{fig:CE_ET_SNR_CS}
\end{figure}

\subsection{\label{subsec:global}Global \texorpdfstring{$U(1)$}{U(1)} strings}

Due to their divergent energy per unit length $\mu(t) \simeq \eta^2 \ln (m_{\rm{scalar}}/H(t))$, understanding the evolution of a network of global strings poses new challenges and uncertainties.
Furthermore, loops now radiate both GWs, radial modes, and Nambu-Goldstone bosons (which in the following we refer to as axions), and the total rate of energy loss can be written as $\Gamma=\Gamma^{\rm{GW}} + \Gamma^a + \Gamma^{\rm{radial}}$. (See \cite{Drew:2019mzc,Drew:2022iqz,Drew:2023ptp} for recent work on axion and radial mode radiation from global strings, particularly in regions of high curvature.) 
The main contribution to the energy loss is axion radiation, which dominates over GW emission by a factor of $\Gamma^{\rm{GW}}/\Gamma^a = G\mu^2/\eta^2$. Thus, the GW background from a network of global strings is significantly suppressed compared to the axion one, and for this reason we do not present it here (see, however,~\cite{Gouttenoire:2019kij,Gorghetto:2021fsn, Chang:2021afa,Gouttenoire:2021jhk,Servant:2023mwt} for studies on the possibility to detect GWs from global strings at NANOGrav, LISA and ET/CE). It is still crucial, however, to understand  the evolution of a network of global strings, since its properties are required to understand the spectrum of axions produced and thus the cosmological constraints which will be imposed on the model (see section \ref{sec:PQ} for a concrete example).

For completeness, we very briefly highlight some of the significant differences between local and global string evolution, and some still debated points regarding global strings.  Global strings satisfy the Kalb-Ramond as opposed to the Nambu-Goto action \cite{PhysRevD.9.2273,DAVIS1988219} from which, assuming the axion radiation force is small, it follows that global string loops remain approximately periodic in time as in the Nambu-Goto case, but with the logarithmic (renormalised) string tension $\mu(t) \simeq \eta^2 \ln (m_{\rm{scalar}}/H(t))$.
Their lifetime is however much shorter than that of local strings.  There is a long standing debate over the spectrum of radiation into axions into each harmonic mode $n$, namely (using analogous notation to Eq.~\eqref{eq:power})
$P_j^{{a}}$ \cite{1987PhLB..195..361H,PhysRevD.32.3172,Battye:1997jk}. The backreaction of axion radiation on kinks and cusps may smooth off these features, and hence also affect GR emission and thus the precise form of $q$ in the GW emission, $P_j^{{\rm GW}}=j^{-q}$, see~\cite{Chang:2021afa, Gorghetto:2021fsn, Baeza-Ballesteros:2023say} for recent results from numerical simulations (in particular~\cite{Baeza-Ballesteros:2023say} suggest $q$ is lower than the standard $5/3$ value for cusps). The GW signal is additionally affected by the behavior of the string network: several recent works~\cite{Gorghetto:2018myk, Gorghetto:2020qws, Buschmann:2021sdq, Saikawa:2024bta} find (logarithmic) deviations from the scaling regime, implying that there can be more strings per Hubble volume at late times in the network evolution (see~\cite{Hindmarsh:2019csc} for a different perspective).

\subsection{\label{sec:DW}Domain Walls}

Domain walls (DWs) are receiving a revived interest currently, mainly because they 
are efficient sources of GWs. As mentioned above, DWs arise whenever a discrete symmetry is 
spontaneously broken leading to disconnected vacuum manifold, $\Pi_0({\cal{M}}) \neq 1$.
The Standard Model (SM) does not lead to DWs but they appear in a variety of SM extensions, remarkably in  axion models. 
In the early universe, phase transitions that break spontaneously a discrete symmetry result in a cosmological 
DW network, a random array of walls formed at all Hubble patches at the same ``formation" time.
DW networks were studied in the past, and it is well known that they have a strong impact on cosmology, see \cite{Vilenkin:2000jqa,Vachaspati:2006zz} for reviews.

The evolution of the DW network is driven to an attractor `scaling' regime with statistically self-similar `scaling', 
essentially described by having $O(1)$ Hubble-sized walls in each Hubble volume at any time $t$. 
This leads to an average network density  
\begin{equation}\label{rhoDW}
    \rho_{DW} = c\, \sigma \, H
\end{equation}
with $\sigma$ the DW tension and $c$ an $O(1)$ parameter that depends on the model.\footnote{This is a faster decay than that of a naive 'DW gas’ (an ensemble of non-interacting DWs) which gain area (and so, mass) as  $a(t)^2$ and would dilute as $1/a(t)$. The scaling regime is sustained instead because the DWs continuously loose energy 
by wall-wall reconnections and by scalar radiation.}

The decay \eqref{rhoDW} is very slow, leading to the so-called ``DW problem". In practice, this means that viable DW models must be equipped with some annihilation 
mechanism\footnote{Or, the DW tension $\sigma$ is tiny, $\sigma\lesssim \text{MeV}^3$ \cite{Zeldovich:1974uw,Lazanu:2015fua}. 
However in this case the GW signal is also small and at ultra-low frequencies \cite{Ramberg:2022irf,Ferreira:2023jbu}.}. 
Various annihilation mechanisms are known, for instance i) a small explicit breaking of the discrete symmetry, ii) an asymmetric population of the degenerate vacua at formation \cite{Larsson:1996sp} or iii) symmetry restoration~\cite{Vilenkin:1981zs}. For simplicity, let us focus on the first mechanism. 

Any explicit breaking leads generically to a vacuum non-degeneracy, $\Delta V$, which acts as a pressure force that drives the DWs, favouring the amount of volume occupied by the true vacuum. This is efficient roughly when the Hubble rate decreases below a certain `annihilation' temperature given by
\begin{equation}
    H(T_{\rm{ann}})\sim \frac{\Delta V}{\sigma}~.
\end{equation}
The details of the annihilation process, including the resulting GW amplitude and spectrum, depend on the mechanism and require intensive numerical simulations.

The evolution of DW networks has been computed in several models using field theory numerical simulations, confirming that the scaling regime is attained.
Results on the produced GWs and their spectrum were confined until recently to networks in the scaling regime, found to be well described by a broken power law with exponents $3$ and $-1$ \cite{Hiramatsu:2010yn,Hiramatsu:2012sc,Hiramatsu:2013qaa,Krajewski:2021jje}. 
While in this work we focus on  GW production only during the scaling regime, we notice that recent progress~\cite{Kitajima:2023cek, Chang:2023rll, Ferreira:2024eru} points to significant emission of GWs throughout the annihilation phase, enhancing the signal by about 1 order of magnitude. We comment below on how our results are affected by the annihilation phase.

For the sake of performing searches and detection prospects, the most important characteristics of the SGWB spectrum are the peak frequency and amplitude. In a model independent fashion one can define, as for PTs,  the fraction of the total energy density in DWs at $T_{\text{ann}}$, namely
\begin{equation}
\label{eq:alphann}
\alpha_{\text{ann}} \equiv  \frac{\rho_{\text{\tiny DW}}}{3 H^2 M_p^2}\Big|_\text{ann}
\end{equation}
The GW spectrum is then given by (see e.g.~\cite{Ferreira:2022zzo})
\begin{align}
\label{eq:omegadw}
&\Omega_\text{\tiny GW,DW}(f)h^2\simeq 10^{-10}\, \tilde{\epsilon}\left(\frac{10.75}{g_{*}(T_\text{ann})}\right)^{\frac{1}{3}}\left(\frac{\alpha_{\rm{ann}}}{0.01}\right)^{2} S\left(\frac{f}{f_{p}^{0}}\right),
\end{align}
where $\tilde{\epsilon}\simeq 0.1-1$ is an efficiency parameter to be extracted from numerical simulations~\cite{Hiramatsu:2013qaa},
and the peak frequency is
\begin{eqnarray} \label{eq:pfreq}
	f_p^0\simeq 10^{-9} \, \text{Hz} \, \left(\frac{g_{*}(T_{\text{ann}})}{10.75}\right)^{\frac{1}{6}} \left(\frac{T_\text{ann}}{10\,\text{MeV}}\right).
\end{eqnarray}
The function $S(x)$ describes the spectral shape of the signal.
A common parametrization is
\begin{equation}
\label{eq:spectrumdw}
S(x)= \frac{4}{x^{-3} + 3x},
\end{equation}
so that the spectral slopes at $f\ll f_{p}^0$ are $3$ as dictated by causality (assuming radiation domination) and $-1$ and for $f\gg f_{p}^0$ as found in most numerical analyses \cite{Saikawa:2017hiv}.\footnote{{Note that numerical results for $\mathbb{Z}_N$ hybrid string-wall networks suggest that the exponent can be flatter at high frequencies, decreasing with $N$ \cite{Hiramatsu:2012sc}.}} 
The spectrum is cut off at frequencies larger than the (redshifted) inverse of the wall width.

In the minimal $Z_2$ model above, one finds \cite{Ferreira:2022zzo}
\begin{equation}
\label{eq:TannDW}
T_\text{ann}\simeq \frac{5\,\text{MeV}}{\sqrt{c}}\left(\frac{10.75}{g_{*}(T_\text{ann})}\right)^{\frac{1}{4}}\left(\frac{\Delta V^{1/4}}{10\,\text{MeV}}\right)^2
\left(\frac{10^5\,\text{GeV}}{\sigma^{1/3}}\right)^{\frac32},
\end{equation}
and 
\begin{equation}
\label{eq:alphann2}
\alpha_{\text{ann}} \simeq c\, \sqrt{\frac{g_{*}(T_\text{ann})}{10.75}}\left(\frac{\sigma^{1/3}}{10^{5}\,\text{GeV}}\right)^3\left(\frac{10\,\text{MeV}}{T_\text{ann}}\right)^{2}.
\end{equation}

\begin{figure}[htbp]
  \centering
\includegraphics[width=0.6\textwidth]{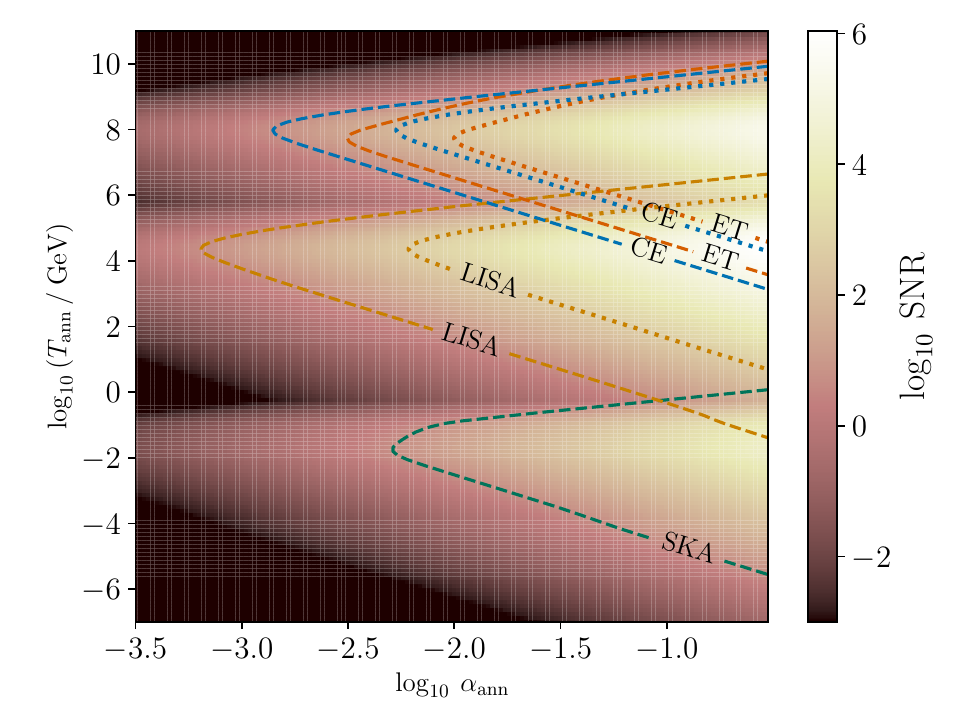}
    \label{fig:DW_SKA+LISA}
  \caption{Expected SNR for DWs for all next generation GW detectors as a function of $T_{\rm{ann}}$ and $\alpha_{\rm{ann}}$.
  The color grid corresponds to the maximal SNR among all the detectors.
  The DW spectra are computed using Eq.~\eqref{eq:omegadw} with $\tilde{\epsilon}=0.7$. 
  Dashed lines: contours with SNR = 5 for each detector. Dotted lines: contour corresponding to the SNR value of the respective astrophysical foregrounds in LISA, CE and ET (see section \ref{sec:detectors}).  To the right of the dotted lines, the DW SNR is above that of the foreground. {For the parameter space between the dotted and the dashed contours, detection can still be possible depending on the particular frequency shape of the SGWB signal}. Note that the region $\alpha_\text{ann}\gtrsim 0.1$ is subject to constraints from PBH overproduction~\cite{Ferreira:2024eru}.} 
  \label{fig:DW_SNR_grid}
\end{figure}

As shown in Fig.~\ref{fig:DW_SNR_grid}, future GW detectors will be a powerful probe of DW network models. Their reach to the SGWB produced by a DW network is shown in terms of the annihilation temperature $T_{\rm{ann}}$ and the fraction of the energy density $\alpha_{\rm {ann}}$.
As expected, ground based detectors are sensitive to annihilation temperatures around $10^8$GeV. However, due to the quite slow $1/f$ decay of the power spectra at high frequency, the range of temperatures with high SNR extends to much lower values. 
In section \ref{sec:PQ} we perform a more specific analysis, focused on the GW signals from axionic DW networks in heavy axion-type QCD axion models.

Before moving to the next section, let us comment on the recent inclusion of the annihilation phase~\cite{Kitajima:2023cek, Ferreira:2024eru}: this can roughly be included by replacing $T_\text{ann}$ in the equations above with the temperature at which the annihilating network stops radiating GWs, $T_\text{GW}\sim 0.5~T_\text{ann}$, and thereby also replacing $\alpha_\text{ann}$ with $\alpha(T_\text{GW})\sim 4\alpha_\text{ann}$.  Importantly, the region of large $\alpha_\text{ann}\gtrsim 0.1$ in Fig.~\ref{fig:DW_SNR_grid} is likely constrained by the overproduction of Primordial Black Holes (PBH) from DW annihilation, see~\cite{Ferreira:2024eru}, although a precise bound is subject to theoretical uncertainties and model dependence.

\section{\label{sec:PQ}Case study: GW backgrounds from axions}

Unlike LISA and PTAs, whose frequency ranges encompass the two fundamental scales of the SM (the EW scale and the QCD scale respectively), the frequency band of ground-based interferometers corresponds to temperatures that are much larger than SM scales, roughly $T\sim(10^7-10^9)~\text{GeV}$ (this range of course depends on the typical timescale of the source, e.g. $\beta^{-1}$ for a first order PT given in \cref{eq:betaoverH}, among other things, see above).  Under the (feasible) assumption that the Universe was reheated to such temperatures after inflation, is there any well-motivated beyond the SM phenomenon which could be probed by ground-based interferometers?

The QCD axion stands out with particular interest: it is motivated as a solution to the strong CP problem of the SM, and a Dark Matter candidate. 
The essential idea in axion models is the Peccei-Quinn (PQ) mechanism, consisting in introducing a new  global $U(1)_{PQ}$ symmetry with two properties: it is spontaneously broken at a high scale $F_a$ and it is anomalous under QCD ~\cite{Peccei:1977hh, Peccei:1977ur}. The simplest realization of  this mechanism is effectively described by an additional complex scalar field 
\begin{equation}
\Phi = \phi \, e^{i \,a / F_a}    
\end{equation}
(and possibly additional heavy color charged fermions~\cite{Kim:1979if, Shifman:1979if} or extra Higgs doublets~\cite{Dine:1981rt, Zhitnitsky:1980tq}, which are not relevant for our discussion). This scalar acquires a VEV, $\langle \phi \rangle \sim F_a$ 
leaving behind a Goldstone boson, the QCD axion~\cite{Weinberg:1977ma, Wilczek:1977pj}, whose interactions with SM particles are suppressed by the axion decay constant $F_a$. 
Astrophysical~\cite{Raffelt:1990yz, MillerBertolami:2014rka,Ayala:2014pea,Dolan:2022kul} (see also~\cite{Bar:2019ifz, Chang:2018rso, Carenza:2020cis}) and CMB observations (see~\cite{Notari:2022ffe} for a recent update) impose a lower bound on $F_a\gtrsim (10^7-10^8)\text{GeV}$ (stronger bounds apply in specific models). 
In `post-inflationary' scenarios (reheating temperature is higher than $F_a$), the axion can be produced non-thermally and behave like cold dark matter; requiring that the axion does not overclose the Universe then leads to an upper bound on $F_a\lesssim (10^{10}-10^{11})~\text{GeV}$~\cite{Gorghetto:2018myk, Gorghetto:2020qws, Buschmann:2021sdq, Saikawa:2024bta}.
We thus see that the viable energy/temperature scale $(10^7-10^8) \text{GeV} \lesssim F_a\lesssim (10^{10}-10^{11})$ GeV has significant overlap with the epoch that can be probed by ground-based interferometers.  

As sketched in Fig.~\ref{fig:PQrecap}, the PQ mechanism leads to two potential sources of GW relic backgrounds from the early universe. The first is from the cosmological $U(1)_{PQ}$ phase transition itself when the temperature $T$ was of order $F_a$. We devote Sec.~\ref{PQ-PT} to this possibility.

The other potential source arises from the axionic topological defects that follow after the PQ transition. The spontaneously broken $U(1)_{PQ}$ leads to a network of global strings. At lower temperature, $T\sim \Lambda_{\mathrm{QCD}}$, non-perturbative QCD effects induce a further breaking of $U(1)_{PQ}$ down to a discrete subgroup  $Z_{N_\text{DW}}$, and domain walls appear. 
The integer $N_\text{DW}$ is the number of DWs attached to each string  (microphysically it is set by the assignment of PQ charges of SM and/or heavy quarks). The subsequent evolution depends dramatically on whether $N_\text{DW}=1$ or $N_\text{DW}>1$. For $N_\text{DW}=1$, the string network is quickly annihilated at $T\sim \Lambda_{\mathrm{QCD}}$ (and a tiny amount of GWs are produced). For $N_\text{DW}>1$, instead, the annihilating effect from the DWs gets compensated and so one enters an epoch with a hybrid string-wall network. An additional small breaking of $U(1)_{PQ}$ is needed to avoid DW domination. Since $U(1)_{PQ}$ is a global symmetry, very minimal addition to the model ({\it e.g.}, gravity) can provide the small vacuum splitting $\Delta V$ for the DW epoch to reach an end \cite{Ferreira:2022zzo}. 
The GW signal from axionic DWs is higher in heavy axion realizations of the PQ mechanism (which have particular interest as they address the so-called axion quality problem \cite{Ferreira:2022zzo}). Remarkably, then, there is a class of interesting QCD axion models with a DW epoch, and so, a sizeable GW signal. We devote Sec.~\ref{PQ-DWs} to them.

\subsection{\label{PQ-PT}GWs from a first-order Peccei-Quinn phase transition}

The spontaneous breaking of the PQ symmetry occurs via a PT, which can be first order, depending on the microphysical implementation of the PQ mechanism and on model parameters. 
In this case, bubble collisions and/or bulk fluid motions can source a GW signal that is strong enough to be observable at future ground-based interferometers. 
A strong first order PT can arise in a variety of PQ models, the options ranging from supersymmetric embeddings of the PQ mechanism to  quasi conformal models, both based on weakly or strong dynamics \cite{VonHarling:2019rgb}, and based both on KSVZ-like \cite{DelleRose:2019pgi} or DFSZ-like \cite{VonHarling:2019rgb} models.

For simplicity we focus now on scenarios with an approximately conformal dynamics. This fixes the potential for the PQ field to take the form
\begin{equation}\label{Vconf}
V= \frac{1}{4}\lambda_\phi(\phi)\,\phi^4,
\end{equation}
with $\lambda_\phi(\phi)$ a slowly varying function of $\phi$.
There are two well known implementations leading to \eqref{Vconf}:  classically scale invariant models where the potential \eqref{Vconf} is induced radiatively \`a la Coleman-Weinberg (CW); and models where the PQ symmetry is embedded in a strongly-interacting 
{conformal field theory}
with a nearly marginal deformation. 
Both ideas have been mostly explored in relation to the EW phase transition (see~\cite{Witten:1980ez}, and~\cite{Iso:2017uuu} for a CW first order PT, ~\cite{Creminelli:2001th, Randall:2006py, vonHarling:2017yew, VonHarling:2019rgb, Agashe:2019lhy, Agashe:2020lfz, Mishra:2024ehr}  for the strongly interacting case). Here we focus  {instead} on the PQ case, as discussed in~\cite{VonHarling:2019rgb} (see also~\cite{DelleRose:2019pgi}). 

In both cases of interest, the first order PT can occur with a significant amount of supercooling, meaning that it completes at temperatures {much smaller than the false vacuum energy}
$T\ll V_0(\varphi_f)^{1/4}$.
In this case, the Universe undergoes a late phase of inflation, since the PQ field is stuck in a local minimum for a long epoch. The transition thus occurs in an inflationary background. 
As discussed  {in section \ref{sec:PT}}, in these cases the mere occurrence of nucleation,  given by the condition $\Gamma(T_n)\simeq H(T_n)^4$ (or, more precisely, by \cref{eq:nucl_one_bubble}), may not be sufficient to ensure that the transition also completes.
To ensure completion, the stronger criterion given in \cref{eq:volume_decrease} on the false vacuum volume decrease at percolation must be imposed. 
We study when the criterion in \cref{eq:volume_decrease} is fulfilled, and find that this occurs in relevant regions of the parameter space.
By accounting for this criterion, we improve on the analysis of~\cite{VonHarling:2019rgb}. 
In some regions of the parameter space, the first order PT completes without dramatic supercooling; in this case, we have assessed completion of the PT by  {imposing} the percolation criterion, i.e.~\cref{eq:percolation_T}. 
On the other hand, in some other regions of the parameter space of the PQ models considered in subsection \ref{sec:strong_dynamics}, the false vacuum volume may start decreasing enough only after percolation. In these cases, completion of the phase transition is questionable. Nonetheless, we include this region of parameter space in our analysis, by imposing a milder criterion, namely that the false vacuum fraction decreases at some temperature, which can be lower than $T_*$.  
In these cases, we identify this temperature with the completion of the PT. 
Note that, in computing $\alpha$ (c.f.~\cref{eq:alpha}), we always use the temperature of completion of the phase transition.

Furthermore, to overcome the difficulties associated to the computation of $\beta/H_*$ in the strongly supercooled regime,  
we compute it at the nucleation time rather than at 
 {percolation time $t_*$ (c.f.~\cref{eq:betaoverH})}. 
 {We find that,}
when $\beta_n/H_n\gtrsim 10$, nucleation and percolation/completion are typically not far from each other, which justifies the approximation. 
For lower values of $\beta_n/H_n$ (which we will only consider in subsection \ref{sec:strong_dynamics}), percolation/completion can occur at temperatures that are significantly lower than $T_n$ (in our examples, this separation will be at most around one order of magnitude). Whenever $\beta_n/H_n\lesssim 10$, our results may be considered as indicative estimates.

\subsubsection{Weakly coupled}

A strong first order PT can arise in a region of parameter space of the DFSZ construction~\cite{Dine:1981rt, Zhitnitsky:1980tq}, which features two Higgs doublets $H_1$ and $H_2$ in addition to the complex PQ field $\Phi$ (the same can occur in KSVZ-type constructions, although only in the presence of extra scalars~\cite{DelleRose:2019pgi}). The potential for the real parts of the $U(1)_\text{EM}$ components of these scalar fields ($h_1, h_2$ and $\phi$ respectively) is given by
\begin{align}
V=&\frac{\lambda_\phi}{4}( \phi^2-v^2)^2
+\frac{1}{2}h_1^2\left( \frac{\kappa_1}{2} \phi^2-\mu_1^2\right)
+\frac{1}{2} h_2^2\left( \frac{\kappa_2}{2} \phi^2+\mu_2^2\right)
-\frac{\kappa_3}{2^{\frac{n}{2}}} \phi^n h_1 h_2 +\nonumber\\
&\frac{\lambda_1}{4}  h_1^4
+\frac{\lambda_2}{4}  h_2^4 +
\frac{\lambda_{12}}{4} h_1^2 h_2^2.
\end{align}
The value of $n$ depends on the PQ charges of $H_1$ and $H_2$, and does not play an important role in our discussion ($n=2$ in the original proposal~\cite{Dine:1981rt, Zhitnitsky:1980tq}, $n=1$ can be easily arranged). All couplings are real in the potential above ($\kappa_3$ can be made real by a field redefinition). 
{Note that with respect to the notation of section \ref{sec:PT}, here we have $v=\varphi_t$ and $\varphi_f=0$.}
In order for the potential above to lead to viable EW symmetry breaking, the determinant of the resulting mass matrix needs to be tuned, which fixes the parameter $\kappa_3$ in terms of the other parameters:
\begin{equation}
v^2 \left(\kappa_1 -2\frac{\mu_1^2}{v^2}\right)\left(\kappa_2 +2 \frac{\mu_2^2}{v^2}\right)\simeq 2 \kappa_3^2\,.
\end{equation}
In order to simplify the analysis, one can neglect the dynamics of $h_2$ (it tracks its minimum during the PT). In this case, the task simplifies to the study of a phase transition in the two-dimensional field space of $h_1$ and $\phi$, where however the loop corrections induced by $h_2$ are taken into account. 

At temperatures $T \gg v$, the fields are both at zero. 
As temperature is lowered, a minimum in the $\phi$ direction appears, which is also the global minimum of the two-field potential if the (tree-level) condition $\lambda_\phi>1/\lambda_1(\mu_1/v)^4$ is imposed. We focus on the case where the PT occurs along the $\phi$ direction (a two-step transition is also possible, see~\cite{VonHarling:2019rgb}). The corresponding potential is of Coleman-Weinberg type when the mass parameters are small, i.e. $\mu_2^2, \lambda_\phi v^2\ll v^2$ (away from this limit, a weaker transition is possible due to a loop-induced cubic term of the form $\sim T^3\phi$). 
This leads to an effective potential of the form \eqref{Vconf} with $\lambda_\phi$ running logarithmically, due to loop corrections.
For simplicity here we focus on the case where $\lambda_\phi = 0$ at tree level and thus it arises only at loop-level. The minimum $F_a\equiv \langle\phi\rangle$ is obtained by solving $\lambda_\phi(\langle\phi\rangle)=-1/4\beta_{\lambda_\phi}(\langle\phi\rangle)$, where $\beta_{\lambda_\phi} = \kappa_2^2/(8\pi^2)$ is the beta function of $\lambda_\phi$. The potential at the minimum is then 
\begin{equation}
V_\text{min}\simeq -\frac{\kappa_2^2}{128\pi^2}F^4_a\,.
\end{equation}
{Note that, with respect to the notation of section \ref{sec:PT}, we here set the zero of the potential at the false vacuum, and $V_{\mathrm{min}}=V_0(\varphi_t)-V_0(\varphi_f)$.}
Tunneling in such a scenario has been first analyzed in~\cite{Witten:1980ez}, where it is shown to proceed via $O(3)$-symmetric bubbles. In our case, the corresponding action is
\begin{equation}
S_B =\frac{S_3}{T}\simeq 7.7\frac{\sqrt{\kappa_2(T)+2 \lambda_\phi(T)}}{-\lambda_\phi(T)},
\end{equation}
where $\lambda_\phi(T)\sim -\beta_{\lambda_\phi}\ln(F_a/T)$. Given the action above, the corresponding values of $\beta_n/H_n$ at nucleation can be numerically computed, as a function of only two parameters $F_a$ and $\kappa_2$ (under our simplifying assumptions on $\lambda_\phi$ and $h_2$).
The transition generically occurs with significant supercooling when $\kappa_2\leq 1.5$. For larger values of $\kappa_2$ on the other hand, the transition can occur during radiation domination, i.e.~before the vacuum energy dominates over the radiation background, when $F_a\simeq 10^8-10^9~\text{GeV}$. 

When nucleation occurs in the supercooled regime, we find that the transition may not complete if $\kappa_2$ is too small, even when the nucleation condition $\Gamma(T_n)\simeq H(T_n)^4$ is fulfilled. Furthermore, even when it completes, the completion temperature is very far from the nucleation temperature when $\kappa_2\lesssim 1$. In order to more reliably assess the reach of ground-based interferometers on this model, we thus focus on some benchmark values of $1\lesssim \kappa_2\lesssim 2$ and $10^{8}~\text{GeV}\leq F_a\leq 10^{10}~\text{GeV}$.
The reheating temperature after the transition (assuming instantaneous reheating) is obtained by equating the radiation energy to the potential energy $-V_\text{min}$, see eq.~\eqref{eq:Treh}:
\begin{equation}
T_\text{reh} \simeq 7\cdot 10^7~\text{GeV}\sqrt{\kappa_2}\left(\frac{F_a}{10^9~\text{GeV}}\right)\left(\frac{106.75}{g_*(T_*)}\right)^{\frac{1}{4}},
\end{equation}
and the parameter $\alpha$ is then found by setting {$V_0(\varphi_f)\rightarrow-V_{\mathrm{min}}$ and $T_f$ by the percolation temperature $T_*$ in Eq.~\eqref{eq:alpha} (for the example values of parameters that we consider below, the volume decrease condition is satisfied at $T_*$). Notice that in the supercooling regime $T_*\ll T_\text{reh}$.}
.

\begin{figure}[t]
  \centering
    \includegraphics[width=0.7\textwidth]{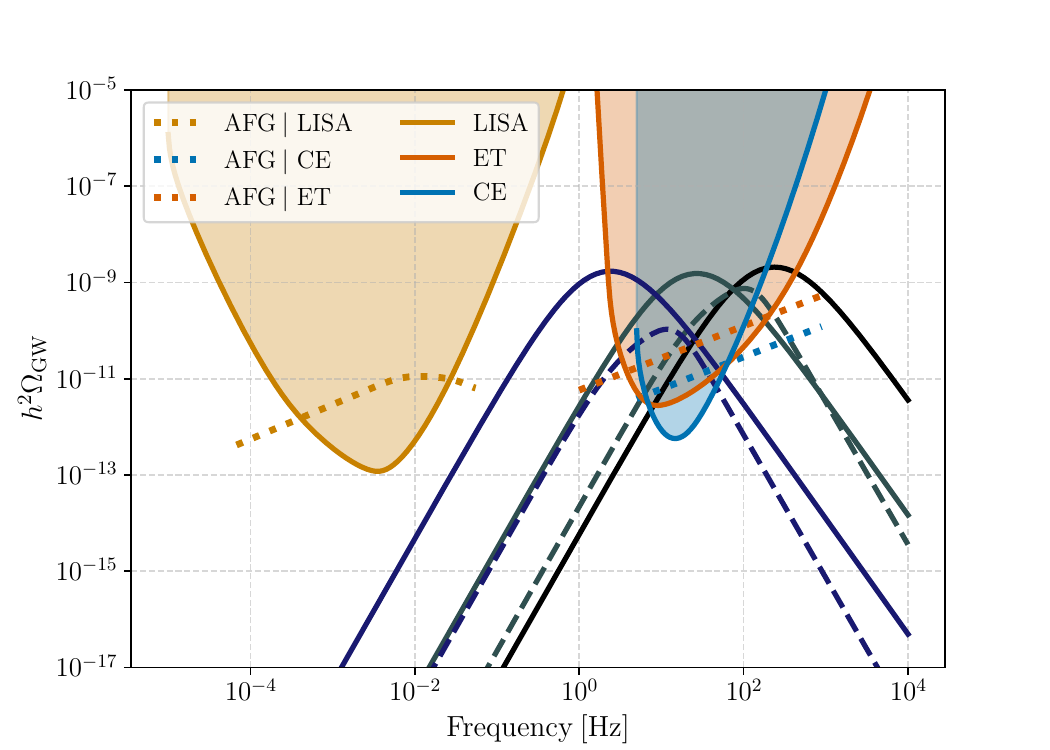}
\caption{Spectra induced by first order {PQ} PT associated with a Coleman-Weinberg type potential, for the benchmark points of Table \ref{tab:benchmark}. Solid lines: strong ($\alpha > 10$) PT; dashed lines: weak PT. Blue, gray, and black spectra have respectively $F_a$ values of $10^8$, $10^9$, and $10^{10}$ GeV. PLISCs corresponding to a SNR threshold of 5 are displayed for ET, CE, and LISA (including the expected galactic foreground). {Additionally, we overlay the corresponding astrophysical foreground for each detector as dotted lines.}}
  \label{fig:CW_PQ_PT_spectra}
\end{figure}

\begin{table}[h]
\centering
\begin{tabular}{|c|c|c|l|c|l|}
\hline
$F_a$  in GeV & $\kappa_2$ &Nucleation occurs in: & $\beta_n/H_n$ & $\alpha$ & $T_\text{reh}$ in GeV \\
\hline
$10^{10}$ & 1.6 & supercooled phase & $\simeq 19.4$ & $\gg 10$ &$\simeq 6.7\cdot 10^8$ \\
$10^{9}$ & 1.5 & supercooled phase &$ \simeq 22.6$ & $\gg 10$  & $\simeq 6.5\cdot 10^7$\\
$10^{9}$ & $1.9$ & radiation dominated phase& $\simeq 36$  & $\simeq 0.7$  &$\simeq 8.1 \cdot 10^7$ \\
$ 10^{8}$& 1.3 & supercooled phase  &$\simeq 21.4$ & $\gg 10$ & $\simeq 6.1\cdot 10^6$\\
$10^{8}$ & 1.85 & radiation dominated phase & $\simeq 42$ &  $\simeq 0.3$ &$\simeq 8\cdot 10^6$ \\
\hline
\end{tabular}
\caption{
Benchmark points {for the PQ phase transition in the weakly coupled case.}}
\label{tab:benchmark}
\end{table}

Table \ref{tab:benchmark} gives the values of $\beta_n/H_n, \alpha$ and $T_\text{reh}$ for some benchmark values of parameters. For these benchmark points, nucleation occurs close to percolation.
The SGWB spectra corresponding to these benchmark points are shown in Fig.~\ref{fig:CW_PQ_PT_spectra}. When nucleation occurs in the supercooled phase, the signal is sourced mostly by bubble and/or relativistic fluid shells collisions, see section \ref{sec:PT}. 
On the other hand, the signal is sourced mostly by bulk fluid motion
(sound waves and turbulence),
when nucleation occurs during radiation domination, as can be the case when the phase transition completes rapidly below the critical temperature. Notice how at least part of the SGWB spectrum is observable at ET/CE for $10^{8}~\text{GeV}\lesssim F_a\lesssim 10^{10}~\text{GeV}$, when $\beta_n/H_n\lesssim O(10-100)$, even for $\alpha\sim 0.1$.

\subsubsection{\label{sec:strong_dynamics}Strong dynamics}

The tunings required in the previous construction are avoided in a different class of models, where a first order PT arises as a new strongly-interacting sector (charged under QCD as well) undergoes confinement. This new sector should have a global  $U(1)_{PQ}\otimes SU(3)_c$  symmetry 
and an  $U(1)_{PQ}-SU(3)_c-SU(3)_c$ anomaly, therefore its constituents are charged under QCD. These models can be investigated by means of holography, when the strongly-interacting sector can be thought of as a gauge theory $SU(N)$ in the large $N$ limit. In this picture, the cosmological evolution of this new sector is similar to that of QCD: at high temperatures, the sector can be thought of as a hot gas of ``gluons" and ``quarks"; below a certain critical temperature a confined phase arises, with ``hadrons" as degrees of freedom. The associated breaking of conformal invariance can be described in terms of a dilaton field $\mu$, which under appropriate circumstances is the lightest field in the spectrum of the new sector. In this case, the confinement phase transition can be studied as a single field problem (see e.g.~\cite{Baratella:2018pxi}).

The corresponding confinement scale $\Lambda_c$ and tunneling rates are determined by a dilaton potential
\begin{equation}
V_{\text{eff}}(\mu)=\frac{N^2}{16\pi^2} \lambda(\mu)\mu^4,
\end{equation}
where $\lambda(\mu)=b_0(\ln(\Lambda_c/\mu)-1/4)$. The bounce action is then given by (see~\cite{Baratella:2018pxi})
\begin{equation}
S_B \simeq \frac{24 N^2}{b_0\log\left(\frac{T_c}{T}\right)}.
\end{equation}
In this case, it is possible to find analytical expressions for the PT parameters $\beta_n/H_n$ and $\alpha$. Here we take $N=3$, as an example value. We are thus left with two parameters only: $b_0$ and $F_a$, similarly to the previous case.

The critical temperature $T_c$ is given by:
\begin{equation}
T_c\simeq 1.4\cdot 10^{10}~\text{GeV}\left(\frac{F_a}{10^{10}~\text{GeV}}\right)b_0^{\frac{1}{4}}.
\end{equation}
The reheating temperature is easily found by matching the effective potential at the minimum with the radiation energy density in the deconfined phase~\cite{Baratella:2018pxi}, and is given by:
\begin{equation}
\label{eq:trhstrong}
T_\text{reh} \simeq 1.4 \left(\frac{106.75}{g_*}\right)^{1/4}b_0^{1/4}~F_a,
\end{equation}
where $g_*$ is the number of relativistic species at $T_*$.

The nucleation temperature $T_n$ is obtained by solving $\Gamma\approx T^4 e^{-S_B}= H^4$. Since the PT is typically strongly supercooled, $H=(\pi^2 g_c/90)^{1/2}T_c^2/M_p$ is constant and corresponds to the Hubble rate during the phase of supercooling. Concretely, we have 
\begin{equation}
T_n = T_c \exp\left[-\frac{y}{2}\left(1-\sqrt{1-\frac{24 N^2}{y^2 b_0}}\right)\right]= T_c \exp[-\mathcal{N}_e]
\end{equation}
where $y \equiv \log(M_p/(a T_c))$, $M_p$ is the reduced Planck mass and $a=\sqrt{\pi^2 g_*(T_c)/90}$ and we have defined the number of efolds $\mathcal{N}_e$ above.
The rate $\beta_n/H_n$ is then found to be:
\begin{equation}
\left. \frac{\beta}{H}\right|_n = \left. T \frac{\d S_B}{\d T}\right|_n - 4 \approx \frac{24 N^2}{\mathcal{N}^2_e b_0}-4.
\end{equation}

We vary $F_a$ between $10^8-10^{10}~\text{GeV}$ and $b_0$ between $0.4$ and $1$. Similarly to the previous case, we find that the transition may not complete even when the nucleation condition is satisfied, due to the inflationary background. However, if the transition completes, according to the volume decrease criterion \cref{eq:volume_decrease}, we find the temperature of completion to be at most roughly one order of magnitude smaller than that of nucleation, unlike the previous case. Furthermore, we checked that the transition completes for the relevant values of $F_a$ whenever $\beta_n/H_n > 5.2$ at the nucleation temperature.

The value of $\alpha$ is always much larger than $10$ for any reasonable value of $b_0$, i.e.~the transition is always supercooled in the relevant range of parameters. Forecasts on the regions of parameter space that will be probed by GW observatories are shown in Fig.~\ref{fig:PQ_results}. In the hatched region the transition does not complete. Close to that region, the GW signal is largest and very loud at ET/CT for $F_a\sim 10^8-10^9~\text{GeV}$. Overall, the transition remains observable at ET/CE for $b_0\lesssim 1$ and for $F_a$ up to $10^{10}~\text{GeV}$. Interestingly, in all this region of parameter space the QCD axion can potentially be the totality of the Dark Matter, depending on the specific model~(see e.g.\cite{Gorghetto:2020qws}). In particular, models with long-lived domain walls allow for axion dark matter with $F_a$ down to $10^8~\text{GeV}$, otherwise $F_a\sim 10^{10}~\text{GeV}$ is needed. In the upper right corner of Fig.~\ref{fig:PQ_results}, the astrophysical foregrounds at ET/CT are also large, and distinguishing a primordial signal is more challenging, but still doable depending on the signal frequency shape.

\begin{figure}[h]
  \centering
  \begin{minipage}[h]{0.49\textwidth}
    \centering
    \includegraphics[width=\textwidth]{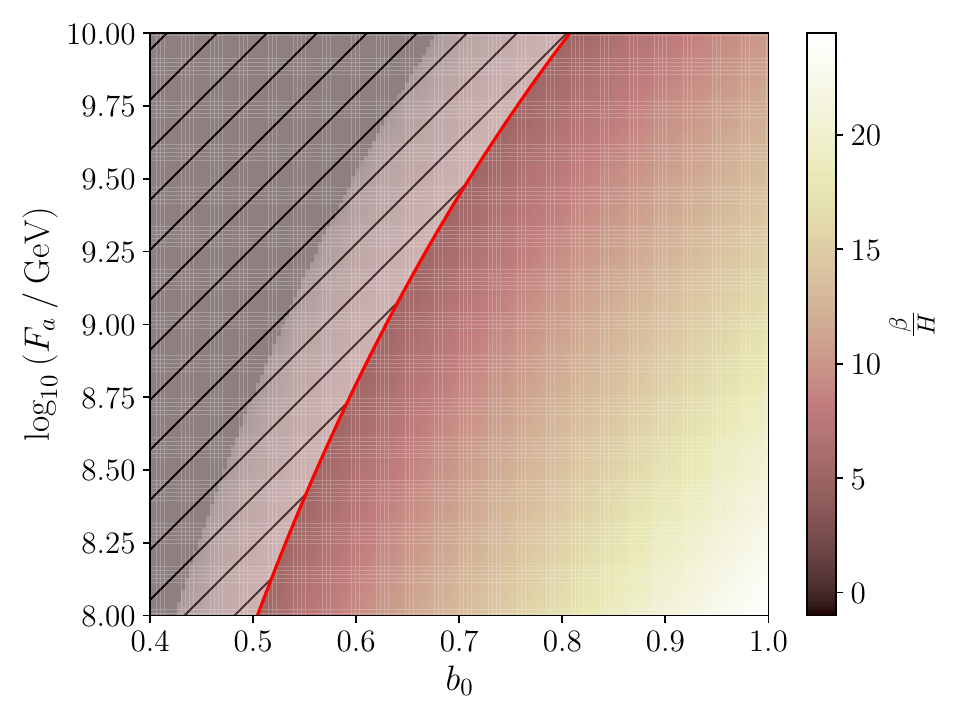}
    \captionof*{figure}{}
    \label{fig:PQ_beta_over_H}
  \end{minipage}
  \hfill
  \begin{minipage}[h]{0.49\textwidth}
    \centering
    \includegraphics[width=\textwidth]{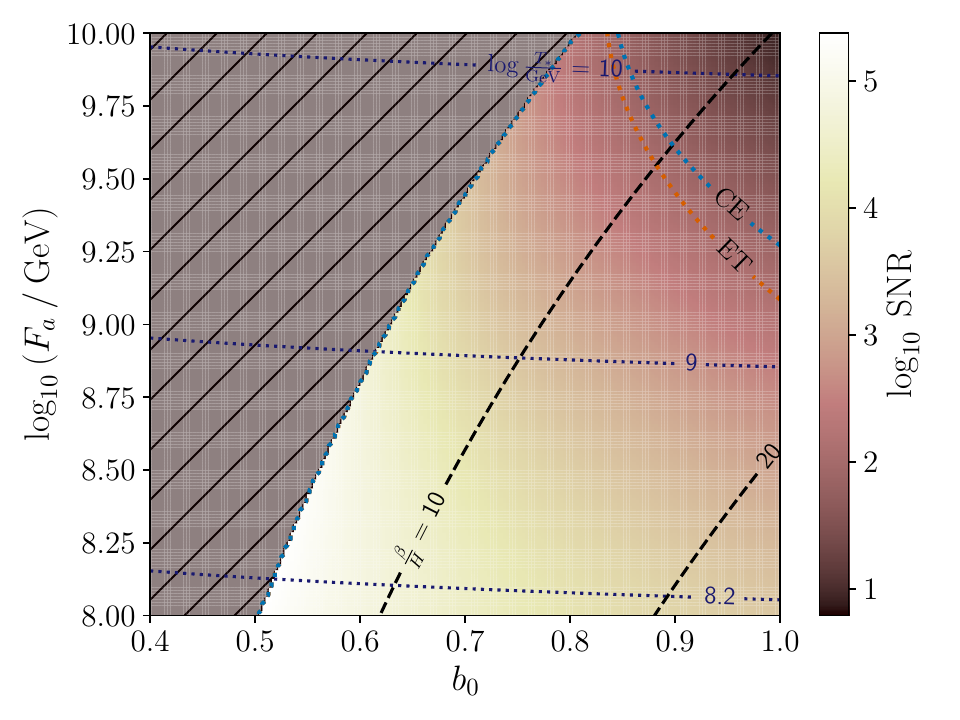}
    \captionof*{figure}{}
    \label{fig:PQ_ET+CE}
  \end{minipage}
  \caption{Left panel: 
  $\frac{\beta}{H}$ at nucleation as a function of $F_a$ and $b_0$. The red line
  corresponds to $\beta_n / H_n = 5.2$, which is the threshold value below which the PT does not complete. Hence, the hatched region of the parameter space is excluded.
  Right panel: $\log_{10}$ of the maximum SNR among CE and ET. Each detector can detect the SGWB signal from the PQ PT with SNR greater than 5, across all parameter space where the PT completes. 
  We overlay lines of constant reheating temperature $T_\text{reh}$ and constant $\beta_n / H_n$, showing the conversion from the parameters $b_0$ and $F_a$. 
  The contours corresponding to the SNR of the foreground from compact binary mergers for each detector are displayed as dotted lines 
  {(in the region enclosed by the contours, the foregrounds have higher SNR than the signal)}.
  For $\beta_n/H_n\lesssim 10$, our results should be taken as indicative estimates.}
  \label{fig:PQ_results}
\end{figure}

\subsection{\label{PQ-DWs}GWs from axionic Domain Wall networks}

Beside the possibility of a first order PT, the PQ mechanism provides another potential source of GWs. The spontaneous breaking of the global $U(1)_{PQ}$ symmetry leads to the formation of a network of topological defects, of hybrid type (see section \ref{sec:Defects}).
First, global cosmic strings are formed at the PQ transition, i.e. at $T\lesssim F_a$. As discussed in Section \ref{subsec:global}, these strings rapidly evolve into a scaling regime (with possible logarithmic corrections) and predominantly emit axions throughout their evolution. 
Then, at a lower energy scale, a network of domain walls can form and emit GWs (see section \ref{sec:DW}), as we now discuss.

\subsubsection{The standard QCD axion}

At $T\sim \text{GeV}$, non-pertubative QCD effects provide an explicit breaking of the PQ symmetry, and therefore induce a potential for the axion field. The original shift symmetry of the axion field is thus broken to a discrete version, due to the periodicity of the QCD vacuum. The axion potential thus exhibits a discrete $\mathbb{Z}_{N_\text{DW}}$ symmetry, $V(a)=V(a+2\pi v/N_\text{DW})$, where $F_a\equiv v/N_\text{DW}$ and the domain wall number, $N_{\rm{DW}}$, depends on the microphysical implementation of the axion model (in particular on the PQ charges of SM and heavy fields). The exact functional form of the axion potential does not play a particularly important role in our discussion, and below QCD confinement it takes the form (the actual form of the potential differs by $O(1)$, see~\cite{GrillidiCortona:2015jxo})
\begin{equation}
V_\text{QCD}(a)\simeq \kappa_\text{QCD}^2\Lambda_\text{QCD}^4\left[1-\cos\left(\frac{a}{F_a}\right)\right],
\end{equation}
where $\kappa_\text{QCD}\simeq \sqrt{m_u/\Lambda_\text{QCD}}$. The axion mass at low temperatures is then $m_a\simeq 5.7~\text{meV}(10^9~\text{GeV}/F_a)$. The residual axion discrete symmetry is spontaneously broken once the axion starts oscillating, i.e. at the temperature $T_\text{DW}$ when the condition $m_a(T)\simeq H(T_\text{DW})$ is met.\footnote{The axion mass rises quickly with temperature above the confinement scale~\cite{Borsanyi:2016ksw} and is constant at lower temperatures.} This condition is satisfied at $T_\text{dw}\simeq \text{GeV}$ for the values of $F_a$ of interest. Correspondingly, domain walls are formed, attached to the previously formed strings. Their tension is $\sigma\simeq 8 m_a F_a^2$ (we neglect the small correction due to pion mixing). 

As described in section \ref{sec:DW}, such walls radiate GWs before disappearing (either because $N_{\rm{DW}}=1$ or because $N_{\rm{DW}}>1$ but an additional explicit breaking of the PQ symmetry is present). The signal would be peaked in the $10^{-7}-10^{-9}$ Hz range, corresponding to the viable annihilation temperatures of the network (see e.g.~\cite{Ferrer:2018uiu}). 

However, in the standard QCD axion cosmology described so far, the amplitude of the signal would be too small for it to be detectable, even by the future SKA. The obstruction is phenomenological: DW annihilation contributes to the production of relic axions, which are stable and behave as Cold Dark Matter.
Requiring that these axions do not overproduce Dark Matter then implies that DWs make up at most a fraction $\alpha_{\text{ann}}\lesssim 10^{-6}-10^{-9}$ of the energy density of the Universe at the time of their annihilation. This leads to an unobservably small GW signal according to~\eqref{eq:omegadw}, see~\cite{Hiramatsu:2012sc}.

\subsubsection{Heavy QCD axions}

This conclusion can be dramatically altered in a certain class of QCD axion models where the axion is much heavier than in the standard case. Indeed, a heavy enough axion would decay to SM particles (gluons and/or photons) before BBN, therefore leaving no dangerous relics behind.

Independently of the specific realisation, a heavy QCD axion scenario assumes an additional contribution to the axion potential
\begin{equation}
		\label{eq:Hpot}
		V_a = V_\text{QCD}(a)+\kappa^2_{\text{H}} \;\Lambda^4_{\text{H}}\left(1-\cos \frac{a}{F_a}\right),
	\end{equation}
where $\Lambda_\text{H}\gg \Lambda_\text{QCD}$ 
and $\kappa_H \leq 1$ depending on details such as the fermion spectrum. Crucially, the additional term should be aligned with the QCD potential (at least up to $\Delta\theta\lesssim 10^{-10}$), otherwise the axion does not solve the strong CP problem. This class of models is motivated by the so-called axion {\it quality problem}~\cite{Georgi:1981pu, Holdom:1982ex, Dine:1986bg, Holman:1992us, Kamionkowski:1992mf, Barr:1992qq, Ghigna:1992iv}: the PQ mechanism is vulnerable to additional sources of explicit symmetry breaking beyond QCD, that would easily displace the minimum of the axion potential. The presence of these explicit sources may be expected on theoretical grounds, since global symmetries are expected not to be exact. The quality problem is quantitatively important because of the smallness of the QCD scale compared to any other higher scale which could provide additional PQ breakings. A heavy axion scenario can significantly alleviate this issue, by raising the scale of the axion potential, which thereby renders the PQ mechanism more robust against additional sources of PQ symmetry breaking. Concrete models, often relying on an additional confining dark sector that the axion couples to, face the challenge of aligning the additional potential with QCD (see~\cite{Holdom:1982ex} and~\cite{Treiman:1978ge, Dimopoulos:1979pp, Tye:1981zy},~\cite{Holdom:1985vx, Flynn:1987rs, Choi:1998ep, Agrawal:2017ksf, Kitano:2021fdl}, \cite{Rubakov:1997vp, Gherghetta:2016fhp, Gherghetta:2020ofz} and \cite{Berezhiani:2000gh, Dimopoulos:2016lvn, Hook:2019qoh} for a partial collection of heavy axion models).

While the axion is not the Cold Dark Matter in this scenario, the hybrid network of axionic topological defects can now offer an alternative probe, via its GW signature~\cite{ZambujalFerreira:2021cte, Ferreira:2022zzo}, when $N_\text{DW}>1$ (which is a rather generic outcome in realistic models). As the axion mass is now $m_a\simeq 10^8~\text{GeV}\left(\frac{10^{12}~\text{GeV}}{F_a}\right)\left(\frac{\Lambda_{\text{H}}}{10^{10}~\text{GeV}}\right)^{2}\kappa_{\text{H}}$, the cosmological evolution of topological defects is shifted to higher temperatures.

The annihilation of the string-wall network is determined by the aforementioned explicit breakings of the PQ symmetry, which can be captured by an additional misaligned potential $V_{b} \simeq - \mu_b^4 \cos \left(\frac{N_{\text{b}}}{N_{\text{DW}}}\frac{a}{F_a} - \delta \right),$ where $N_b$ should be $1$ or co-prime  with $N_\text{DW}$ in order to induce annihilation. For illustration we take the representative values $N_b=1, \kappa_\text{H}=1$  and $N_\text{DW}=6, \delta=0.3$  in the following (see \cite{Ferreira:2022zzo} for other choices). The annihilation temperature is then 
\begin{eqnarray}
		\label{eq:Tann}
		T_\text{ann}\simeq 10^{7}\text{GeV} 
		\sqrt{\frac{10^{11}\text{GeV}}{F_a} }
		\left(\frac{\Lambda_{\text{H}}}{10^{10}\text{GeV}}\right)\left(\frac{r}{0.003}\right)^2,
	\end{eqnarray}
 where the parameter $r\equiv \mu_b/\Lambda_\text{H}$.

 \begin{figure}[h]
  \centering
    \includegraphics[width=0.75\textwidth]{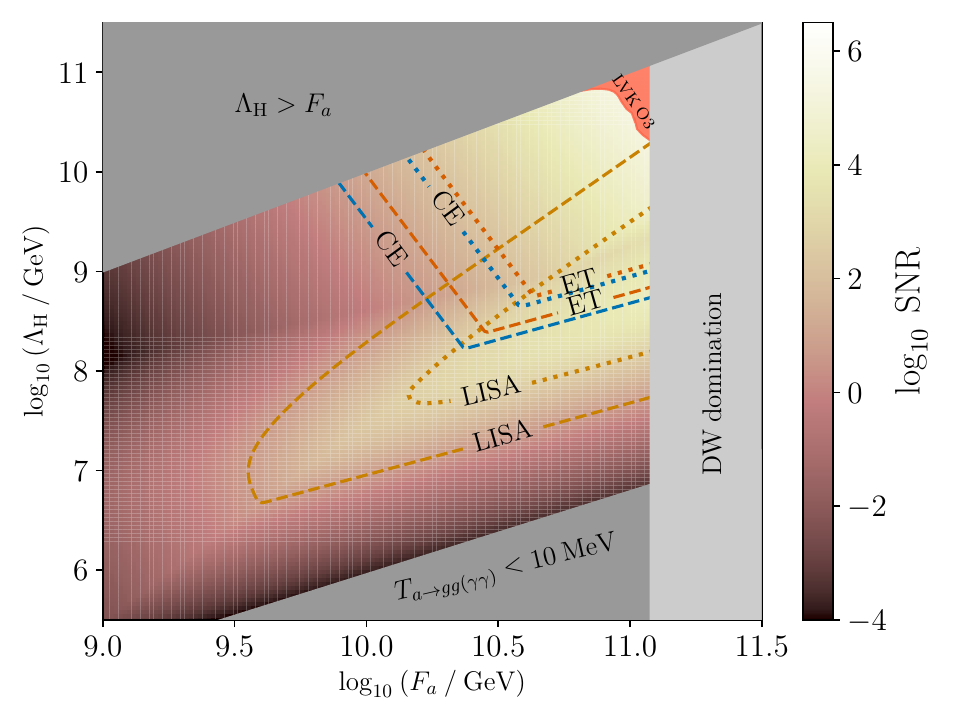}
  \caption{Expected SNR of the DW SGWB from the heavy QCD axions for LISA and 3G Earth-based detectors, assuming $\kappa_{\rm H} = 1$, $\alpha_s = 0.1$, $N_\text{DW} = 6$ and $r = 0.002$, corresponding to $\Delta \theta \simeq 2.5 \times 10^{-12}$. The maximum individual SNR among detectors is depicted in the color grid. Contour lines representing SNR = 5 for each detector are displayed in dashed lines. Additionally, the contour corresponding to the SNR value of the respective astrophysical foregrounds for LISA, CE and ET  are depicted as dotted lines: in the region on the left and below these lines, the SNR of the foreground is larger than that of the cosmological signal (see section \ref{sec:detectors} for details on the expected foregrounds). Four exclusion zones are overlaid: the three gray regions correspond to $\Lambda_{\rm H} > F_a$, DW domination and $T_{a \rightarrow gg(\gamma\gamma)} < 10\: \mathrm{MeV}$, while the red region is the 95\% exclusion region from the LVK O3 run \cite{Jiang:2022svq}.}
  \label{fig:Heavy_QCD_DW}
\end{figure}
 
Most importantly, it is now possible for domain walls to make up a large fraction of the energy density of the Universe at the time of their annihilation
\begin{equation}
		\label{eq:fraction}
		\alpha_\text{ann}\simeq 0.1~\left(\frac{F_a}{10^{11}~\text{GeV}}\right)^2\left(\frac{0.003}{r}\right)^{4}.
\end{equation}
Of course, $V_b$ also causes a displacement of the axion from the QCD minimum by $\Delta \theta \simeq r^4/N_\text{DW}$, for a natural value $\delta\sim O(1)$. Such a displacement induces electric dipole moments for the neutron and the proton and upcoming measurements are able to probe down to $\Delta \theta\lesssim 10^{-13}$~\cite{Abel:2018yeo, Filippone:2018vxf}.

Most of the energy density in domain walls is released into non-relativistic quanta of the constituent field at annihilation (a small fraction goes instead to GWs), i.e. into axions, which can then decay to light degrees of freedom; for the model of interest these are SM gluons (if the decay occurs above QCD confinement) and photons. The decay rate is $\Gamma_{a\rightarrow gg (\gamma\gamma)}=1/(64\pi)(c_{g, \gamma}\alpha_{s,\text{em}})/(2\pi)^2 m_a^3/F_a^2$, where $c_{g}=1$ and $c_\gamma$ is a model-dependent coupling to photons~(see e.g.~\cite{Zyla:2020zbs} for a review in standard DFSZ and KSVZ constructions) and $\alpha_{s,\text{em}}$ is the (strong) electromagnetic coupling constant. The decay temperature is found by setting $\Gamma\simeq H$:
	\begin{align}
	\label{eq:tdecG}
	T_{a\rightarrow gg (\gamma \gamma)}&\simeq 3\cdot 10^9~\text{GeV}~\alpha_{s (\text{em})}~\kappa_{\text{H}}^{3/2} \left(\frac{\Lambda_{\text{H}}}{10^{10}~\text{GeV}}\right)^3\left(\frac{10^{11}~\text{GeV}}{F_a}\right)^{\frac{5}{2}},
	\end{align}
and is typically above QCD confinement for values of parameters of interest for this work. Decay to photons is thus subleading with respect to decay to gluons. When decay is not efficient at $T_\text{ann}$, non-relativistic axions can come to dominate the universe before their decay becomes efficient. By setting the initial energy carried by the non-relativistic quanta to equal the energy density in domain walls at the time of annihilation, and redshifting as $\sim a^{-3}$, one can determine the temperature at which non-relativistic axions start dominating the universe
\begin{align}
	T_{\text{MD}}\simeq 1.2\cdot 10^7~\text{GeV}~C_1\left(\frac{106.75}{g_{*,\text{ann}}}\right)^{\frac{1}{4}} \left(\frac{F_a}{10^{11}~\text{GeV}}\right)^{\frac{3}{2}}\left(\frac{\Lambda_{\text{H}}}{10^{10}~\text{GeV}}\right)\left(\frac{0.003}{r}\right)^2,
	\end{align}
(see the appendix of~\cite{Ferreira:2022zzo} for more details). Of course an intermediate phase of matter domination occurs only if $T_{a\rightarrow gg (\gamma \gamma)}<T_{\text{MD}}$, that is when 
 \begin{equation}
\Lambda_\text{H}\lesssim 7\cdot 10^{8}~\text{GeV}\left(\frac{F_a}{10^{11}~\text{GeV}}\right)^2\left(\frac{0.003}{r}\right), 
 \end{equation}
otherwise  the axions decay as soon as they are radiated from the walls. If this intermediate phase of matter domination occurs, then the amplitude of the GW signal from DW annihilation~\eqref{eq:omegadw} is suppressed by a redshift factor:
 \begin{equation}
	\nonumber \Omega_{\text{gw,MD}}h^2\simeq\Omega_\text{\tiny GW,DW}(f)h^2\left(\frac{g_{*}(T_{a\rightarrow gg(\gamma\gamma)})}{g_{*}(T_{\text{MD}})}\right)^{\frac{1}{3}}\left(\frac{T_{a\rightarrow gg(\gamma\gamma)}}{T_{\text{MD}}}\right)^{\frac{4}{3}},
	\end{equation}
 and the peak frequency~\eqref{eq:pfreq} is shifted to lower frequencies: 
\begin{align}
f_{p,\text{MD}}^0\simeq f_p^0\left(\frac{T_{a\rightarrow gg(\gamma\gamma)}}{T_{\text{MD}}}\right)^{\frac{1}{3}}.
\end{align} 
Finally, the amplitude and frequency of GW signal sourced by heavy axion DWs is found by plugging \eqref{eq:Tann} and \eqref{eq:alphann} above into \eqref{eq:omegadw} and \eqref{eq:pfreq}, and accounting for matter domination when appropriate.

Fig.~\ref{fig:Heavy_QCD_DW} shows the sensitivity reach of future detectors to this scenario (dashed contours). We have fixed $r=0.002$ and $N_\text{DW}=6$ as representative values, which correspond to $\Delta\theta=2.5\cdot 10^{-12}$, within the reach of future electric dipole moment measurements. 
The gray-shaded regions are excluded/constrained by (for further details, see~\cite{Ferreira:2022zzo}): a) in the upper left corner, the axion effective theory is not valid; b) in the lower right corner, the axion decay can alter BBN predictions; c) for $F_a\gtrsim 10^{11}~\text{GeV}$ DWs dominate the universe before their annihilation becomes efficient, most likely leading to a universe incompatible with observations (see also the discussion in~\cite{Vilenkin:2000jqa}). (Note that the region $F_a\gtrsim 4\times 10^{10}~\text{GeV}$ is expected to be subject to constraints from PBH production \cite{Ferreira:2024eru}.) The orange-shaded region is the current exclusion limits by the LVK O3~\cite{Jiang:2022svq}. In the left part of the plot below the dotted contours astrophysical foreground are large and may thus obstruct the detection of a cosmological signal. Nonetheless, in most of the sensitivity region of ET, CE and LISA a cosmological signal from DWs is distinguishable, and these detectors will thus extend the exploration of heavy axion models very significantly, providing the exciting possibility of detecting a signal correlated with measurement of electric dipole moments.

\subsubsection{First order PT at ground-based detectors + DW signal at PTA}

So far we have discussed GWs from a first order PQ phase transition in the standard QCD axion model, and from domain walls in the heavy axion scenario. One may wonder whether both signals can coexist in a heavy axion scenario. The answer is positive, though it is hard to arrange for the peaks of both GW signals to lie in the ground-based frequency band (see \eqref{eq:Tann} and \eqref{eq:trhstrong}). On the other hand, a viable heavy axion model with $F_a\lesssim 10^8~\text{GeV}$ and $m_a\simeq \text{GeV}$ could have a first order PT {leading to a SGWB} in the frequency band of ground-based interferometers, and DW annihilation occurring around the QCD scale for appropriate values of $\mu_b$, such that the corresponding GW signal would be peaked in the PTA frequency band (and may provide an interpretation of the current detection). In this case, the SGWB would be characterized by two peaks, one at nHz and the other one at $1-10$ Hz, with the high frequency tail of the DW signal possibly also observable at LISA. Several cosmological bounds apply to this scenario, which may also be probed at laboratories and colliders, as discussed in~\cite{Ferreira:2022zzo}.

\section{\label{sec:conc}Conclusion}

The aim of this review was to quantify the ability of 3G ground based detectors, working together with LISA and SKA, to probe cosmological SGWBs sourced by PTs in the early universe. PT and symmetry breaking are ubiquitous in quantum field theories and particle physics beyond the SM and, as we have discussed, they can lead to copious GW production across different frequency bands. Among the most interesting possible sources of GWs are (i) first order PTs as reviewed in section \ref{sec:PT}, and (ii) topological defects of different kinds, as reviewed in section \ref{sec:Defects}. Figure \ref{fig:CS+DW_spectra_PTA_values} shows {examples of} characteristic spectra from these sources, {together with the PLISCs of the detectors across the frequency band, accounting for foregrounds}. 
{The broad and complementary frequency span of 3G ground based detectors, LISA, and PTA, will be a fundamental asset for probing the variety of SGWB signals that can arise from PTs in the early Universe. 
For example, the coincident detection of a SGWB from the same generation process across the frequency bands would offer enhanced information on the process itself, and would fortify its interpretation as a primordial signal. 
Furthermore, the wider the frequency coverage, the broader the time span in the early Universe that one can access, see \cref{eq:s}. \cref{fig:CS+DW_spectra_PTA_values} clearly shows that the combination of future GW detectors opens a new window of exploration of the early Universe, spanning 11 decades in frequency and therefore in energy. 
However, \cref{fig:CS+DW_spectra_PTA_values} also shows that the presence of foregrounds highly influences the sensitivity of some detectors, the worst case being PTA with the SKA, for which accounting for the SMBHB foreground changes the sensitivity by about 3 orders of magnitude. }

Measuring a cosmological SGWB requires {indeed} a thorough understanding of both the astrophysical foregrounds relevant to each detector (given their frequency band and sensitivity), and the detection method applicable to each different detector type. In section \ref{sec:detectors} we have presented and summarised these points for each of the detectors we consider here, namely CE, ET, LISA and SKA. Furthermore we have explained how the SNR for a cosmological SGWB is calculated for each. Our results are summarised in the PLISCs of figure \ref{fig:CS+DW_spectra_PTA_values}, as well as in table \ref{tab:detectors} where we highlight the relevant astrophysical foregrounds for each detector and their corresponding SNRs. Our results provide realistic sensitivities to cosmological signals for SKA and LISA. For ET/CE, rather than deriving new sensitivity curves, we have used estimates of the astrophysical foregrounds from~\cite{Branchesi_2023, Bellie:2023jlq} to understand to what extent astrophysical foregrounds can limit the detection prospects of cosmological backgrounds, see Fig.~\ref{fig:generalFOPT} and Fig.~\ref{fig:DW_SNR_grid}.

{In section \ref{sec:PT}, we have focused on 
GW production by a first order PT.
We have first reviewed the PT dynamics and defined the relevant stages of its progression, and the associated 
parameters such as the PT duration and the size of the bubbles at percolation. 
We have then reviewed the phenomena associated with the PT capable of sourcing a SGWB, namely: bubbles and/or relativistic fluid shells collisions for strongly first order PTs; bulk fluid motion generated at percolation in the form of sound waves and (M)HD turbulence for PTs of intermediate strength; and sound waves alone for weak PTs. 
We have shown how the scaling of the SGWB signals with the source parameters, provided by detailed analyses of the aforementioned sourcing mechanisms, can be interpreted based on simple models for the source UETC, namely, constant-in-time and stationary. 
We have then presented placeholders for the SGWB spectra from the three generation mechanisms, highlighting the assumptions behind them.
Evaluating the SGWB signal from a first order PT is a complicated endeavour, requiring particle physics and statistical physics as well as non-equilibrium phenomena, fluid dynamics and cosmology, and further progress is needed to clarify and overcome all the caveats that enter in this process. 
Nevertheless, we have concluded by 
showing the reach of future 3G detectors (as well as LISA and PTA with the SKA) to the compelling case of SGWBs generated by first order PTs, based on the state of the art of our knowledge. 
The result is presented in \cref{fig:generalFOPT} as a scan on the SNR of the SGWB signal in CE, ET, LISA and SKA, as a function of the parameter space of first order PTs (characterised by the transition temperature $T_*$ and its duration $\beta/H_*$, for  different fixed strengths $\alpha=0.05, 1, \gg 1$), including the astrophysical foregrounds. 
The figure clearly shows the potential of the considered detectors to probe first order PTs in full complementarity with respect to the PT energy scale (represented by its temperature). 
{Therefore, when all these observatories will be in operation, they will offer the opportunity to probe a wide range of energy scales in the early Universe.
3G detectors allow to explore new, high energy scales for first order PTs with respect to the PTs predicted within the SM, probed rather by LISA (EW scale) and PTA (QCD scale). 
Particularly interesting is the possibility
to probe 
relatively weak ($\alpha=0.05$) and fast ($\beta/H_*\simeq 100-1000$) PTs, predicted by many BSM models.}
Fig.~\ref{fig:generalFOPT} also shows that, in a large region of parameter space, the signal will have higher SNR than that of the expected astrophysical foregrounds. For strong first order PTs ($\alpha \gg 1$), some regions of parameter space are already excluded by LVK O3 data, and others could account for the observed PTA signal.

In section \ref{sec:Defects} we have focused on topological defects which may be formed in PTs provided a symmetry is broken. Their existence depends on the properties of the vacuum manifold ${\cal{M}}$, and the argument for defect formation is based solely on causality: in other words, they may form whether the PT is first order or not.  For local $U(1)$ strings, whose tension is given by $\eta^2$, where $\eta$ the energy scale of the PT, we reviewed in section \ref{sec:LocalCS} the shape of the spectrum which covers orders of magnitude in frequency, and whose amplitude is proportional to $(G\mu)^2$. As discussed, understanding the detailed properties of the spectral shape still requires further theoretical understanding. In that section we have shown that, with a precise estimation of the astrophysical foreground, CE  should be able to probe strings with tensions $G\mu 
\gtrsim 10^{-16}$ (Model A) and $G\mu \gtrsim 10^{-18.7}$ (Model B). For ET, one has respectively $G\mu \gtrsim 10^{-14.5}$ and $G\mu \gtrsim 10^{-17.4}$. 
Both ET and CE will be able to search for individual bursts emitted by cosmic string cusps and kinks, but if none are detected, then constraints on the tension of cosmic loops is less competitive than those from the SGWB.

In section \ref{sec:DW}, we discussed domain walls, namely  planar-like topological defects which, unlike line-like strings, cannot exist indefinitely: indeed, they must annihilate in order not to dominate the energy budget of the universe. We briefly reviewed the properties of DWs and the SGWB spectrum they produce.  Figure \ref{fig:DW_SNR_grid} shows the expected SNR for domain walls for all next generation GW detectors, as a function of the annihilation temperature $T_{\rm{ann}}$ and $\alpha_{\rm{ann}}$,  the fraction of the total energy density in DWs at $T_{\rm{ann}}$. Ground based detectors in particular will be sensitive to $T_{\rm{ann}} \sim 10^{8}$GeV, and $\alpha_{\rm{ann}} \gtrsim 10^{-2}$ (and up to $10^{-3}$ depending on the understanding of the astrophysical foregrounds).

In section~\ref{sec:PQ} we then focused on the PQ mechanism as a well-motivated beyond the SM scenario that can potentially be probed by ET and CE {(see also \cite{Yang:2024npd}), as well as LISA}. GWs can arise in this mechanism from the PQ phase transition, if it is of first order, and/or from its topological byproduct, that is the axionic string-domain wall network. First, we reviewed UV realizations where the PQ sector is approximately scale-invariant, within weakly or strongly coupled frameworks. The PT is then expected to be of first order, and typically exhibits a significant amount of supercooling. The corresponding GW signal is thus mostly sourced by bubble {and/or relativistic shells} collisions (although bulk fluid motion can be efficient in certain regions of the parameter space) and is particularly loud at ET/CE, since the scale of the PT $\sim F_a$ is constrained to be around $10^8-10^{10}~\text{GeV}$ from cosmological and astrophysical bounds on the QCD axion. We presented a forecast of the regions of parameter space in strongly-coupled PQ models which will be probed by 3G detectors in Fig.~\ref{fig:PQ_results}, whereas we presented spectra for benchmark points in the case of weakly coupled models. Our analysis improves significantly over previous work~\cite{VonHarling:2019rgb}, in that we have taken into account more stringent conditions on the completion of PTs in vacuum dominated eras, beside the calculation of the SNR of the signals. In both cases, we have shown that significant regions of parameter space will be probed by 3G detectors. {In particular, the GW signal is loudest at 3G detectors when the axion decay constant lies towards the lower end of the traditional QCD axion window, i.e. $F_a\sim 10^8-10^9~\text{GeV}$, although an observable signal can be obtained also for larger decay constants. This corresponds to QCD axion masses in the range $m_a\sim \text{meV}-0.1~\text{eV}$. For significantly lighter masses the signal would be peaked at frequencies larger than the detectors sensitivity regions.}  {Concerning dark matter, the detection of a SGWB from the early Universe at 3G detectors, if interpreted as coming from the PQ mechanism, would suggest that the axion decay constant is compatible with the QCD axion being the Dark Matter. In this case, the Dark Matter would be produced either by cosmic strings and the misalignment mechanism (for $F_a\lesssim 10^9-10^{10}~\text{GeV}$) or by domain walls (for smaller values).} 

{Secondly and independently of the order of the PQ phase transition, we have investigated GWs from axionic domain walls. While the signal is too small to be observable in the standard QCD axion scenario, we showed that DWs can source a large GW signal in {heavy QCD axion models}, {where the axion mass is an additional independent parameter}.
While UV realizations of these models require additional ingredients, the scenario can potentially alleviate the axion quality problem, while also leading to heavy DWs and large GW backgrounds. The relevant GW abundance and frequencies depend on the axion mass and decay constant, and additionally on the scale at which the axionic discrete symmetry is broken to allow a viable cosmology without DWs at late times. Our results are presented in Fig.~\ref{fig:Heavy_QCD_DW} and show that the GW signal is louder than astrophysical foregrounds in a large region of parameter space of these models. {In particular, the GW signal from domain walls is detectable at 3G detectors when $10^{10}~\text{GeV}\lesssim F_a \lesssim 10^{11}~\text{GeV}$, and the axion is very heavy, $10^{6}~\text{GeV}\lesssim m_a\lesssim 10^{11}~\text{GeV}$. The signal from domain walls is louder for larger values of the mass and decay constant.}  Very interestingly, the region that can be probed via GWs in Fig.~\ref{fig:Heavy_QCD_DW} also leads to observable electric dipole moments for the neutron and the proton, thereby providing a complementary channel to test heavy axion models. {Concerning axion dark matter, the important difference from the first order PT scenario is that the detection of a GW signal from DWs at 3G detectors, together with the observation of electric dipole moments, would give support to scenarios where the axion is unstable and thus cannot be the dark matter.}}

{The possibility of observing GW backgrounds is informative even in the case of null detection. For the case of the QCD axion, this translates into constraining realizations of the PQ mechanism which predict a first order PT (the simplest being scenarios with approximate scale invariance reviewed above) and heavy axion models with long-lived domain walls, as can arise when $N_\text{dw}>1$. Existing observations from LVK have already started constraining those scenarios~\cite{Badger:2022nwo, Jiang:2022svq}.}

{{Here} we have focused on PTs as a well understood {and physically well motivated}
phenomenon that sources GWs. {However,} perhaps the most exciting perspective is that, by probing early Universe temperatures as high as $\gtrsim 10^8~\text{GeV}$, 3G detectors may reveal imprints of totally new phenomena which occurred in the primordial universe. 
{As summarized over the review, reaching this goal will require improving the characterization and understanding of the instruments and the astrophysical foregrounds. In parallel, the characterization of the primordial sources, which often require heavy numerical computations, still requires significant improvement. This includes the understanding of the associated signals, in the form of SGWBs in various observatories, but also possibly as dark matter, dark radiation, primordial black holes or even electric dipole moments as in the case of axions. These additional signals will of course play a crucial role in model discrimination. }}

\section*{Acknowledgements}

We would like to thank Jos\'{e} Juan Blanco-Pillado, Francesco Iacovelli, Michele Maggiore, Gijs Nelemans, Alberto Roper-Pol, G\'eraldine Servant and  Peera Simakachorn for useful discussions and comments. CC is supported by the Swiss National Science Foundation (SNSF Project Funding grant \href{https://data.snf.ch/grants/grant/212125}{212125}). D.A.S is grateful to CERN for hospitality whilst this work was in progress. The work of F.R.~is supported by the grant RYC2021-031105-I from the Ministerio de Ciencia e Innovación (Spain). F.R. and O.P. acknowledge support from the Departament de Recerca i Universitats from Generalitat de Catalunya to the Grup de Recerca ‘Grup de Fisica Teorica UAB/IFAE’ (Code: 2021 SGR 00649) and the Spanish Ministry of Science and Innovation (PID2020-115845GB- I00/AEI/10.13039/501100011033). IFAE is partially funded by the CERCA program of the Generalitat de Catalunya. H.Q-L thanks Institut Polytechnique de Paris for funding his PhD.

\appendix

\section{\label{app: ORF}Overlap Reduction Function}

In this appendix we provide the the definition of the overlap reduction function used in Sec.~\ref{sec:detectors} to compute the sensitivity of the GW detectors, \cite{Allen:1996vm, Thrane_Romano}.
The response of a detector located at $\Vec{x}$ to a SGWB is the correlation \cite{Thrane_Romano}
\begin{align}
    \label{eq: general detector response}
    h(t,\Vec{x}) & = 
    \int_{-\infty}^{+\infty} \d {t^{\prime}} \int \d^{3}y R^{\mu\nu}({t^{\prime}},\Vec{y})h_{\mu\nu}(t-{t^{\prime}},\Vec{x}-\Vec{y}) \\
    &= 
    \int_{-\infty}^{+\infty} \d f \int \d^{2} \Omega_{\hat{k}} \: R^P(f, \hat{k}) h_P(f, \hat k) e^{i 2 \pi f (t - \hat k \cdot \vec{x} / c)},
\end{align}
where {in the second equality} we have used Eq.~\eqref{eq: h for SGWB},  and $R^P(f, \hat{k})$ is the detector response to a sinusoidal plane wave propagating along direction $\hat k$: see \cite{Romano:2016dpx} for its expression for the different GW detectors considered here.
For two interferometers $I,J$ situated at $\Vec{x}_I,\Vec{x}_J$ respectively, and on using Eq.~\eqref{eq: definition of S_h}, it follows that
\begin{align}
    \left\langle h_{I}(t) h_{J}(t^{\prime})\right\rangle &= \int \d^{2} \Omega_{\hat{k}} \int \d^{2} \Omega_{\hat{k}^{\prime}} \int_{-\infty}^{+\infty} \d f \int_{-\infty}^{+\infty} \d f^{\prime} \nonumber \\
    & \quad \quad  \left[\left\langle\tilde{h}_{P}(f, \hat{k}) \tilde{h}_{P^{\prime}}^{*}(f^{\prime}, \hat{k}^{\prime})\right\rangle R_{I}^{P}(f, \hat{k}) R_{J}^{P^{\prime}}(f', \hat{k}^{\prime})^* \times e^{i 2 \pi f\left(t-\frac{\hat{k} \cdot \vec{x}_{I}}{c}\right)} e^{-i 2 \pi f^{\prime}\left(t^{\prime}-\frac{\hat{k}^{\prime} \cdot \vec{x}_{J}}{c}\right)}\right] \\
    & =\int \d^{2} \Omega_{\hat{k}} \int_{-\infty}^{+\infty} \d f \: \frac{1}{2} S_{h}(f) \times \frac{1}{8 \pi} \sum_{P} R_{I}^{P}(f, \hat{k}) R_{J}^{P}(f, \hat{k})^* e^{-i 2 \pi f\frac{ \hat{k} \cdot \vec{\Delta x}_{IJ}}{c}} \times e^{i 2 \pi f\left(t-t^{\prime}\right)} \\
    & = \frac{1}{2} \int_{-\infty}^{+\infty} \d f  \: S_{h}(f) \Gamma_{I J}(f) e^{i 2 \pi f\left(t - t^{\prime}\right)} \label{eq:SgammaIJ}
\end{align}
where $\vec{\Delta x}_{IJ} = \vec{x}_I - \vec{x}_J$, and the factor of $1/2$ is due to the one-sided definition of $S_h$ in Eq.~\eqref{eq: definition of S_h}. 
The overlap reduction function (ORF) $\Gamma_{IJ}(f)$ between the two detectors $I$, $J$ is thus given in terms of their respective response functions by
\begin{equation}
    \label{eq: ORF definition}
    \Gamma_{I J}(f) = \frac{1}{8\pi} \sum_P \int \d^{2} \Omega_{\hat{k}} R_I^P(f, \hat k) R_J^P(f, \hat k)^* e^{- i 2\pi f \frac{\hat k \cdot \vec{\Delta x}_{IJ}}{c}}.
\end{equation}
Note that one often introduces the normalized ORF, defined as  \cite{Allen:1997ad, Thrane_Romano, Romano:2016dpx}
\begin{equation}
    \label{eq: normalized ORF}
    \gamma_{IJ}(f) = \frac{5}{\sin^2(\beta)} \: \Gamma_{IJ}(f),
\end{equation}
where $\beta$ is the opening angle of the interferometers, so that $\gamma_{IJ}(0) = 1$ for two co-located and co-aligned interferometers with the same opening angle $\beta$. Figure \ref{fig: ORF CE}
shows the normalized overlap reduction function between the two Cosmic Explorer detectors with $\beta = \pi/2$, assumed to be located at Hanford and Livingston sites. It vanishes starting from $f\sim 80$ Hz and above, which limits the sensitivity of the CE network to these frequencies. 
\begin{figure}[htbp]
    \centering
    \includegraphics[width=0.55\textwidth]{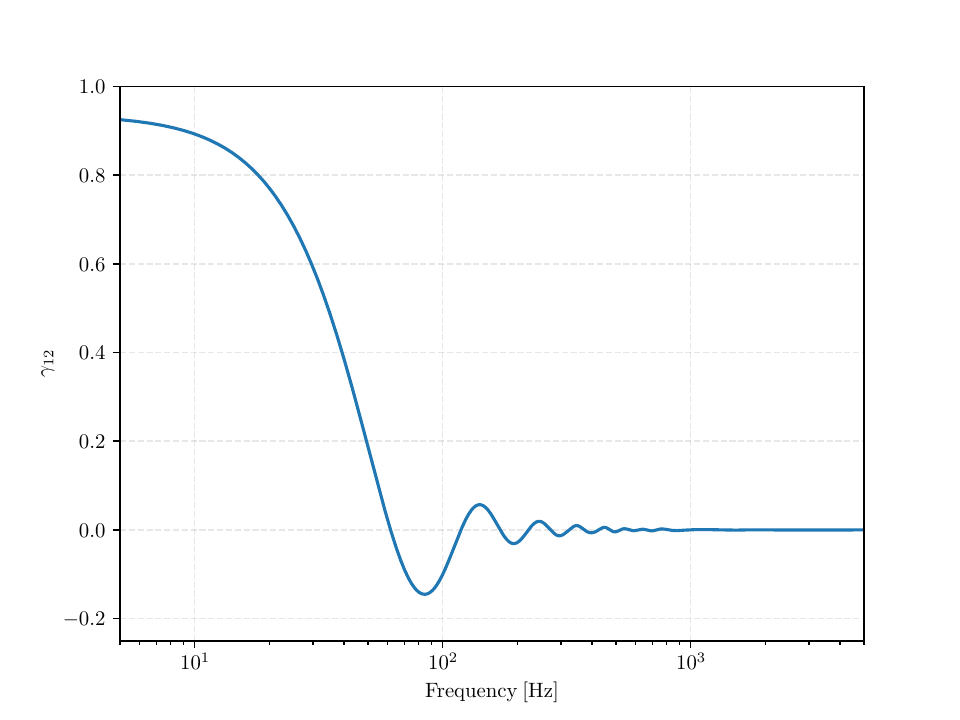}
    \caption{Normalized overlap reduction function between the two Cosmic Explorer detectors assumed to be located at Hanford and Livingston sites.  }
    \label{fig: ORF CE}
\end{figure}

On setting $I=J$ is then straightforward to obtain the response function $\mathcal{R}(f) = \Gamma_{II}(f)$ of one detector leading to
\begin{equation}
    \label{eq: response function def}
    \mathcal{R}(f) = \frac{1}{8\pi} \sum_P \int \d^{2} \Omega_{\hat{k}} \left| R_I^P(f, \hat k) \right|^2.
\end{equation}

Note that for Earth-based interferometers, in the limit $f L / c \ll 1$, where $L$ is the interferometer arm length, the response function is frequency independent as \cite{Romano:2016dpx}
\begin{equation}
    \label{eq: Earth based response}
    R_{I}^{P}(f, \hat{k}) \simeq e_{a b}^{P}(\hat{k}) D_I^{a b},
\end{equation}
where $D_{I}^{a b}=\frac{1}{2}\left(\hat{X}_{I}^{a} \hat{X}_{I}^{b} - \hat{Y}_{I}^{a} \hat{Y}_{I}^{b}\right)$, $\hat X$ and $\hat Y$ being unit vectors pointing in each interferometer arm direction.

\section{\label{app: ET SNR}Einstein Telescope SNR computation}

In this appendix, we derive Eq.~\eqref{eq: S_eff ET}, used to compute the SNR of a cosmological SGWB in ET.

Considering the triangular 10km ET configuration, the method for detecting the SGWB that we adopt would consist in cross-correlating the data streams from the three identical V-shaped interferometers labeled by $I, J \in [1,3]$. As a consequence all overlap reduction function $\Gamma_{IJ}(f)$ ($I \neq J$) are equal thanks to the integration over the sky in Eq.~\eqref{eq: ORF definition}.
Furthermore, since these three interferometers are co-located, they will be affected by common noise sources (such as seismic noise), causing their respective noise realizations to be correlated. Consequently, one cannot use Eq.~\eqref{eq: signal corr with 2 detectors} to isolate the signal component in Eq.~\eqref{eq: correlation 12 expectation value}. 
Instead, we utilize a ``cross-correlation excess" statistic, which presupposes knowledge of the cross-correlated noise,
\begin{equation}
    \label{eq: excess cross corr stat}
    \stat_{IJ} = \int_{-\frac{\Tobs}{2}}^{\frac{\Tobs}{2}} \d t \int_{-\frac{\Tobs}{2}}^{\frac{\Tobs}{2}} \d t' \left[ d_I(t) d_J(t') - \frac{1}{2}N_{IJ}(t' - t)\right]Q(t - t'),
\end{equation}
where 
\begin{equation}
    \label{eq: cross correlated cov matrix}
    \langle n_I(t) n_J(t') \rangle = \frac{1}{2} N_{IJ}(t' - t).
\end{equation}
It is then straightforward to show that
\begin{equation}
    \label{eq: cross corr stat expv}
    \langle \stat_{IJ} \rangle = \frac{\Tobs}{2} \int_{-\infty}^{+\infty} \d f \: \Gamma_{IJ}(f) S_h(f) \tilde Q (f).
\end{equation}
To compute its associated SNR, we proceed by calculating the cross-correlation of this statistic (in the weak-signal regime) between two pairs of interferometers $(I,J)$ and $(K,L)$ (which may be identical),
\begin{align}
    \left\langle \stat_{IJ}  \stat_{KL}\right\rangle &= \int \d t_1...\d t_4 \left\langle 
    \left[ d_I(t_1)d_J(t_2) - \frac{1}{2}N_{IJ}(t_1 - t_2) \right] \left[ d_K(t_3)d_L(t_4) - \frac{1}{2}N_{KL}(t_3 - t_4) \right] \right\rangle Q(t_1-t_2) Q(t_3-t_4)
    \nonumber
    \\
    &= \int \d t_1...\d t_4\left[\left\langle n_{I}(t_1) n_{J}(t_2) n_{K}(t_3) n_{L}(t_4)\right\rangle 
     - \frac{1}{4}N_{IJ}(t_1-t_2) N_{KL}(t_3-t_4)\right] Q(t_1-t_2) Q(t_3-t_4)
        \nonumber\\
     &= \frac{1}{4}\int \d t_1...\d t_4\left[ 
     N_{IK}(t_1-t_3) N_{JL}(t_2-t_4) + N_{IL}(t_1-t_4) N_{JK}(t_2-t_3)\right] Q(t_1-t_2) Q(t_3-t_4),
      \label{eq: cross corr correlation derivation}
\end{align}
leading, in the frequency domain, to
\begin{equation}
    \label{eq: cross corr correlation result}
    \left\langle \stat_{IJ}  \stat_{KL}\right\rangle = \frac{\Tobs}{4} \int_{-\infty}^{+\infty} \d f\left[\tilde{N}_{IK}(f) \tilde{N}_{JL}^{*}(f)+\tilde{N}_{IL}(f) \tilde{N}_{JK}^{*}(f)\right]\left|\tilde{Q}(f)\right|^{2}.
\end{equation}
Here we have introduced the Fourier transform of the auto/cross-correlation functions $N_{IJ}$, which we factorize as
\begin{equation}
    \label{eq: interf pair PSD}
    \tilde{N}_{IJ}(f) = \Gamma^{(n)}_{IJ}(f) P_n(f),
\end{equation}
where $P_n$ is the one-sided noise PSD of one V-interferometer defined in Eq.~\eqref{eq: noise PSD def}, and we have introduced $\Gamma^{(n)}_{IJ}$ that accounts for the correlation of the noise amongst detectors. It satisfies $\Gamma^{(n)}_{II}(f) = 1$, to recover the one-sided detector noise PSD $P_n$. We now follow \cite{Branchesi_2023} and consider that $\Gamma^{(n)}_{IJ}$ is constant across frequency, taking $\Gamma^{(n)}_{IJ}(f) = 0.2$, for $I \neq J$. 

Contrary to the case of uncorrelated detector noises, one cannot simply add up the SNR squared of the individual detector pairs as their associated statistics is clearly correlated from Eq.~\eqref{eq: cross corr correlation result}. Instead, one can derive the way to linearly combine these estimators in order to minimize the variance of the resulting statistics while preserving their expectation value (unbiased estimator). This is a well known problem already derived in \cite{Allen:1997ad, Romano:2016dpx}. To simplify the notation, we denote with $A$ and $B$ interferometer pairs $(I,J)$ and $(K,L)$ (note again that $A$ and $B$ do not need to be distinct) so that $\stat_A = \stat_{IJ}$.
We now define a statistic that linearly combines the one from individual interferometer pairs $A$
\begin{equation}
    \label{eq: general form of cross corr stat}
    \stat_{\rm{cross}} \equiv \sum_{A=1}^{3} \lambda_A \stat_A.
\end{equation}
The $\lambda_A$ are chosen to minimize the variance of the estimator $\langle \stat_{\rm{cross}} \rangle$ and are given by (see Eq.~(6.14) of \cite{Romano:2016dpx})
\begin{equation}
    \label{eq: minimizer of stat}
    \lambda_A = \left[ \sum_{C=1}^3 \sum_{D=1}^3 (C^{-1})_{CD}\right]^{-1} \sum_{B=1}^3 (C^{-1})_{AB},
\end{equation}
where we have introduced the inverse of the statistics covariance matrix, $C_{AB} = \left\langle \stat_A \stat_B  \right\rangle$ which, by symmetry, in the weak-signal regime, takes the form
\begin{equation}
    \label{eq: stats_cov_matrix_in_WSR}
    C =  
    \begin{bmatrix}
    \mathcal{A} & \mathcal{T} & \mathcal{T} \\
    \mathcal{T} & \mathcal{A} & \mathcal{T} \\
    \mathcal{T} & \mathcal{T} & \mathcal{A}
    \end{bmatrix}.
\end{equation}

We can then use the fact that $C$ is symmetric (and so $C^{-1}$) to see from Eq.~\eqref{eq: minimizer of stat} that $\lambda_A = \frac{1}{3}$ for all detector pairs $A$. The estimator  verifies $\left\langle \stat_{\rm{cross}} \right\rangle = \left\langle \stat_A \right\rangle$, with its explicit form given in Eq.~\eqref{eq: cross corr stat expv}. Its variance is then
\begin{equation}
    \label{eq: variance total cross estimator}
    \left\langle \stat_{\rm{cross}}^2 \right\rangle = \frac{\mathcal{A} + 2\mathcal{T}}{3}.
\end{equation}
Thanks to our assumption that the cross-correlated noise is an order of magnitude smaller than the auto-correlation power (namely $\Gamma_{IJ}^{(n)} = 0.2$ for $I \neq J$), we only keep the linear terms in $\Gamma_{IJ}^{(n)}$ from Eq.~\eqref{eq: cross corr correlation result} leading to,
\begin{equation}
    \label{eq: mathcal A value of C}
    \mathcal{A} \simeq \frac{\Tobs}{4} \int_{-\infty}^{+\infty} \d f \: P_n^2(f) \tilde{Q}(f)^{2},
\end{equation}
and
\begin{equation}
    \label{eq: mathcal T value of C}
    \mathcal{T} \simeq \frac{\Tobs}{4} \int_{-\infty}^{+\infty} \d f \:  \Gamma^{(n)}_{12} P_n^2(f) \tilde{Q}(f)^{2}.
\end{equation}

The SNR can then be derived (in the weak signal regime) as
\begin{equation}
    \label{eq: SNR def for stat}
    \mathrm{SNR} \simeq \frac{\left\langle \stat_{\rm{cross}} \right\rangle}{\sqrt{\left\langle \stat_{\rm{cross}}^2 \right\rangle}} = \sqrt{N \Tobs} \frac{\int_{-\infty}^{+\infty} \d f \: \Gamma_{12} S_h(f) \Tilde{Q}(f)}{\sqrt{\int_{-\infty}^{+\infty} \d f \: \left[ 1 + (N-1)\Gamma^{(n)}_{12}\right] P_n^2(f)\Tilde{Q}^2(f)}},
\end{equation}
where we write the SNR for a general number of $N$ equivalent colocated interferometers. For ET, $N=3$. Using the standard method from \cite{Allen:1997ad, Romano:2016dpx}, one can choose the filter function $\Tilde{Q}$ to maximize the SNR. One finds
\begin{equation}
    \label{eq: maximizing filter}
    \tilde{Q}(f) = \Lambda \times \frac{\Gamma_{12} S_h(f)}{\left[ 1 + (N-1)\Gamma_{12}^{(n)} \right]P_n^2(f)},
\end{equation}
where $\Lambda$ is an arbitrary normalization factor. Substituting this filter into Eq.~\eqref{eq: SNR def for stat}, one finally finds
\begin{equation}
    \label{eq: final ET SNR}
    \mathrm{SNR} = \sqrt{2 \Tobs \int_0^{+\infty} \d f \frac{N \Gamma_{12}^2 S_h^2(f)}{\left[ 
1 + (N-1)\Gamma^{(n)}_{12} \right]P_n^2(f)}},
\end{equation}
leading to the effective strain noise PSD
\begin{equation}
    \label{eq: general S_eff for ET}
    S_{\mathrm{eff}} = \sqrt{\frac{1 + (N-1)\Gamma_{12}^{(n)}}{N \Gamma_{12}^2}} P_n
\end{equation}
written in Eq.~\eqref{eq: S_eff ET} for the case of a triangular ET detector with $N=3$ and $P_n = P_{\rm{ET-L}} / \sin^2(\pi / 3)$. This should be compared with Eq.~(A.7) of \cite{Branchesi_2023}.

\bibliography{ref}

\end{document}